**Ultraviolet Radiation from Evolved Stellar Populations:**

**II. The Ultraviolet Upturn Phenomenon in Elliptical Galaxies**

Ben Dorman, Robert W. O'Connell, and Robert T. Rood

Dept. of Astronomy
University of Virginia, P.O. Box 3818, Charlottesville, VA 22903-0818, U.S.A.

*Received* _______

# *Abstract*

We present an analysis of the far-ultraviolet upturn phenomenon (UVX) observed in elliptical galaxies and spiral galaxy bulges. Our premise is that the UV radiation from these systems emanates primarily from extreme horizontal branch (EHB) stars and their progeny. Such objects have Zero Age Horizontal Branch envelope masses $M_{env}^0 \lesssim 0.05 M_\odot$. Local examples of EHB stars exist in some globular clusters and in the Galactic disk field and serve both as a guide and constraint. We re-derive the broad-band UV colors $1500 - V$ and $2500 - V$ for globular clusters and elliptical galaxies from the available satellite data and investigate color-color and color-line strength correlations. There are several important distinctions between clusters and galaxies. They do not occupy a single $Mg_2$-color sequence. Clusters can be bluer than any galaxy in $15 - V$ and $25 - V$, implying larger hot star populations, but galaxies are significantly bluer than clusters in $15 - 25$ at a given $15 - V$. We attribute this primarily to the effect of metal abundance on the mid-UV (2500 Å) light. It also implies that the UVX in galaxies is not produced by metal poor subpopulations similar to the clusters.

We develop a simple spectral synthesis formulation for all phases of single star evolution from the ZAMS to the white dwarf cooling track that requires only one or two parameters for each choice of age and abundance. We provide the ingredients necessary for constructing models with arbitrary HB morphologies in the age range $2 < t < 20$ Gyr and for 6 metallicities in the range $-2.26 < $ [Fe/H] $< 0.58$; we also consider the effect of enhanced $Y$ in metal rich models. UV properties of the models are predicted using the Kurucz (1991) atmospheres. The maximum lifetime UV output is produced by EHB stars with $M_{env}^0 \sim 0.02 M_\odot$, and can be up to 30 times higher than for post-asymptotic-giant-branch (P-AGB) stars. The ultraviolet output of old populations is governed primarily by the distribution of $M_{env}^0$, $P(M_{env}^0)$, on the ZAHB. The UV output is not very sensitive to [Fe/H] or to $Y$, but it can change very rapidly with $M_{env}^0$. Thus, it is extremely sensitive to the precise nature of giant branch mass loss. Because this process is not well understood physically, we choose to leave mass loss as an implicit free parameter. Our models use simple descriptions of $P(M_{env}^0)$ to bracket the colors produced from any real distribution of stars.

Our models accurately predict the range of UV colors observed for the globular clusters, given known constraints on their age, abundances, and HB morphologies. Clusters with "blue HB" morphologies do not require the hotter EHB stars to explain their UV colors, although a small EHB population is consistent with our models. The largest known population of these stars in a cluster, as a fraction of the total HB, is $\sim 20\%$ in $\omega$ Cen. For [Fe/H] $\gtrsim -0.5$, however, blue HB stars will be rare. As a consequence, we find that models with [Fe/H] $\gtrsim 0$ which do not contain EHB stars cannot reproduce the colors of most of the galaxies. However, only small EHB fractions are required: $\lesssim 5\%$ for the bulk of the E galaxies and $\sim 20\%$ for those with the strongest UVX. These results are independent of the assumed [Fe/H]. The EHB fraction required for most galaxies is comparable to the fraction of hot subdwarfs in the Galactic disk. Most of these are EHB stars, and their existence considerably strengthens the case for EHB populations as the source of elliptical galaxy UV light. The models also predict that the fraction of the far-UV light from P-AGB stars, which are spatially resolvable in nearby galaxies, is $\sim 70\%$ and $\sim 20\%$ for moderate UVX and strong UVX systems, respectively.

We find that $25 - V$, but not $15 - V$, is sensitive to the age and abundance, though these cannot always be cleanly distinguished. The galaxy colors place strong limits of $\langle$[Fe/H]$\rangle > -0.5$ and $< 15\%$ on the contribution of globular cluster-type populations to the V light. Galaxy colors are consistent with solar-abundance models with ages in the range $6 - 14$ Gyr. However, the $25 - V$ colors of the galaxies other than the strong UVX systems are too blue to be consistent with [Fe/H] $> 0.2$ for any age. This may be additional evidence that [Mg/Fe] $> 0$ in elliptical galaxies. UV colors for M32 are consistent with the solar abundance, intermediate age ($4 - 6$ Gyr) population inferred from optical/IR observations.

We discuss several implications of the observations and the models, including the the question of light metal vs. iron peak enhancements in galaxies, whether the UV color-Mg$_2$ correlation is continuous or discrete, effects of helium abundance on the UVX, and the key question of whether red giant branch mass loss can be large enough to produce the necessary EHB population in the strong UVX galaxies.

*Subject headings:* galaxies: stellar content — clusters: globular — stars: AGB and post-AGB — stars: horizontal branch — stars: evolution —stars: Population II — ultraviolet: galaxies— ultraviolet: globular clusters

## 1    Introduction

The ultraviolet upturn ("UVX") observed in the spectra of elliptical galaxies and spiral bulges has emerged as an important issue in our understanding of stellar populations and thus the use of elliptical galaxies as tracers of the evolution of the universe. It was first recognized as an unanticipated rise in the spectral energy distribution for $\lambda < 2000$ Å in observations of the M31 bulge from the OAO-2 satellite (Code 1969; see also Code & Welch 1979). Soon after the initial discovery, Hills (1971) pointed out that the shape of this UV upturn was consistent with a hot thermal spectrum, and he suggested as the probable origin post asymptotic giant branch (P-AGB) stars, during their high luminosity ($\log L/L_\odot \sim 3 - 4$) transit to become white dwarfs. The UVX has since been found to be present in all elliptical galaxies, and to vary in amplitude by more than an order of magnitude. It has attracted considerable theoretical attention in recent years (see Greggio & Renzini 1990 [GR] and references therein), seeking to identify the hot component responsible for the UV radiation.

Candidate UV-bright stellar objects for this population include massive stars, very low-mass stars in advanced stages of stellar evolution, and low-mass products of binary interactions (see GR; O'Connell 1993). We review the observational constraints on these possibilities in the next two sections. Currently, the most likely candidates seem to be the Extreme Horizontal Branch (EHB) stars and their post-HB descendants. These stars have very thin hydrogen envelopes, and do not reach the thermally-pulsing stage of the AGB. In the first paper of this series (Dorman, Rood, & O'Connell 1993, hereafter Paper I) we presented a large grid of evolutionary tracks of these EHB stars with various metallicities and envelope masses. Here, we use this grid to model the observed UV colors of galaxies and globular clusters and to investigate the hypothesis that EHB stars and their progeny are indeed responsible for the UV upturn phenomenon.

Our grid of evolutionary tracks is defined by the composition parameters $Z$ and $Y$, the core mass $M_c^0$, and by $M_{env}^0$, the envelope mass of the object on the Zero Age Horizontal Branch (ZAHB). Of these, the parameter with the strongest influence on the integrated lifetime UV output of a star is $M_{env}^0$. In a real stellar population, the distribution function of $M_{env}^0$ will be determined by mass loss on the red giant branch (RGB). The observed HB morphologies in the Galactic globular clusters imply that a typical star loses $\sim 0.10 - 0.30\, M_\odot$ before arriving on the ZAHB. There is significant dispersion in the mass lost from the individual stars of a given cluster (Iben & Rood 1970).

Unfortunately, the mass-loss process is not well understood. Modelers usually assume a functional dependence based on the Reimers (1975) formula, which is derived from dimensional analysis rather than a well-grounded physical theory (see §8). Since the physics here is so uncertain, we have chosen to leave mass loss as an implicit free parameter in our models. In this respect our approach is more circumspect than studies which adopt explicit descriptions for stellar mass loss (e.g., Bertelli, Chiosi, & Bertola 1989; Barbaro & Olivi 1989; Bruzual & Charlot 1993; Magris & Bruzual 1993; Bressan, Chiosi, & Fagotto 1994). These studies also attempt to predict the UV/optical spectrum of galaxies in some detail. It is difficult in such complex models to isolate the effects of age and metallicity from those of mass loss and to evaluate the dependence of the results on the assumptions. Here, we prefer to consider only broad-band UV colors and to investigate their variation with individual model ingredients and adopted parameters. We concentrate on addressing the following question: what distributions of $M_{env}^0$ are consistent with the observed UV fluxes of galaxies, given the other available constraints on their ages and abundances?

To synthesize the UV colors of galaxies, we use the model grid of Paper I for the HB, EHB, and AGB-manquè phases of low-mass stellar evolution together with Schönberner's tracks for P-AGB stars and our own isochrones for the pre-He flash phases. We compute the integrated energy output of all phases at UV and optical wavelengths. As a proving ground for our models, we apply them to the Galactic globular clusters. We find good agreement between the observations and synthetic models (for appropriate choices of age, abundance, and HB morphology) in the UV colors $m_\lambda(1500\,\text{Å}) - V$ and $m_\lambda(2500\,\text{Å}) - V$, as defined in §4. We will refer to these UV colors for the remainder of this paper as $15 - V$ and $25 - V$. Turning to the galaxies, we find the fraction and type of the post-RGB population necessary to explain the range of $15 - V$ colors observed in the elliptical galaxies, independent of assumptions about RGB mass loss. Our principal conclusion is that, for $Z \gtrsim Z_\odot$, enhanced mass loss in a small fraction ($\lesssim 0.20$) of the stars arriving on the ZAHB can produce sufficient far-UV flux to explain even the strongest UV upturns. The models satisfy all the available constraints provided by the data and thus make a strong case for the hypothesis that EHB stars and their progeny are indeed responsible for the UV upturn. We also show the importance of correctly accounting for the mid-UV flux from pre-HB phases of evolution and find that the mid-UV color $25 - V$ can be used to place bounds on the ages and metallicities of galaxies.

The paper falls into two main sections. In §2, 3, and 4, we discuss the available data on UV radiation from old stellar populations (clusters and galaxies). §5, 6 and 7 describe our procedure for building models, present numerical results, and compare them with the observations. In §8 we discuss the known limitations of UV synthesis modeling, the effects of possible non-solar abundance ratios, and the implications of our results for stellar mass loss in galaxies.

## 2    The UVX Phenomenon in Galaxies

### 2.1    Observational Properties

In this section, we briefly describe the observational characteristics of the UVX phenomenon. More details can be found in several recent reviews (O'Connell 1993; GR; Burstein *et al.* 1988, hereafter B3FL). The most comprehensive tabulations of data are in B3FL and Longo *et al.* (1989).

The salient observed features of the UVX phenomenon are these:

(1) *Incidence*— It is found in the nuclear regions of almost all normal Es, S0s, and spiral bulges observed to date. By 'normal', we mean those which are neither AGN's nor show obvious signs of recent star formation.

(2) *Strength*— The strength of the UVX, as measured by the color $15 - V$, varies by about an order of magnitude, or 2.5 mag (B3FL, Longo *et al.* 1989, and Figure 1). This degree of variation is much greater than that found in optical and infrared broad-band colors. In the most extreme cases the UVX is estimated to contribute ~2–3% of the bolometric luminosity (GR, B3FL, Renzini & Buzzoni 1986).

(3) *Abundance Correlation*— As first pointed out by Faber (1983), the magnitude of the UV upturn is positively correlated with the absorption line index $Mg_2$ (*cf.* Figure 1), i.e., it is stronger in more strong-lined systems. This is reversed in sense from the well-known dependences of $(U - B)$ or $(B - V)$ colors on metal abundance in old populations. The $Mg_2$ – UV correlation was confirmed by the study of B3FL using IUE observations of the nuclear regions of M31 and bright elliptical galaxies. The UV flux also shows a correlation with the velocity dispersion of the galaxies. Since, however, the correlation with $Mg_2$ has less scatter, it is held to indicate a causal relationship between abundance, at least of the $\alpha$-elements, and the strength of the UV upturn. Nevertheless, the scatter in the UVX at a given $Mg_2$ index is still appreciable, especially among the strongest-lined systems (B3FL). The relationship between the abundances of magnesium and other elements is discussed in §8.

(4) *Spatial Structure/Resolution* —Longo *et al.* (1989) noted that the ellipticals with stronger UVX components tend to be those with 'boxy', rather than 'disky', isophotes. These boxy types also have the strongest $Mg_2$ features. Other studies have indicated that boxy ellipticals are more frequently radio sources and X-ray bright (Bender et al. 1989). This result and the observed scatter with $Mg_2$ may indicate the influence on the UVX of parameters other than metal abundance (age, helium abundance, [$\alpha$/Fe], etc.) or that more than one process is involved.

IUE observations of brighter sources (e.g. Bertola *et al.* 1980; Oke, Bertola, & Capaccoli 1981; Welch 1982; O'Connell, Thuan, & Puschell 1986) showed that the UVX is spatially extended on scales of $\sim 10\,''$. More recently, vacuum UV images of four objects (M31, M32, M 81, and NGC 1399) have been obtained with the Ultraviolet Imaging Telesope (UIT) on the Astro-1 mission (O'Connell *et al.* 1992), which confirm the radial structure of the UVX. These ellipticals and spiral bulges have smooth, de Vaucouleurs-like UV profiles to below the sky background level at $r \gtrsim 80\,''$. None of the clumpiness normally associated with recent massive star formation is detected. In the M31 bulge, the image excludes the presence of individual main sequence stars hotter than B1 V.

M31, M81, and NGC 1399 show remarkably strong UV color gradients with amplitudes up to $\sim$1 mag in $15 - 25$ for $r \lesssim 80\,''$. There was preliminary evidence for UV color gradients in the IUE spectra of M31 (Deharveng *et al.* 1982 and Welch 1982) and in Virgo ellipticals observed with a sounding rocket experiment (Kodaira *et al.* 1990). The UV colors become strongly redder outward. If metal abundances decline outward, then the sense of these gradients is the same as the overall UVX/Mg$_2$ correlation.

The dwarf elliptical M32 has the reddest $15 - V$ in B3FL and a small UV color gradient. Its distinctive features could be related to the fact that M32 appears to contain a younger ($\sim 5$ Gyr) population than the other objects (see §7). If so, this again suggests the influence of parameters other than abundance on the UVX.

The more luminous UVX sources have apparently been resolved on recent HST Faint Object Camera exposures obtained of the M31 bulge. In a preliminary analysis of these data, King *et al.* (1992) detect $\sim 150$ resolved objects, which reach $m_\lambda(1750\,\text{Å}) \sim 20$ over the central $44\,''$. These are likely to be P-AGB stars. No massive OB stars appear to be in the field: Palomar Schmidt observations of the surrounding area show no concentration of B stars toward the center, whereas the diffuse UV radiation in the HST image does show a gradient in this sense. The percentage of the absolute flux unaccounted for by the resolved objects has some uncertainty because of the large red-leak of the HST UV filters.

(5) *Spectral Shape and Features* —IUE spectroscopy (e.g., B3FL) indicated that the far-UV spectral slope of the UVX is roughly constant from object to object and corresponds to $T_{\text{eff}} \gtrsim 20,000$ K. As pointed out by B3FL, such steep slopes are not found in systems known to have massive star populations. The UV spectra of star forming galaxies are flatter, apparently because of a wider range of temperature than found in the UVX sources. Even for the brightest sources (e.g. M31; see Welch 1982), however, the quality of the IUE spectra are not good enough to detect absorption features with confidence.

Far-UV (900–1800 Å) spectra of NGC 1399 and M31 were obtained by the Hopkins Ultraviolet Telescope (HUT) during the Astro-1 mission (Ferguson *et al.* 1991). These have better S/N than IUE spectra because of the larger HUT entrance aperture. NGC 1399 is the galaxy with the strongest UV upturn ($15 - V \sim 2$) found to date. Its far-UV spectral shape and the weak C IV $\lambda$1550 absorption feature are inconsistent with massive metal-rich OB stars as the source of the UVX. The turnover of the spectrum at $\lambda < 1200$ Å, in a region not accessible to IUE, indicates that $T_{\text{eff}} \lesssim 25000$ K.

The HUT spectrum of M31 (Ferguson & Davidsen 1993) differs significantly from that of NGC 1399. Regardless of the extinction assumed, the two energy distributions do not match. For the adopted extinction, M31 is hotter for $\lambda < 1200$ Å but cooler for longer wavelengths. Weak lines of C IV, C II, and Si II are detected. The lines appear too weak to originate in a massive OB star population with high metallicity; they could be consistent with either metal-poor stars or with post-HB phases which have lost their outer envelopes and exposed a carbon-depleted layer. Ferguson and Davidsen argue that at least two distinct types of hot, low mass stars produce the UV spectral energy distribution of M31.

## 2.2    Theoretical Considerations

Current observational evidence thus disfavors an explanation for the UVX in terms of massive stars, albeit without excluding it in all systems. The recent literature, as well as the early suggestion by Hills (1971), has focussed on candidates from old stellar populations, the two major categories being the products of binary evolution and single stars in advanced stages of evolution. In the former class (see GR for a detailed discussion) are (1) the post-red giant products of close binary systems in which extreme mass transfer prevents the red giant component from reaching the helium core flash; (2) mass transfer binaries in which one star does reach the helium flash (Mengel, Norris, & Gross 1976); (3) white dwarfs burning accreted hydrogen in their outer layers; and (4) mergers between low mass white dwarfs (*cf.* Iben 1990) forming helium stars. Of these, the first is unable to produce enough UV radiation to explain the strongest UV upturns. The second possibility appears at present to require a fine-tuning of binary orbital parameters that would make the formation of a large number of objects implausible. It must be said, though, that this scenario has not been exhaustively explored to date.

In the single star category are a variety of candidates. Most widely discussed have been 'normal' P-AGB stars, i.e., stars with exhausted helium cores evolving to the white dwarf cooling track after the loss of their envelopes at the tip of the AGB. Their UV output represents a lower bound to the flux from any evolved stellar population, because they are inevitable products of stellar evolution. If the remaining envelope expelled by these stars is $\gtrsim 0.1 - 0.2 M_\odot$ and removed sufficiently rapidly, they become central stars of planetary nebulæ. Owing to the strong decrease of their lifespan with increasing core mass, the total UV energy emitted by an ensemble of P-AGB stars will be significantly lower if brighter, more massive objects are produced. This would naturally be the case for intermediate-age rather than old populations. The characteristic mass of such objects is, however, determined by the mass loss processes on the AGB and RGB.

Brocato *et al.* (1990) and GR showed that even the least massive P-AGB stars modeled at that time (Schönberner 1983) were too short-lived to produce sufficient flux to account for the strongest UVX galaxies. They suggested that stars with still smaller cores on the AGB — having suffered greater degrees of mass loss on the first red giant branch — were possible candidates. They observed that stars with sufficiently small $M_{env}^0$ on the ZAHB will leave the AGB before reaching the thermally pulsing stage. These evolve instead directly to the white dwarf cooling track, and are termed post-Early AGB (P-EAGB) stars. However, theoretical models indicate that only a very small range of masses evolves in this way (see Paper I). GR also emphasized the importance of stars that undergo slightly more mass loss on the RGB – referred to as AGB-manqué stars – which never reach the AGB at all but spend their entire helium shell-burning lifetimes ( $\gtrsim 20$ Myr) at $T_{eff} \gtrsim 25000$ K. We refer to the HB progenitors of both of these types of later evolutionary behavior as "extreme horizontal-branch" or EHB stars, which we defined in Paper I to mean objects that do not reach the thermally pulsing stage while still on the AGB. These have ZAHB temperatures $\gtrsim 16000$ K. Because the term "blue" HB (BHB) is often used to refer to all HB stars hotter than the RR Lyrae strip, we refer to stars that lie blueward of the instability strip but still reach the AGB after core helium exhaustion as "intermediate blue HB" (IBHB) stars. These stars live near the ZAHB for $\sim 100$ Myr with $8000 \lesssim T_{eff} \lesssim 16000$ $K$. Castellani & Tornambè (1991) presented evolutionary sequences for examples of both P-EAGB and AGB-manqué stars, demonstrating that their lifetime-integrated UV energy output can be more than an order of magnitude greater than that from Schönberner's post-AGB models. A fuller review of the literature on the evolution of hot, low-mass stellar models to 1993 appears in Paper I. More recent calculations have been performed by Castellani *et al.* (1994) and Bressan *et al.* (1994).

In the recent HST observations of M31 by King *et al.* (1992), the resolved sources are almost certainly P-AGB stars; several, in fact, appear to be surrounded by planetary nebulæ. The resolved sources account for $\sim 18\%$ of the total FUV light. For NGC 1399, the HUT observations of Ferguson and Davidsen (1993) imply that the characteristic temperature of the UV radiation is too cool to be consistent with the classic P-AGB models. They argue that EHB, AGB-manqué, and P-EAGB stars are the most likely sources for the UVX. As for M31, they favor a model in which 65% of the $\lambda1400$ light comes from hot P-AGB stars, with the remainder originating from the lower mass objects. Modeling details aside, it appears that the UVX is a composite population, with a mixture of HB and various post-HB types that varies from system to system.

In Paper I we presented an extensive grid of model evolutionary sequences for HB and post-HB evolution for metallicities between [Fe/H] = $-2.26$ and +0.58. We noted that all models with $M_{\rm env}^0 \lesssim 0.04\,M_\odot$, irrespective of abundance, are copious UV emitters, as first pointed out by Ciardullo & Demarque (1978). However in order to reach such small envelope masses in metal-rich populations, for which RGB tip masses are larger at fixed age, RGB mass loss must be enhanced over what is inferred from globular clusters. As another possibility, stars with strongly enhanced envelope helium abundance are less massive at the same age than their lower $Y$ counterparts (GR; Horch, Demarque, & Pinnsoneault 1992). In that case, the UVX stars could be explained with similar mass loss to that inferred for the globular clusters. We consider these various possibilities for the production of EHB stars in the concluding section of this paper.

## 3 Template Populations for the UVX

### 3.1 Metal Poor Globular Clusters

As stated earlier, the goal of this study is to find the fraction of EHB stars that explain the galaxy observations. As a starting point, in this section we examine populations of EHB stars in our own Galaxy in the hope of gaining some insight into their formation as a function of composition and other factors.

The globular clusters have long served as population templates, although their metallicity range barely overlaps that of the brighter galaxies. For our purposes, they provide samples of homogeneous, coeval, old stellar populations which can be used to test synthetic UV colors. A sample of UV-bright P-AGB stars has been identified with ground-based telescopes in metal-poor Galactic globular clusters (see de Boer 1987; Zinn, Newell, & Gibson 1972). Clusters have also been studied in the UV from the ANS satellite (van Albada, deBoer and Dickens 1981, hereafter ABD), OAO-2 (Welch and Code 1980, hereafter WC80), IUE (for a review, see Castellani & Cassatella 1987), and the Ultraviolet Imaging Telescope (Hill *et al.* 1992; Landsman *et al.* 1992; Parise *et al.* 1994; Whitney *et al.* 1994). Spatially-resolved UV observations, mainly with UIT, have isolated a number of individual EHB, AGB-manquè, and related stars in M79, NGC 1851, and $\omega$ Cen; the largest sample, not surprisingly since it is also the most luminous object, is in $\omega$ Cen (Whitney *et al.* 1994).

The far-UV flux from globulars is correlated with the HB morphology (ABD; WC80). The spectral energy distributions (SED) of all of the clusters have much shallower slopes for $\lambda \lesssim 2000$ Å than the galaxies (e.g. O'Connell 1993). The UV flux from clusters thus emanates largely from IBHB stars, i.e., the "normal" blue HB stars with $M_{\rm env}^0$ large enough to reach the AGB. The hardest far-UV spectral energy distributions among the clusters are observed in those with "extended blue HB tails" (*cf.* Fusi Pecci *et al.* 1993), such as NGC 6752, M79, M13, and $\omega$ Cen. Unfortunately, ground-based surveys are not often complete down to the blue extreme of the HB, which has $V \gtrsim V_{TO}$. The relative numbers of RGB and HB stars in these "blue tail" clusters do not imply that a large number of very HB stars have so far escaped detection. In $\omega$ Cen the number of EHB stars counted on the UIT image is about 20%, while for M79 the number is $\lesssim 10\%$. The recent HST observations of the core of M15 (de Marchi & Paresce 1994) also find only a small sample of EHB objects. Taken with the shape of the far-UV SED, these observational constraints suggest that the globular cluster UV flux is usually not dominated by EHB stars.

As is well known, the HB morphology in globular clusters is determined by at least two factors, of which only the first — metallicity[1] — is securely identified. The most metal rich clusters (e.g., 47 Tuc, NGC 6637) have very red HBs and radiate weakly in the far-UV. The so-called "second parameter effect" (van den Bergh 1967) refers to the variations in the HB morphology at fixed [Fe/H]. The most famous manifestations of this phenomenon, e.g., the contrast between the HBs of NGC 288 and NGC 362 (see e.g., Bolte 1989; VandenBerg, Bolte, & Stetson 1990) are taken as evidence that in some cases the difference in HB morphology is due to age. The older cluster (NGC 288) has a smaller turnoff mass, and thus if similar mass loss processes operate on the RGB the result will be a bluer HB. The cause of the difference in stellar distribution between clusters with "blue" and "very blue" HB morphology, as exemplified by the pair NGC 288 and M13, is not so easy to interpret; perhaps it is confusing to refer to these variations also as an instance of the second parameter phenomenon. That age is not the only factor contributing to the presence of extreme blue tails is clear from HUT observations of the cluster NGC 1904 (M79). The HUT far-UV spectrum is inconsistent with the gaussian-type distribution of stellar masses on the HB (Dixon *et al.* 1994) that is almost invariably used (e.g., Lee, Demarque, & Zinn 1990, 1994; Rood 1973) to produce synthetic models that fit the morphology of "normal" cluster horizontal branches well.

Clusters may be subject to phenomena that are unlikely to affect the stellar populations of galaxies. For example, a study by Buonanno, Corsi, & Fusi Pecci (1985) suggested a link between the HB morphology and the structural and dynamical properties of the cluster (see Fusi Pecci *et al.* 1993; van den Bergh 1993; Djorgovski *et al.* 1991). In particular, the presence of extended blue HB tails is found to be more prevalent in the clusters with central cusps in their surface brightness profiles. These systems either have high central concentration, or thought to be "post core collapse" (PCC) objects, in a phase after a violent collapse in which the central density oscillates (for a recent explicit calculation, see Drukier 1993). Djorgovski & Piotto (1992) conducted an IUE study of clusters classified as EB (Extremely Blue) on the basis of the ANS data. Their results show that, at least for the central $20''$ (the IUE aperture), the PCC clusters have radial gradients in their UV radiation. However the cluster NGC 6093 (M80), which satisfies a King model, also has an extended HB tail and thus provides a counterexample to the general rule. Firm evidence for a link between dynamics and the presence of a blue HB tail would imply that stellar encounters can be a major factor in shaping the population of stars in late stages of evolution. However, the centers of galaxies have lower density than found in either of these types of clusters, and are less likely to have populations affected by dynamical evolution.

## 3.2    Metal Rich Clusters

Most metal-rich clusters are faint in the UV because their HB morphology is dominated by cool stars close to the RGB, in accord with theoretical expectations that blue HB stars are rarely produced at higher abundances. The notion that mass loss might increase enough with metallicity to produce some EHB stars suggests an investigation of metal-rich Galactic globulars. Rich, Minniti, & Liebert (1993, hereafter RML) gathered data from IUE for disk globular clusters. Despite observational difficulties arising from the high reddening toward many of these objects ($0.2 < E(B - V) < 0.6$), implying 2–5 mag of extinction in the UV), they obtained positive detections of UV radiation from the clusters NGC 6388 ([Fe/H] $= -0.48$), NGC 6441 ([Fe/H] $= -0.24$) and NGC 6637 ([Fe/H] $= -0.59$). These were also observed, and the first two detected, by ANS. None of these clusters are known to harbor special objects such as X-ray binaries that could also contribute to the UV flux. The far-UV flux received in all cases is too small to emanate from even a single post-AGB star, and various model fits conducted by RML imply the presence of small numbers of EHB stars. Liebert, Saffer, & Green (1994) also recently discovered four EHB objects which are almost certainly members of the old open cluster NGC 6791 ([Fe/H] $\sim$ +0.2).

---

[1]    Theoretical investigations (Dorman, Lee, & VandenBerg 1991; Dorman 1992a) show that at low metallicity enhanced CNO abundance leads to red HB morphologies, while at high metallicity both CNO and iron-peak elements lead to red HB stars. Thus the "first parameter" really arises from two separate effects.

In summary, most metal-poor clusters contain few "true" EHB stars, and their far-UV flux can be accounted for by the observed P-AGB type objects and large populations of "blue HB" stars. $\omega$ Cen is the only known case where the fraction of the HB population in EHB stars is as large as 10%; NGC 6752 may provide another example, based on its optical CM diagram. There is some evidence for the production of very hot HB stars in the relatively metal-rich globulars. Unfortunately, statistics on such relatively small populations will never be good. A better sample of such objects is available among the field stars of our Galactic disk, to which we now turn.

## 3.3    Hot Subdwarfs in the Field

More than 1200 hot subluminous stars have so far been discovered in the Galactic field population. The majority were found by the Palomar-Green survey (Green, Schmidt & Liebert 1986) of UV-excess objects. The sdB stars, together with somewhat hotter objects termed either sdOB or sdB-O, constitute 40% of the catalog. They occupy a narrow range of surface gravities with log $g \sim 5-5.5$, with $20000 < T_{\rm eff} < 40000$ K. The stars denoted as sdO —13% of the sample— are distinguished by their Balmer line profiles, and have surface temperatures extending to about 65000 K. They have a large range in surface gravity indicative of a range in luminosity. 20% of the remainder of the catalog consists of hot white dwarfs, with the remainder consisting of cataclysmic variables and extragalactic objects. The bulk of the hot objects in the catalog are thus evolved stars. Saffer (1991) finds that the kinematics of both sdB and sdO populations implies that they belong to the thin disk population. He also finds that the observed frequency of the subdwarfs corresponds to about 2% of the post-RGB population. Their disk membership makes it likely that they originate from a metal-rich population; their surface chemistry is known to be affected by mass loss, diffusion and radiative levitation, making their original abundance impossible to determine.

The observational characteristics of hot subluminous stars are reviewed by Heber (1992) and Saffer *et al.* (1994). The sdB/sdOB stars have surface temperatures and gravities that coincide with the properties of model EHB stars, as first demonstrated by Greenstein & Sargent (1974). The narrowness of the range in surface gravities is consistent with the notion that they are drawn from a single (post-helium flash) mass population of about 0.5 $M_\odot$. Hence *they almost certainly provide a large sample of EHB stars in our own Galaxy, making plausible the idea that these objects are a normal constituent of old stellar populations*.

As for the sdO stars, there are several subclasses with different temperatures, gravities and surface abundances; these may arise from different progenitors. However, their ranges of surface temperatures and gravities are consistent with those of the AGB-manqué progeny of the EHB. One can compare relative numbers of each type of star with the theoretical lifetimes of EHB model tracks. In the PG survey, the approximate number of sdB and sdOB candidate stars is 800. The numbers of sdO stars of all subclasses is about 200. A crude comparison of the relative lifetimes of EHB stars and their AGB-manqué progeny predicts a ratio of about 1:5. When account is taken of contamination of the sample by misclassifications and by alternative origins (such as binary mergers), the agreement is reasonable. We therefore suggest (see also Heber 1992) that *the bulk of the sdO stars are evolved from EHB stars*.

The majority of the hot subdwarfs appear to be single stars that have suffered high rates of mass loss earlier in their lifetimes. Considerable discussion has addressed alternative binary-star origins for the subdwarfs (e.g. Mengel *et al.* 1976, Green *et al.* 1986, Iben 1990, Saffer 1991, Heber 1992 & references therein). These mechanisms can plausibly generate a subset of the hot subdwarfs. It is unclear whether such mechanisms could be responsible for the correlations seen in the galaxy UVX phenomenon.

To summarize, EHB stars and their AGB-manqué progeny appear in globular clusters and in the Galactic field. Although the metallicity of the field subdwarfs is uncertain, they belong to the disk population, which is largely near solar abundance and of intermediate ($\lesssim$ 10 Gyr) age. While the subdwarfs do not provide conclusive evidence that other galaxies produce EHB stars, their presence is certainly highly suggestive of this conclusion.

## 4 Integrated UV Observations of Globular Clusters and Galaxies

### 4.1 Cluster Colors

In this section we compare the available interpreted UV photometry for galaxies with that of globular clusters. We construct the following colors from the data collected by the ANS, OAO-2 and IUE, and from the optical ($V$−band) spectrum: $15 - V$, $25 - V$, and $15 - 25$. Data for the clusters have been selected from the table given by de Boer (1985), based on the earlier presentations by ABD and WC80. More recent data from UIT and IUE have also been included for a few clusters.

The observations from ANS and OAO-2 differed in aperture, passband, and instrument sensitivity. ANS had a square aperture of 2.5 ′, sensitivity at passbands at 1500, 1800, 2200, 2500, and 3300 Å, and bandwidths of 100-150 Å (van Duinen *et al.* 1975). OAO-2 had a 10 ′ circular aperture and made observations centered around 1550, 1910, 2460, 2980, 3320, and 4260 Å with bandwidths of 200-500 Å (Code *et al.* 1970). The cluster UV colors are thus based on narrower-band measurements than the galaxy data constructed from the IUE spectra in §4.2.

We define the $15 - V$, $15 - 25$, and $25 - V$ colors of the clusters observed by ANS using the flux through the 1500 and 2500 Å filters. The OAO colors are defined using the 1550 and 2460 Å filters. The IUE $15 - 25$ colors are defined using the values tabulated by RML, who summed the flux in 50 Å intervals centered around 1500, 1800, and 2500 Å. The IUE aperture is only 20 ″ × 10 ″, and thus IUE observations do not properly represent the integrated UV radiation.

Five clusters were observed by ANS but not detected in the far-UV: 47 Tuc, NGC 6356, NGC 6624, NGC 6637, and NGC 6779; these appear only in the figures showing mid-UV colors. 47 Tuc was detected by OAO-2; the larger aperture almost certainly included a P-AGB star (see de Boer 1985). The ANS observation of NGC 6441 is suspected to be contaminated by radiation from a foreground star, but we have included it to demonstrate the agreement in $15 - 25$ between ANS and the IUE observation by RML, which is not thought to be so affected. NGC 6624 was positively detected in the UV by RML; 95% of the UV radiation emanates from the X-ray binary 4U 1820-30 (King *et al.* 1993). We have included 22 of the ANS observations in our tabulation and our figures. The other clusters are $\omega$ Cen, NGC 6397, M79, M22, and M55. The first two of these observations were affected by pointing errors. For M79, the ANS flux disagrees with the OAO-2 measurement, which is consistent with the recent UIT data (see below). De Boer (1985) derived large scaling factors for the last two, and we have taken this as a sign of unreliability for those specific observations.

The lower sensitivity of OAO-2 compared to ANS placed many of the clusters below the detection threshold in the shortest wavelength passbands, given that the absorption $A_\lambda/E(B-V) \gtrsim 8$ for $\lambda < 2000$Å. The bluest cluster observed by OAO-2 in the far-UV, NGC 6254 (= M10), has been excluded here because the two observations at 1550 Å differed by a factor of 3. After removing the observations regarded as uncertain, 13 OAO-2 clusters remain, of which only six were detected at 1550 Å. WC80 state that the standard deviation of repeated measurements at 1550 Å is 43%, uncomfortably large for our purposes; we have not listed this value in the table.

The $V$ magnitudes in the ANS and OAO-2 apertures have been estimated from the database of concentric aperture photometry of Peterson (1987). This gives a more reliable measure of $15 - V$, $25 - V$ colors than can be derived directly from the Table A2 of de Boer, who derived scaling factors that estimate the ultraviolet flux from the entire cluster. These scaling factors appear to overestimate the ultraviolet flux relative to $V$ in a number of specific cases. Observations from Kron & Gordon (1985), Kron & Mayall (1960), and Chun & Freeman (1979) were interpolated to a radius of equal area to the ANS aperture (6.25 square arcmin) and to 10$'$ for the OAO-2 observations. Photometry covering the appropriate range of apertures exists for all of the ANS clusters and 5 of the OAO-2 clusters we selected; for the others, the $V$-magnitude within a 10$'$ diameter aperture is probably well-represented to within $\sim 0.2$ by the total cluster $V$ magnitude. In these cases the $V$ magnitudes were drawn from Peterson (1993).

The 1500Å integrated flux for $\omega$ Cen from UIT is a new value (Whitney 1994, private communication) measured in a 10$'$ diameter aperture. The $15 - V$ color is about 0.2 mag bluer than that given by Smith *et al.* (1993) because of a lower estimate of the sky background in the UIT frame and a revised absolute flux calibration (see Whitney *et al.* 1994)[2]. The $V$-magnitude in that aperture is interpolated from the data of Gascoigne & Burr (1956). Since the UIT near-UV image was contaminated by daylight, we have used the OAO-2 measurement (10$'$ aperture) at 2460 Å for the 2500Å flux. The OAO-2 1500 Å flux is, however, about $2\,\sigma$ less than the UIT measurement.

The integrated fluxes for M79 from UIT are also new values, kindly derived by Hill (1994, private communication). The far-UV flux compares well with the OAO-2 observation. The most appropriate comparison with the UIT A1 filter is an average of the OAO-2 2460 Å and 2980 Å fluxes; it is this value that appears in our table. Both the ANS 1500 and 2500 fluxes are significantly lower than the equivalent UIT measurements.

The monochromatic magnitude system we use is defined by

$$m_\lambda = -2.5 \log(\langle f_\lambda \rangle) - 21.1,$$

where $\langle f_\lambda \rangle$ is the mean flux in the band in units of ergs s$^{-1}$ cm$^{-2}$ Å$^{-1}$.

The cluster data are summarized in Tables 1. The first two columns identify the cluster; the third column gives the observing satellite. The fourth through ninth columns list the UV colors and their estimated errors. The metallicity values in the tenth column are from Zinn and West (1984), if tabulated there, and from Harris and Racine (1979) otherwise. The Mg$_2$ values and uncertainties in the 11th and 12th column are from Burstein *et al.* (1984), for 7 of the clusters, and derived from the [Fe/H] value using the relation given by Brodie and Huchra (1990) for the others. The relation we use to estimate Mg$_2$ from [Fe/H] is that derived from the Burstein *et al.* (1984) data:

$$\mathrm{Mg}_2 = \frac{1}{\alpha}([\mathrm{Fe/H}] + \beta),$$

where $\alpha = 12.67$, $\beta = 2.34$. Reasonable changes in either the calibration or the metallicities, provided that they do not much affect the abundance ranking of the clusters, do not change our comparisons of the clusters to the galaxies. Finally, the 13th column gives the reddening; values are again from Zinn & West (1984) and Harris & Racine (1979), with the exception of 47 Tuc, where it is from Hesser *et al.* (1987) and $\omega$ Cen, where it is taken from Whitney *et al.* (1994).

## 4.2 Galaxy Colors

The colors for the galaxies listed in Table 2 are derived from the tabulated B3FL spectra, with the exception of NGC 1399, for which a newer spectrum was kindly provided by Buson, Bertola, & Burstein (1993, private

---

[2] The aperture used by Smith to derive the integrated far-UV magnitude was incorrectly printed in his Table 2; the aperture diameter was 20$'$.2, not 27$'$.2 as stated in the paper.

TABLE 1

Ultraviolet Data for Globular Clusters

| NGC | Name | Platform | $15-V$ | $\sigma_{15-V}$ | $15-25$ | $\sigma_{15-25}$ | $25-V$ | $\sigma_{25-V}$ | [Fe/H] | $Mg_2$ | $\sigma_{Mg_2}$ | $E(B-V)$ |
|---|---|---|---|---|---|---|---|---|---|---|---|---|
| 104...... | 47 Tuc | ANS | ... | ... | ... | ... | 3.80 | 0.02 | −0.71 | 0.129 | 0.014 | 0.04 |
|  |  | OAO | 4.64 | ... | 1.50 | ... | 3.14 | 0.08 |  |  |  |  |
|  |  | IUE | ... | ... | 1.53 | ... |  |  |  |  |  |  |
| 362...... |  | OAO | ... | ... | ... | ... | 1.93 | 0.08 | −1.27 | 0.084 | 0.014 | 0.04 |
|  |  | IUE | ... | ... | 0.11 | ... |  |  |  |  |  |  |
| 1261...... |  | OAO | ... | ... | ... | ... | 1.93 | 0.08 | −1.31 | 0.081 | 0.014 | 0.00 |
| 1851...... |  | OAO | 1.89 | ... | 0.10 | ... | 1.78 | 0.08 | −1.36 | 0.077 | 0.014 | 0.10 |
| 1904...... | M79 | OAO | 1.70 | ... | 0.23 | ... | 1.47 | 0.08 | −1.69 | 0.051 | 0.014 | 0.00 |
|  |  | UIT | 1.79 | ... | 0.34 | ... | 1.44 | ... |  |  |  |  |
| 2298...... |  | ANS | 2.31 | 0.07 | 0.50 | 0.06 | 1.82 | 0.09 | −1.85 | 0.039 | 0.014 | 0.15 |
| 5024...... | M53 | ANS | 3.22 | 0.07 | 0.95 | 0.06 | 2.27 | 0.09 | −2.04 | 0.039 | 0.015 | 0.00 |
| 5139...... | ω Cen | OAO | 2.27 | ... | 0.65 | ... | 1.62 | 0.08 | −1.60 | 0.058 | 0.014 | 0.15 |
|  |  | UIT | 1.29 | 0.06 | −0.33 | 0.10 | 1.62 | 0.08 |  |  |  |  |
| 5272...... | M3 | ANS | 3.41 | 0.04 | 1.39 | 0.02 | 2.01 | 0.05 | −1.66 | 0.042 | 0.008 | 0.01 |
|  |  | OAO | 2.96 | ... | 0.99 | ... | 1.96 | 0.08 |  |  |  |  |
| 5897...... |  | OAO | ... | ... | ... | ... | 1.61 | 0.08 | −1.68 | 0.052 | 0.014 | 0.13 |
| 5904...... | M5 | OAO | 1.96 | ... | −0.05 | ... | 2.01 | 0.08 | −1.40 | 0.064 | 0.006 | 0.03 |
| 6093...... | M80 | ANS | 2.13 | 0.22 | −0.11 | 0.18 | 2.24 | 0.29 | −1.68 | 0.052 | 0.014 | 0.19 |
| 6121...... | M4 | OAO | ... | ... | ... | ... | 2.52 | 0.08 | −1.33 | 0.080 | 0.014 | 0.45 |
| 6205...... | M13 | ANS | 1.63 | 0.00 | 0.03 | 0.00 | 1.59 | 0.00 | −1.65 | 0.043 | 0.004 | 0.02 |
|  |  | OAO | 1.68 | ... | 0.29 | ... | 1.39 | 0.08 |  |  |  |  |
| 6266...... | M62 | ANS | 2.00 | 0.07 | −0.34 | 0.07 | 2.35 | 0.10 | −1.14 | 0.095 | 0.014 | 0.53 |
| 6341...... | M92 | ANS | 2.07 | 0.04 | 0.63 | 0.02 | 1.44 | 0.05 | −2.24 | 0.013 | 0.005 | 0.02 |
| 6352...... |  | IUE | ... | ... | 0.40 | ... | ... | ... | −0.51 | 0.144 | 0.014 | 0.24 |
| 6356...... |  | ANS | ... | ... | ... | ... | 4.14 | 0.12 | −0.62 | 0.174 | 0.010 | 0.22 |
| 6388...... |  | ANS | 3.22 | 0.22 | 0.28 | 0.16 | 2.94 | 0.27 | −0.48 | 0.147 | 0.014 | 0.38 |
|  |  | IUE | ... | ... | 0.51 | ... |  |  |  |  |  |  |
| 6402...... | M14 | OAO | ... | ... | ... | ... | 3.18 | 0.08 | −1.39 | 0.075 | 0.014 | 0.57 |
| 6441...... |  | ANS | 2.34 | 0.26 | 0.08 | 0.18 | 2.26 | 0.32 | −0.24 | 0.166 | 0.014 | 0.44 |
|  |  | IUE | ... | ... | 0.04 | ... |  |  |  |  |  |  |
| 6541...... |  | ANS | 2.31 | 0.07 | 0.60 | 0.05 | 1.71 | 0.09 | −1.02 | 0.104 | 0.014 | 0.09 |
| 6624...... |  | ANS | ... | ... | ... | ... | 3.94 | 0.05 | −0.34 | 0.154 | 0.009 | 0.24 |
|  |  | IUE | ... | ... | 0.54 | ... |  |  |  |  |  | 0.24 |
| 6626...... | M28 | ANS | 1.82 | 0.41 | 0.10 | 0.29 | 1.73 | 0.50 | −1.08 | 0.099 | 0.014 | 0.42 |
|  |  | IUE | ... | ... | 0.21 | ... |  |  |  |  |  |  |
| 6637...... |  | ANS | ... | ... | ... | ... | 3.83 | 0.05 | −0.59 | 0.138 | 0.014 | 0.10 |
|  |  | IUE | ... | ... | 0.76 | ... |  |  |  |  |  |  |
| 6681...... | M70 | ANS | 1.74 | 0.04 | 0.04 | 0.04 | 1.70 | 0.05 | −1.51 | 0.066 | 0.014 | 0.07 |
| 6715...... | M54 | ANS | 2.47 | 0.04 | 0.29 | 0.04 | 2.18 | 0.05 | −1.55 | 0.062 | 0.014 | 0.13 |
| 6752...... |  | ANS | 1.34 | 0.04 | −0.18 | 0.02 | 1.52 | 0.05 | −1.54 | 0.063 | 0.014 | 0.04 |
| 6779...... | M56 | ANS | ... | ... | ... | ... | 1.96 | 0.05 | −1.94 | 0.032 | 0.014 | 0.23 |
| 6864...... | M75 | ANS | 2.59 | 0.22 | 0.41 | 0.16 | 2.18 | 0.27 | −1.32 | 0.081 | 0.014 | 0.17 |
|  |  | OAO | ... | ... | ... | ... | 1.56 | 0.08 |  |  |  |  |
| 7078...... | M15 | ANS | 2.36 | 0.07 | 0.67 | 0.05 | 1.69 | 0.09 | −2.15 | 0.017 | 0.007 | 0.10 |
| 7099...... | M30 | ANS | 2.56 | 0.04 | 0.80 | 0.04 | 1.76 | 0.05 | −2.13 | 0.017 | 0.014 | 0.06 |

communication). The columns in Table 2 give the NGC number, Mg$_2$ index, $E(B-V)$, and the calculated intrinsic $15-V$, $25-V$, and $15-25$ colors respectively. The last two columns are empty for systems without LWP spectra in the IUE catalog. The table includes the 24 galaxies marked as 'quiescent' systems by B3FL, plus NGC 1052 and NGC 4278, whose UV spectra are probably not affected by their low-level nuclear activity. The two galaxies with the weakest Mg$_2$ index are M32 and M85 (= NGC 4382). The former is a Local Group compact dwarf elliptical with no evidence for activity or star formation. The latter is a luminous Virgo S0 galaxy which is part of an interacting pair and shows evidence for star formation and internal extinction (see Kinney *et al.* 1993 and references therein). We exclude this object from our analysis in later sections of this paper.

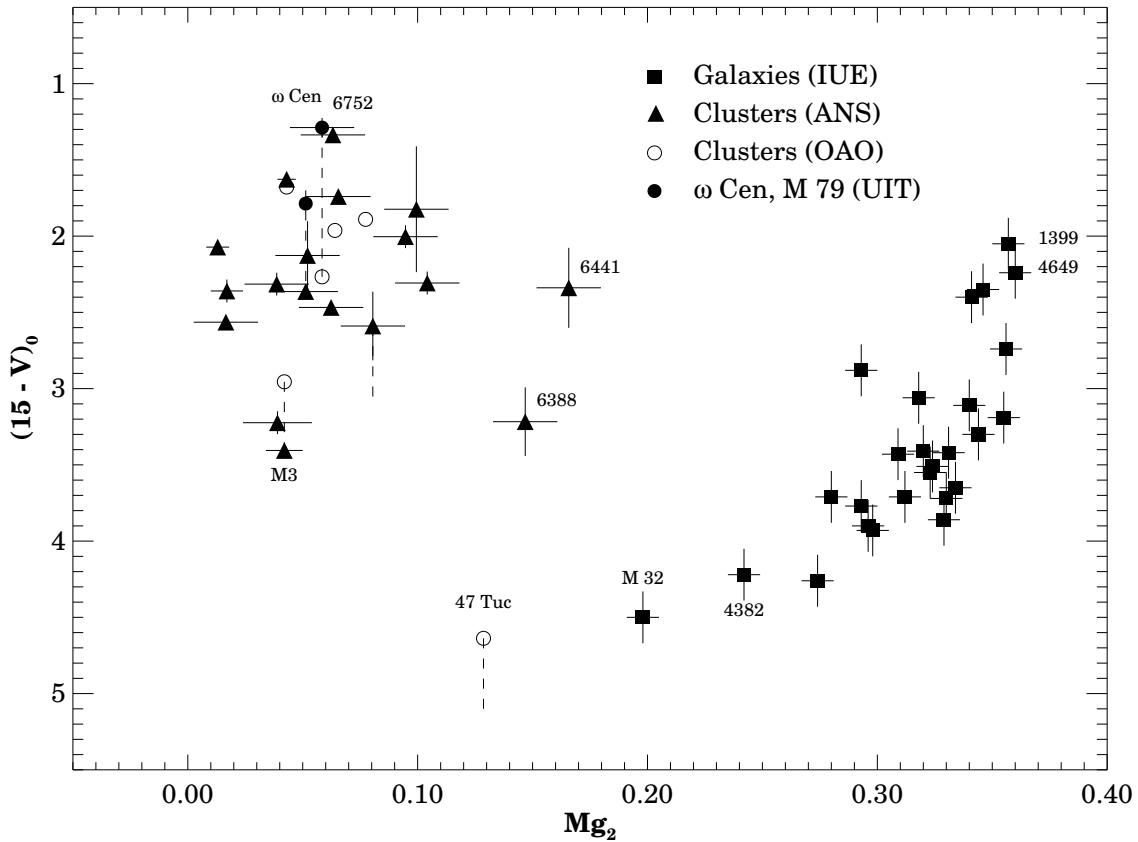

**Figure 1:** *Observations of $15-V$ colors of galaxies and clusters plotted against the absorption line index* Mg$_2$. *The galaxy data, derived from IUE SWP observations, are from B3FL, with the exception of NGC 1399 which is from Buson, Bertola, & Burstein (1993, private communication). The UV cluster data observed by the ANS and OAO-2 satellites are from de Boer (1985), with $V$-magnitudes derived from Peterson (1987, 1993). The UIT data for M79 and $\omega$ Cen are from Hill (1994, private communication), and Whitney (1994, private communication). Dashed lines join observations of the same cluster from more than one satellite, except where the cluster was only detected from one of the observing platforms, in which case the line has no observation marking its lower end.*

TABLE 2

GALAXY COLORS

| NGC | Messier | Mg$_2$ | $E(B-V)$ | $(15-V)_0$ | $(25-V)_0$ | $(15-25)_0$ |
|------|---------|--------|----------|------------|------------|-------------|
| NGC 221 . . . . . . . | M 32 | 0.198 | 0.077 | 4.495 | 3.246 | 1.249 |
| NGC 224 . . . . . . . | M 31 | 0.324 | 0.077 | 3.462 | 3.728 | −0.266 |
| NGC 584 . . . . . . . | | 0.298 | 0.030 | 3.930 | . . . | . . . |
| NGC 1052 . . . . . . . | | 0.318 | 0.015 | 3.071 | 3.459 | −0.387 |
| NGC 1399 . . . . . . . | | 0.357 | 0.000 | 2.040 | 2.731 | −0.691 |
| NGC 1404 . . . . . . . | | 0.344 | 0.000 | 3.258 | 3.700 | −0.442 |
| NGC 1407 . . . . . . . | | 0.341 | 0.040 | 2.400 | . . . | . . . |
| NGC 2768 . . . . . . . | | 0.296 | 0.040 | 3.900 | . . . | . . . |
| NGC 2784 . . . . . . . | | 0.334 | 0.165 | 3.650 | . . . | . . . |
| NGC 3115 . . . . . . . | | 0.309 | 0.025 | 3.397 | 3.628 | −0.231 |
| NGC 3379 . . . . . . . | | 0.329 | 0.013 | 3.817 | 3.835 | −0.018 |
| NGC 4111 . . . . . . . | | 0.274 | 0.000 | 4.260 | . . . | . . . |
| NGC 4125 . . . . . . . | | 0.312 | 0.010 | 3.724 | 3.596 | 0.129 |
| NGC 4278 . . . . . . . | | 0.293 | 0.025 | 2.886 | 3.302 | −0.416 |
| NGC 4374 . . . . . . . | M 84 | 0.323 | 0.033 | 3.473 | 3.707 | −0.233 |
| NGC 4382 . . . . . . . | M 85 | 0.242 | 0.010 | 4.209 | 3.070 | 1.140 |
| NGC 4406 . . . . . . . | M 86 | 0.330 | 0.028 | 3.675 | 3.644 | 0.031 |
| NGC 4472 . . . . . . . | M 49 | 0.331 | 0.000 | 3.390 | 3.696 | −0.306 |
| NGC 4494 . . . . . . . | | 0.293 | 0.015 | 3.770 | . . . | . . . |
| NGC 4552 . . . . . . . | M 89 | 0.346 | 0.035 | 2.317 | 3.120 | −0.803 |
| NGC 4621 . . . . . . . | M 59 | 0.355 | 0.018 | 3.143 | 3.316 | −0.173 |
| NGC 4649 . . . . . . . | M 60 | 0.360 | 0.008 | 2.201 | 3.226 | −1.025 |
| NGC 4697 . . . . . . . | | 0.320 | 0.010 | 3.410 | . . . | . . . |
| NGC 4762 . . . . . . . | | 0.280 | 0.010 | 3.678 | 3.432 | 0.245 |
| NGC 4889 . . . . . . . | | 0.356 | 0.013 | 2.712 | 3.218 | −0.506 |
| NGC 5846 . . . . . . . | | 0.340 | 0.035 | 3.110 | . . . | . . . |

It is important to bear in mind that the galaxies were observed with IUE through a small aperture of size $10''\times20''$, and the colors represent just the nuclear values. Note also that because of strong UV color gradients (O'Connell *et al.* 1992), the UVX strengths for the more distant objects (e.g., NGC 4889) are underestimated compared to nearer sources. The 1500 Å entry is the average of the flux from 1250 to 1850 Å, while that at 2500 Å is calculated from the fluxes between 2200 and 2800 Å inclusive. The fluxes tabulated in Tables 2A,B of B3FL are not, as B3FL state in their paper, corrected for extinction, so we made extinction corrections to the spectra using the Savage and Mathis (1979) reddening law. For the galaxies in which the reduced spectrum is positive in all bins, the extinction correction was applied point-by-point to the spectra. Otherwise, the mean flux received in the two bands was adjusted by an average extinction correction. In this case the extinction curve was weighted with the high S/N spectrum of NGC 4649, which has a low value of the reddening ($A_B = 0.03$). The results for $15 - V$ in the table differ little from those given by B3FL, the largest discrepancy being 0.077 mag for NGC 4374, with all the others less than 0.05 mag. $E(B-V)$ given in the second column of the table is from the reddening atlas of Burstein & Heiles (1984), who adopted $A_B = 4\,E(B-V)$. The Mg$_2$ values are those quoted in B3FL.

## 4.3    Color-Color and Color-Line Strength Correlations

Figures 1 and 2 exhibit the extinction-corrected colors from Tables 1 and 2 plotted against the Mg$_2$ index. Observations of the same clusters from different spacecraft are joined by dotted lines. The error bars in Mg$_2$ for the BFGK clusters are shown. Where [Fe/H] has been used to derive Mg$_2$, the error bar shown in the figures is the measure of the error Brodie & Huchra (1990) give for the calibration (their $\sigma_M$) and does not include errors in the measurement of [Fe/H], which are typically a few tenths of a dex. Error bars for the colors are plotted for the ANS data; the errors for the OAO data amount to ∼ 0.5 mag using the estimates of WC80 and are omitted in the diagram for the sake of clarity. For the galaxies the error bars plotted are the mean errors quoted by B3FL. The $15 - 25$ color of NGC 6266 ( = M62) is anomalously blue; the color is uncertain as the cluster is the most highly reddened in the sample and is also affected by differential reddening (Caloi, Castellani, & Piccolo 1987).

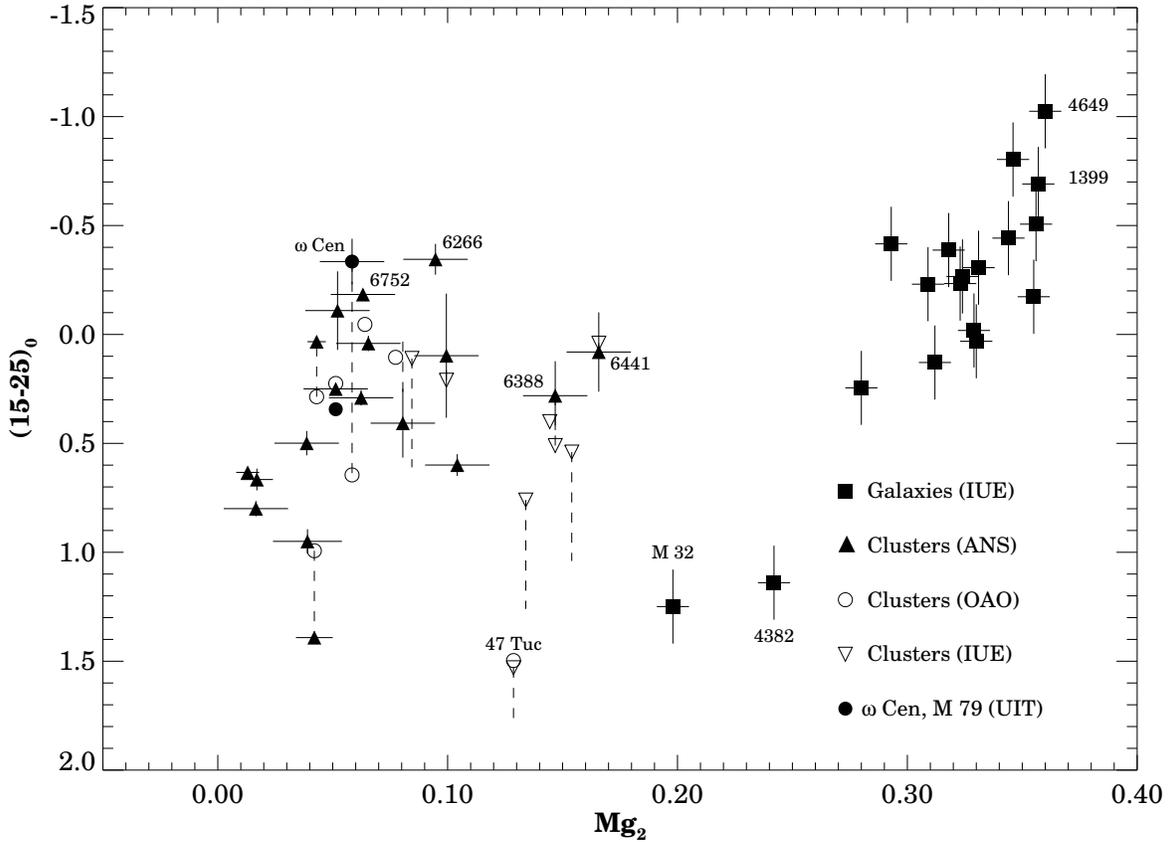

**Figure 2:** *Observations of* 15 − 25 *colors of galaxies plotted against the absorption line index* Mg$_2$. *The galaxy data are from B3FL, and derived from IUE SWP observations. The cluster data are from the same sources as in Fig.1, with the addition of the IUE observations of Rich, Minniti, & Liebert (1993). The 25 magnitude for the ω Cen used with the UIT observation is from OAO-2.*

The (15 − V) color index can be considered an indicator of the 'specific UV luminosity', i.e., it measures the *production* of hot stars in the emitting population. The 15 − 25 color is related to the *temperature* of the hot population. However, since the 2500 Å flux from the turnoff and subgiant branch (SGB) can be significant, the 15 − 25 color can only be understood by modeling the earlier phases together with the hot star contribution. This composite nature of the UV spectra has been stressed by B3FL.

It is immediately apparent from Figure 1 that the globular clusters and the galaxies do not occupy a single sequence. Among galaxies, the specific UV luminosity is largest in the most metal-rich objects, while the most metal-rich clusters (with red HB morphology) are fainter in the UV than those with lower metallicity. In 15 − V, the cluster colors correlate with their known HB morphologies (WC80; ABD) and can be bluer than any of the galaxies. The bluest clusters all have "extended blue HB tails", with stars hotter than 12,000 K (e.g., M13, M79, NGC 6752). In addition, they are all intermediate in metallicity ([Fe/H] ∼ −1.5), whereas the most metal-poor globulars (M92, M15, and M30, which have [Fe/H] ∼ −2) are fainter in the far-UV. In M13 and similar objects, the entire HB phase contributes to the UV. In contrast M3, which has a red HB despite similar [Fe/H] (Kraft *et al.* 1993), is 2 mag fainter in 15 − V. The location of the clusters in Fig. 1 can thus be seen as a sequence in which the relative UV output decreases with increasing metallicity (the 'first parameter' effect), upon which is superposed a large 'second parameter scatter' at intermediate metallicities.

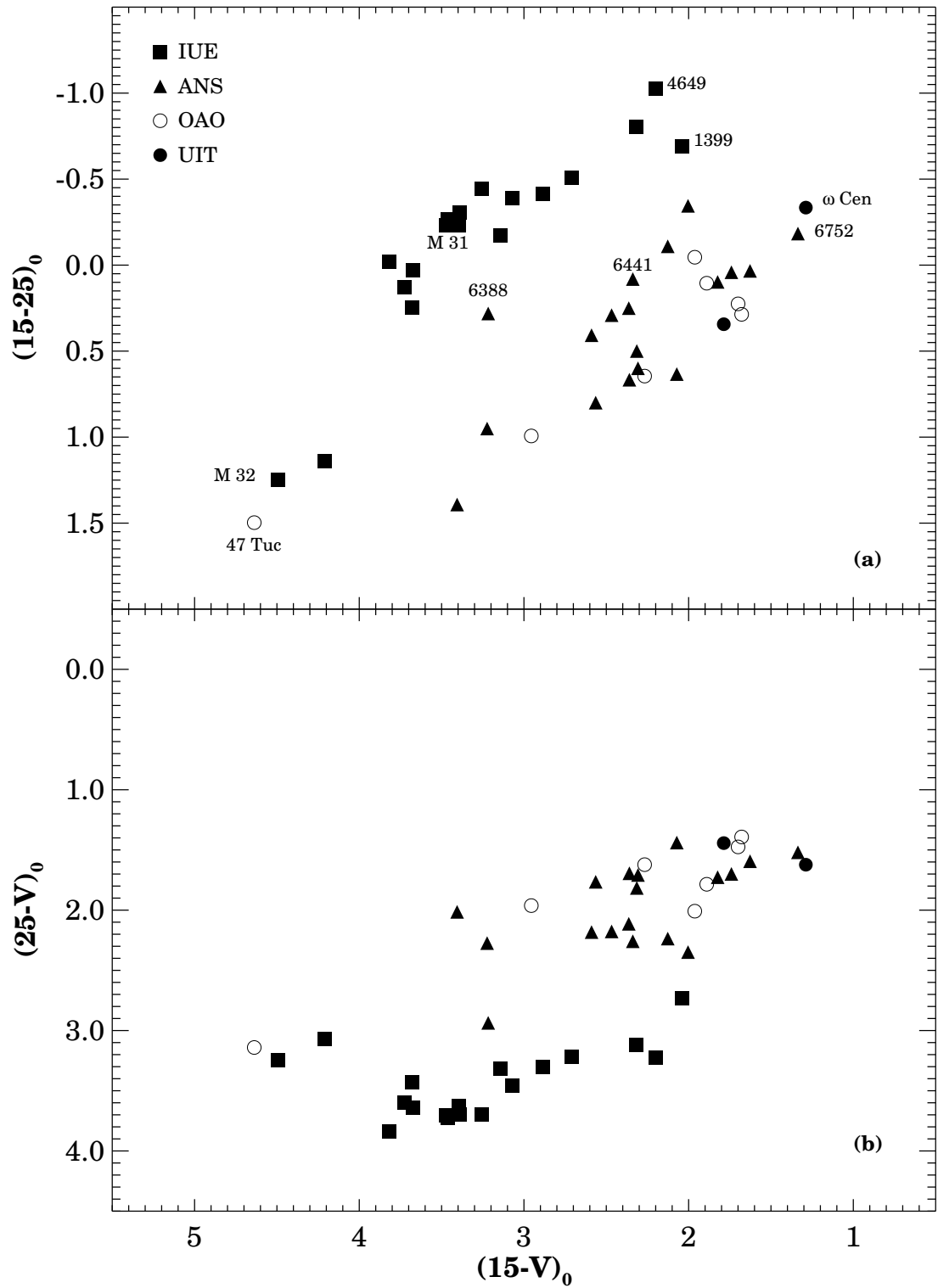

**Figure 3:** *Observations of galaxies and globular clusters plotted in the ultraviolet two-color diagrams. The data are the same as shown in Figures 1. Panel (a): Data in $15 - V$ vs. $15 - 25$. The galaxies and clusters separate into two almost parallel sequences, with the abundance parameter increasing left to right in the galaxy sequence, and right to left in the clusters. Objects discussed in the text are marked by NGC number or other designation. (b) The same data plotted in the $15 - V$ vs. $25 - V$ diagram used later for most of the analysis.*

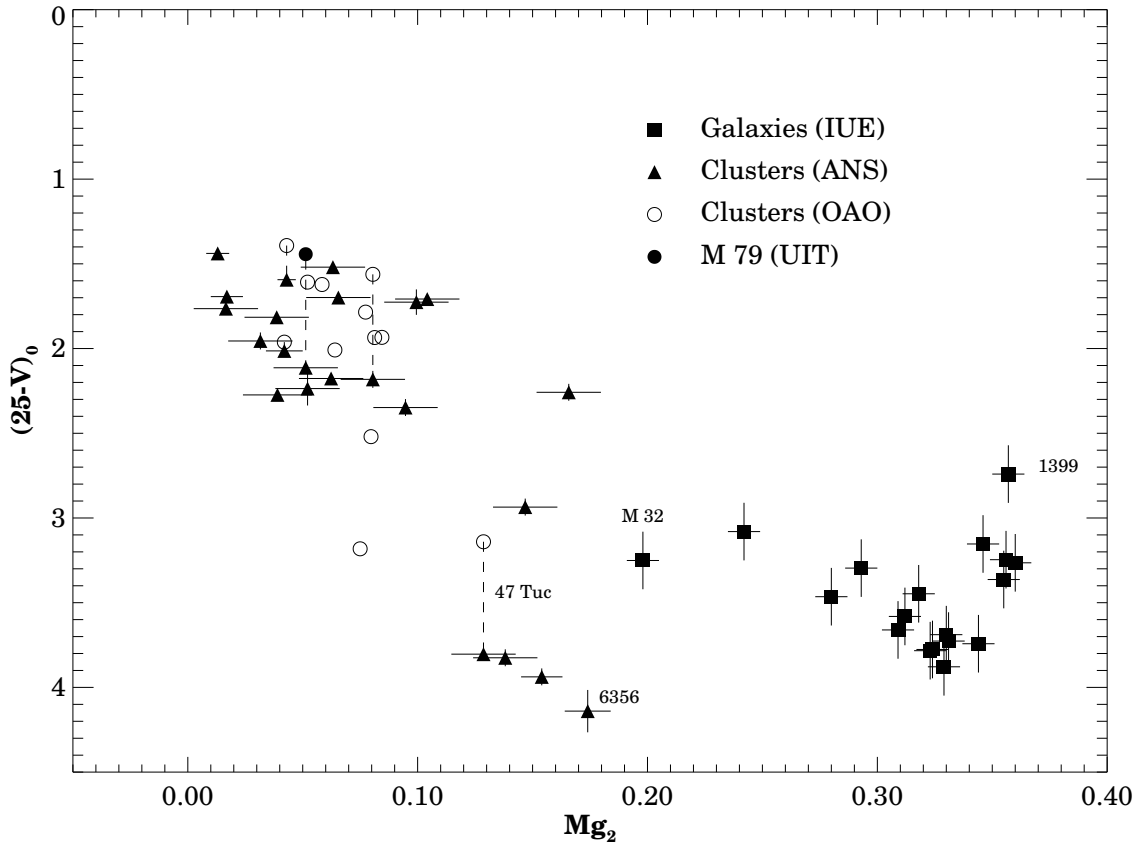

**Figure 4:** *Observations of $25 - V$ colors of galaxies and clusters plotted against the absorption line index* $Mg_2$. *The source data for this diagram are the same as in Figures 1-3.*

In the $15 - 25$ color index, the strong UVX galaxies are bluer than the clusters. This is partly a consequence of the steeper SEDs of the galaxies below 2000 Å (e.g. O'Connell 1993) and partly because the contribution at 2500 Å from the main sequence, SGB, and IBHB stars is greater in the clusters. The obvious conclusion is that *the UVX in galaxies does not arise from a metal-poor subpopulation similar to the globular clusters.* Lee (1993) has suggested that the UVX in galaxies arises from the "blue HB" of a minority metal-poor population and that its correlation with the observed line strengths is not caused by abundance but by a variation in age, with the larger UVX galaxies being older. This hypothesis appears to be inconsistent with the data. The $15 - 25$ colors of the UVX galaxies can be bluer than any cluster, and their $25 - V$ colors (below) are also not consistent with the existence of large minority metal-poor populations.

The contrast between the galaxies' and the clusters' UV colors is brought into sharp focus by the two color diagrams shown in Fig 3 (a) and (b). We have omitted the error bars in this diagram for clarity. The galaxies and the clusters form two distinct, almost parallel sequences of points. The metallicity index in the galaxy sequence increases from left to right (the UVX-$Mg_2$ correlation), while in the clusters it increases from right to left (1st Parameter effect). An increase (decrease) in the fraction of hot stars present mainly produces a rightward (leftward) shift in the location of an object in this diagram. The $15 - 25$ color crudely represents the mean temperature of all stars radiating significantly at wavelengths shorter than 2800 Å. Any explanation for the UVX must be able to reproduce this behavior.

Figure 4 exhibits $25 - V$ *vs.* $Mg_2$. Both the hot components and the pre-HB phases contribute flux around 2500 Å. As expected (and explicitly demonstrated by B3FL – see their Fig. 8) the 2500 Å contribution from the cool phases decreases with increasing metallicity. For the low abundances of the metal-poor clusters, $[Fe/H] < -1.5$ ($Mg_2 < 0.05$), the mid-UV flux from the pre-HB and the more advanced evolutionary stages is comparable. The galaxies are all relatively faint at 2500 Å. However, the slight downward trend of the mid-UV flux with $Mg_2$ is reversed in the strongest UVX galaxies. NGC 1399 appears to have particularly strong mid-UV flux.

# 5    Spectral Synthesis Formulation

In order to model the colors of an old stellar population that contains a hot component in the later stages of stellar evolution, it is necessary to perform three separate tasks. First, we must calculate the integrated fluxes from the main sequence, subgiant and giant phases. Second, we must model the integrated flux from stars in advanced stages of evolution, i.e., in the HB and the post-HB stages. Third, we must determine the number of stars in the post-RGB phases. The formulæ presented here parallel the treatment of this problem given by Renzini & Buzzoni (1986) in their derivation of the Fuel Consumption Theorem (see also Tinsley 1980). This formulation differs by taking into account in greater detail the changes in luminosity and effective temperature of the model components with evolution.

## 5.1    Early (pre-He Flash) Stages of Evolution

The first of these tasks is achieved by constructing isochrones, and by adopting a mass function $\psi(\mathcal{M})$, which represents the number of stars formed in the mass interval $[\mathcal{M}, \mathcal{M} + d\mathcal{M}]$. Define the integrated luminosity from a stellar population, in a passband centered at wavelength $\lambda$, by

$$\mathcal{L}_\lambda = \mathcal{L}_\lambda^{\mathrm{E}} + \mathcal{L}_\lambda^{\mathrm{L}}. \tag{5.1}$$

$\mathcal{L}_\lambda^{\mathrm{E}}$ represents the contribution from the evolutionary phases prior to the helium flash at the RGB tip, and $\mathcal{L}_\lambda^{\mathrm{L}}$ represents that from the later phases. Apart from relatively young ($\sim 2$ Gyr) metal-poor populations, only the late evolutionary phases contribute significantly at far-UV wavelengths ($\lambda \lesssim 2000$ Å). At longer wavelengths, both early and late phases are important, and the contribution from later phases is dominated by flux from bright AGB and red HB stars.

The integrated flux at $\lambda$ from the early phases can be calculated from the formula

$$\mathcal{L}_\lambda^{\mathrm{E}}(t, Y, Z) = \int_{\mathrm{MM}}^{\mathrm{RGT}} L_\lambda(\mathcal{M}; t, Y, Z)\psi(\mathcal{M})d\mathcal{M}. \tag{5.2}$$

The integral limits are from the hydrogen-burning minimum mass (MM) to the red-giant tip (RGT); the lower limit is a function of abundance $Z$, while the upper limit is a function of age $t$ and $Z$. The isochrones give $L_{\mathrm{bol}}$, $T_{\mathrm{eff}}$, and the surface gravity $g$ for fixed $t$ and composition $Y, Z$. To convert bolometric luminosities to fluxes received in a given passband, we use bolometric corrections $\alpha_\lambda(g, T_{\mathrm{eff}}; Z)$ derived from the Kurucz (1991) grid of synthetic stellar fluxes, interpolated at each required value of $T_{\mathrm{eff}}$ and $g$ for each fixed $Z$. That is,

$$L_\lambda(\mathcal{M}, t, Y, Z) = \alpha_\lambda(T_{\mathrm{eff}}, g; Z)L_{\mathrm{bol}}(\mathcal{M}; t, Y, Z), \tag{5.3}$$

where $\alpha_\lambda$ is the ratio of the flux (in $\mathrm{erg\,s^{-1}\,cm^{-2}\,\AA^{-1}}$) in the passband $\lambda$, divided by the total flux ($= \sigma T_{\mathrm{eff}}{}^4$).

## 5.2 Late (post-He Flash) Stages of Evolution

For each individual post-RGB evolutionary track with a ZAHB envelope mass $M_{\mathrm{env}}^0$, the energy radiated over the lifetime of a single star is

$$E_\lambda(M_{\mathrm{env}}^0; Y, Z) = \int_{\mathrm{ZAHB}}^{\mathrm{WD}} L_\lambda(M_{\mathrm{env}}^0, \tau; Y, Z) d\tau \qquad (5.4)$$

The integral over time runs from the ZAHB to the 'death' of the star as a low-luminosity white dwarf, although the contribution from the white dwarf cooling track is small because the stars dim relatively rapidly. The explicit time dependence of $E_\lambda$ can be neglected as it arises only from the variation in $M_c$ with age, which is very small for $t \gtrsim 5$ Gyr. The integral is an important quantity for population syntheses because, under suitable conditions, it is directly proportional to the luminosity radiated by a population of stars with masses equal to that of the track (see next section).

## 5.3 Evolutionary Flux into Late Stages of Evolution

In a large stellar system, the number of red giants that may undergo a helium core flash in an interval of time comparable to the HB lifetime is large, while the probability of any particular star reaching the RGB tip is small. This circumstance meets the definition of a Poisson process, in which the input rate of stars into the HB has a finite mean. The size of the post-RGB population will approximate a "steady state" if the input rate to the HB phase varies on a timescale much longer than the duration of the post-RGB phases. If the evolutionary path from the ZAHB location to the white dwarf stage is divided into time intervals $\{t_j\}$, then the number of stars $\Delta N_j$ in the interval $\Delta t_j = t_j - t_{j-1}$ is

$$\Delta N_j \approx \dot{N}(t; Y, Z)\Delta t_j, \qquad (5.5)$$

where $\dot{N}(t)$ is the number of stars passing through the helium flash in unit time. Since the difference in mass between the turnoff and the RGB is small [3], $\dot{N}(t)$ is approximately equal to the number of stars leaving the main sequence in unit time, which can be found by evaluating the mass function and the time derivative of the mass at the main sequence turnoff.

$$\dot{N}(t) = \psi(\mathcal{M})\frac{d\mathcal{M}}{dt}\bigg|_{TO} \qquad (5.6)$$

The mass function $\psi(\mathcal{M}) = dn/d\mathcal{M}$ is usually taken to be a power law, $\psi(\mathcal{M}) = \mathcal{A}\mathcal{M}^{-(1+x)}$. $\dot{N}(t)$ is proportional to the mass of the system, which enters through the normalization constant $\mathcal{A}$. The relative size of the post-He flash population is insensitive to $x$.

The luminosity received at $\lambda$ from the stars with envelope mass $M_{\mathrm{env}}^0$ evolving off the ZAHB is given by

$$\mathcal{L}_\lambda(M_{\mathrm{env}}^0; Y, Z) = \dot{N}(t)\sum_j L_\lambda(M_{\mathrm{env}}^0, t_j; Y, Z)\,\Delta t_j. \qquad (5.7)$$

---

[3]       The mass difference $\delta M_{RGB-MSTO}$ can be as large as $\sim 0.1 M_\odot$ for ages of order 10 Gyr (compare Tables 4 and 5). This is arguably significant, but probably does not introduce an error in our calculations larger than other approximations used in deriving the results.

For sufficiently large stellar populations this sum (5.7) can be approximated by the integral (5.4), i.e.,

$$\mathcal{L}_\lambda(M_{\mathrm{env}}^0; Y, Z) = \dot{N}(t) E_\lambda(M_{\mathrm{env}}^0; Y, Z). \tag{5.8}$$

The condition for this continuum approximation to be valid is that the input rate is sufficiently large, i.e., $\dot{N}(t) \gg 1/\Delta t_j$. This condition ensures that the most significant rapid phases (e.g., the P-AGB phase but not the shell flash epochs) are populated. The continuum approximation is certainly justified for galaxies ($M \sim 10^{9-12} M_\odot$), but for objects the size of globular clusters ($\sim 10^5 M_\odot$), the most rapid evolutionary phases such as the P-AGB will not be properly represented. A single "UV-bright" star may produce a sizeable fraction of the total UV radiation from the cluster (e.g., UV-5 in NGC 1851; see Parise *et al.* 1994).

Now let the HB morphology be described formally by a function $P(M_{\mathrm{env}}^0)$, which is the fraction of stars arriving on the ZAHB with envelope mass $M_{\mathrm{env}}^0$. The normalization is

$$\int_0^{M_{\mathrm{RGB}} - M_c} P(M_{\mathrm{env}}^0) dM_{\mathrm{env}}^0 = 1,$$

where $M_c$ is the ZAHB core mass. Then the integrated post-RGB luminosity is

$$\mathcal{L}_\lambda^{\mathrm{L}}(Y, Z) = \dot{N}(t) \int_0^{M_{\mathrm{RGB}} - M_c} P(m) E_\lambda(m; Y, Z) dm$$

$$= \dot{N}(t) \mathcal{E}_\lambda^{\mathrm{L}}. \tag{5.9}$$

The integral over the envelope mass distribution ranges from its maximum value, equal to the RGB tip mass (at the appropriate age) minus the corresponding $M_c$, to zero at the helium main-sequence model with mass $M_c$.

In order to compare the results from this formulation to observable quantities, we define the 'specific evolutionary flux' as the arrival rate of stars on the horizontal branch per unit integrated $V$ luminosity, which is given by

$$\dot{n}(t) = \frac{\dot{N}(t)}{\mathcal{L}_V}. \tag{5.10}$$

A difficulty arises in evaluating this quantity theoretically because the contribution $\mathcal{L}_V^{\mathrm{L}}$ of the later phases to $V$ depends somewhat on the assumed HB morphology. If the HB morphology is red and the bulk of the stars evolve up the AGB, $\mathcal{L}_V^{\mathrm{L}}$ is typically 20% of the total $V$-flux (see Buzzoni 1989). This may seem surprising since HB and bright AGB stars are rare: however, recall that the upper RGB contributes about half of $\mathcal{L}_V^{\mathrm{E}}$, and the bright giants have similar lifetimes. On the other hand, the quantity

$$\dot{n}_0(t) = \frac{\dot{N}(t)}{\mathcal{L}_V^{\mathrm{E}}}, \tag{5.11}$$

which is related to $\dot{n}(t)$ by

$$\dot{n}(t) = \frac{\dot{n}_0(t)}{\left(1 + \dot{n}_0(t) \mathcal{E}_V^{\mathrm{L}}\right)}, \tag{5.12}$$

may be easily derived from the isochrone component alone (see Table 6).

$\dot{n}(t)$ may also be derived semi-empirically from spectral synthesis analysis (e.g., O'Connell 1976, 1980). We can write

$$\dot{n}(t) = \frac{\phi(M_V)}{\mathcal{L}_V} \frac{\mathrm{d}\,M_V}{\mathrm{d}\,t}\bigg|_{TO},$$ (5.13)

where

$$\phi(M_V) = -\left(\frac{2.5}{\ln 10}\right)\mathcal{L}_V\,\psi(\mathcal{M})\left[\left(\frac{\partial\,\mathcal{M}}{\partial\,L_V}\right)_t + \left(\frac{\partial\,\mathcal{M}}{\partial\,t}\right)_{L_V}\left(\frac{\partial\,t}{\partial\,L_V}\right)\right]$$ (5.14)

is the luminosity function per unit magnitude (see Bergbusch & VandenBerg 1992), evaluated at the main-sequence turnoff. The magnitude derivative at the turnoff can be extracted from stellar isochrones (Ciardullo & Demarque 1977; Bergbusch & VandenBerg 1992), and for $t \gtrsim 5$ Gyr it is found to be

$$\frac{\mathrm{d}\,M_V}{\mathrm{d}\,t}\bigg|_{TO} \approx \frac{1}{t_9}\ \mathrm{mag}\ /\mathrm{Gyr},$$ (5.15)

The spectral synthesis models allow an estimate of the ratio

$$\left(\frac{\phi(M_V)}{\mathcal{L}_V}\right) = \begin{cases} 0.2, & \text{for a mean gE SED, age 9 Gyr (O'Connell 1976)} \\ 0.08 & \text{for M 32, age 5 Gyr (O'Connell 1980)} \end{cases}$$

and thus

$$\dot{n}(t) \approx 0.02\ \mathrm{stars}\ /\mathrm{Gyr}\,/L_V^{\odot}.$$ (5.16)

This semi-empirical estimate is a factor of about 2 lower than either our theoretical estimates (in Table 6) or those derived by Renzini & Buzzoni (1986), who obtained 0.04 stars /Gyr /$L_V^{\odot}$ for populations of age $\sim 10$ Gyr. This is good agreement considering the different techniques and uncertainties involved. However, predicted photometric results are sensitive to $\dot{n}(t)$. If, for example, $\dot{n}(t)$ changes by a factor of $\sim 2$, the predicted $m_\lambda - V$ color changes significantly, by $\sim 0.75$ mag. Note that $\dot{n}(t)$ is a weak function of age and composition (Table 6; Renzini & Buzzoni 1986). We use the values in Table 6 for subsequent computations.

# 6    Model Construction and Ingredients

Our synthetic population models use the HB and post-HB evolutionary tracks presented in Paper I. The contribution of pre-HB stellar evolutionary phases is derived from isochrone calculations spanning the age range 2 to 20 Gyr, which have been constructed from evolutionary tracks for all of the compositions for which we have HB and post-HB calculations. We have used the Kurucz (1991) database of model stellar spectra to convert stellar temperatures and luminosities to fluxes radiated in UV passbands and in the $V$ filter. Although such synthetic spectra are not as realistic as observed spectra (e.g., the IUE libraries of Wu *et al.* 1992 or Fanelli *et al.* 1992), they cover a larger range of $Z$ and $Y$ than can be obtained empirically. They are satisfactory for our purposes, considering the precision of the available cluster and galaxy photometry. We present tables of all the relevant quantities produced from our models so that they may be used to construct other models of interest.

## 6.1 Isochrones

The evolutionary tracks for pre-HB phases of evolution were calculated using our version of the code described by VandenBerg (1983; 1992), and differ in input physics from our calculations of the later evolutionary phases only by the use (for [Fe/H] $\leq 0$) of the pressure grids described by VandenBerg (1992) rather than the scaled-solar $T - \tau$ relation. Isochrones constructed from these tracks express $\log T_{\rm eff}$, $\log L/L_\odot$ and the mass along loci of fixed age, using the central hydrogen abundance on the main sequence, the arclength from the turnoff to the base of the giant branch, and the arclength from that point to the RGB tip, to parameterize the tracks. We use the Salpeter mass function $\psi(\mathcal{M}) \propto \mathcal{M}^{-2.35}$ to compute integrated quantities from the isochrones. We have compared the fluxes we derive to similar quantities computed from the isochrones of Bergbusch & VandenBerg (1992), and they agree to within about 10% for isochrones with similar input parameters. This agreement is despite our use of Kurucz color transformations for the $V$-band (theirs are a modified version of the transformations presented by VandenBerg & Bell 1985). These tests give us confidence that our isochrones are sufficiently reliable for the task of producing integrated fluxes for comparison with the available photometry and spectrophotometry.

A typical set of metal-rich isochrones is illustrated in Figure 5, for $Z = 0.04$ ([Fe/H] $= 0.38$). The evolutionary sequences used to construct them are indicated by dotted lines on the plot. The evolutionary tracks have not been calibrated to solar values at the accepted solar age. The size of the error in the results may be appreciated by noting that the sequence with [Fe/H] $= 0$, $Y = 0.27$, $M = 1\,M_\odot$ (computed with mixing length to pressure scale height ratio $l/\mathcal{H}_P = 1.5$) has $\log L/L_\odot = 0$ and $T_{\rm eff} = 5816$K at $t = 4.3$ Gyr, an age difference from the accepted value of $\sim 200 - 300$ Myr and a temperature offset of about 40K. Also, the effects of convective overshooting, which may be important for the two younger models (i.e., in those tracks which possess "convective hooks" prior to their main sequence turnoff), are not included. Our main focus is in any case upon older populations, which are not affected by this consideration.

For all of the isochrones, the Kurucz model stellar spectra were used to derive mean fluxes through passbands centered on 1500, 2500, and 3300 Å, with bandwidth 600 Å. The flux in the $V$ filter was found using the transmission profile given by Bessell (1990). Beyond the low temperature end of the Kurucz database ($T_{\rm eff} < 3500$ K, $0 < \log g < 5$) the fluxes were approximated by constant extrapolation of the values at the boundary. The ultraviolet fluxes are combined with the results for the $V$ filter to produce specific UV fluxes, and presented as colors, i.e.,

$$m_\lambda - V = -2.5 \log_{10} \left[ (\mathcal{L}_\lambda / \mathcal{L}_V) \right]^E$$

in Tables 3 a) – c). The colors are given for each of the 8 sets of composition parameters for which calculations were presented in Paper I and for ages between 2 and 20 Gyr.

As one check of the reliability of our calculations, the Kurucz database was also used to generate the fluxes in the $B$ filter, using the Bessell filter profiles, and compared to observed galaxy integrated colors. The fluxes in $B$ were normalized so that the Kurucz model spectrum of Vega (A0 V) gave $B - V = 0$. For $Z = Z_\odot$ we obtain $0.95 < (B - V)_0 < 1.05$ for $8 < t < 20$ Gyr, and $(B - V)_0 = 0.81, 0.86$ for $t = 4, 6$ Gyr (compare Worthey 1994). These synthetic colors are in agreement with canonical values (de Vaucouleurs 1961) for E/S0 galaxies and for spiral galaxy bulges for ages $\gtrsim 8$ Gyr.

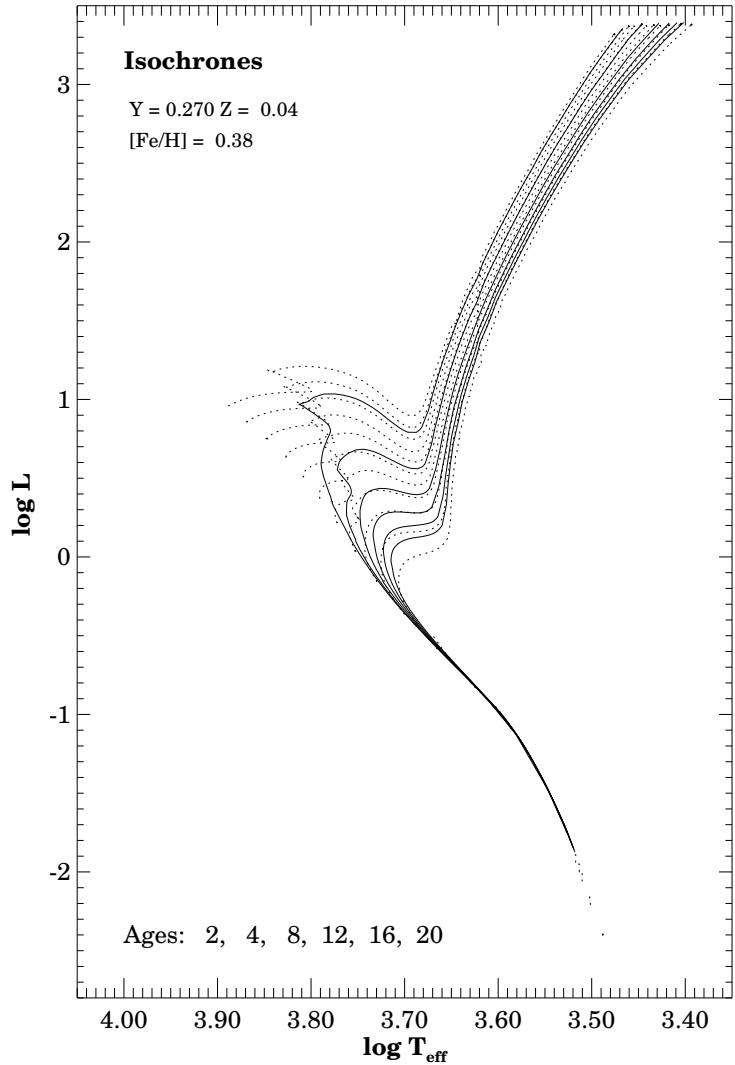

**Figure 5:** *Sample of theoretical isochrones for metal rich stars. The diagram shows isochrones (solid curves) for ages 2–20 Gyr for the composition $Y = 0.27$, $Z = 0.04$ ([Fe/H] = 0.38). The evolutionary tracks used to construct the isochrones are shown as dotted lines. The mass range of the tracks is $0.20 < M/M_\odot < 1.90$.*

Tables 4 and 5 give turnoff masses and red giant tip masses (in the absence of mass loss) also derived from the isochrones. These tables are useful for quantifying the amount of mass that must be lost from red giants to populate the EHB. Upper mass limits for EHB stars are given in Paper I (Table 1), and are in the range 0.52–0.54 $M_\odot$ for all compositions. Table 5 then implies that production of an EHB population requires stars of age 12 Gyr to shed 0.3–0.6 $M_\odot$ on the giant branch as the metallicity ranges from [Fe/H] = −2.26 to +0.58; for 6 Gyr, the mass loss must be 0.5–0.8$M_\odot$. No matter the age or abundance, the presence of EHB stars evidently requires that a large fraction of the initial stellar mass must be lost.

TABLE 3

SYNTHETIC INTEGRATED COLORS FOR PRE-HE FLASH ISOCHRONE SEQUENCES

(a) 15 − V

| [Fe/H] | Z | Y | Age (Gyr) | | | | | | | | | |
|---|---|---|---|---|---|---|---|---|---|---|---|---|
| | | | 2 | 4 | 6 | 8 | 10 | 12 | 14 | 16 | 18 | 20 |
| −2.26[a] | | 0.235 . . . . . . . | 0.635 | 2.28 | 3.57 | 4.20 | 4.55 | 5.02 | 5.56 | 5.47 | 5.74 | 6.07 |
| −1.48[b] | | 0.235 . . . . . . . | 1.65 | 3.58 | 4.63 | 5.13 | 5.69 | 5.91 | 6.41 | 6.40 | 6.65 | 6.92 |
| −0.47[c] | | 0.247 . . . . . . . | 4.49 | 6.34 | 7.60 | 8.37 | 8.84 | 9.55 | 9.82 | 10.3 | 10.8 | 11.2 |
| 0.00 | 0.017 | 0.270 . . . . . . . | 7.18 | 9.56 | 10.7 | 11.5 | 12.0 | 12.4 | 12.8 | 13.3 | 13.4 | 13.8 |
| 0.38 | 0.04 | 0.270 . . . . . . . | 10.5 | 12.4 | 13.1 | 13.6 | 14.2 | 14.7 | 15.0 | 15.5 | 16.0 | 16.2 |
| 0.58 | 0.06 | 0.270 . . . . . . . | 11.8 | 13.5 | 14.3 | 14.8 | 15.5 | 16.0 | 16.6 | 17.0 | 17.6 | 18.1 |
| 0.43 | 0.04 | 0.339 . . . . . . . | 10.2 | 12.1 | 12.8 | 13.4 | 14.0 | 14.2 | 14.7 | 15.1 | 15.3 | 15.6 |
| 0.71 | 0.06 | 0.450 . . . . . . . | 11.1 | 12.7 | 13.3 | 13.7 | 14.3 | 14.7 | 15.0 | 15.4 | 15.9 | 16.3 |

(b) 25 − V

| [Fe/H] | Z | Y | Age (Gyr) | | | | | | | | | |
|---|---|---|---|---|---|---|---|---|---|---|---|---|
| | | | 2 | 4 | 6 | 8 | 10 | 12 | 14 | 16 | 18 | 20 |
| −2.26[a] | | 0.235 . . . . . . . | 0.485 | 0.862 | 1.15 | 1.29 | 1.34 | 1.53 | 1.77 | 1.61 | 1.72 | 1.85 |
| −1.48[b] | | 0.235 . . . . . . . | 0.844 | 1.18 | 1.50 | 1.63 | 1.89 | 1.88 | 2.09 | 2.01 | 2.10 | 2.22 |
| −0.47[c] | | 0.247 . . . . . . . | 1.66 | 2.18 | 2.61 | 2.77 | 2.86 | 3.14 | 3.16 | 3.33 | 3.53 | 3.77 |
| 0.00 | 0.017 | 0.270 . . . . . . . | 2.47 | 3.08 | 3.42 | 3.83 | 4.00 | 4.20 | 4.40 | 4.71 | 4.69 | 4.87 |
| 0.38 | 0.04 | 0.270 . . . . . . . | 3.42 | 4.23 | 4.56 | 4.79 | 5.08 | 5.44 | 5.52 | 5.81 | 6.09 | 6.15 |
| 0.58 | 0.06 | 0.270 . . . . . . . | 3.98 | 4.82 | 5.14 | 5.43 | 5.79 | 6.02 | 6.38 | 6.50 | 6.78 | 7.07 |
| 0.43 | 0.04 | 0.339 . . . . . . . | 3.34 | 4.08 | 4.38 | 4.68 | 5.06 | 5.12 | 5.40 | 5.68 | 5.69 | 5.90 |
| 0.71 | 0.06 | 0.450 . . . . . . . | 3.76 | 4.45 | 4.70 | 4.92 | 5.26 | 5.50 | 5.57 | 5.81 | 6.08 | 6.33 |

(c) 33 − V

| [Fe/H] | Z | Y | Age (Gyr) | | | | | | | | | |
|---|---|---|---|---|---|---|---|---|---|---|---|---|
| | | | 2 | 4 | 6 | 8 | 10 | 12 | 14 | 16 | 18 | 20 |
| −2.26[a] | | 0.235 . . . . . . . | 0.382 | 0.505 | 0.569 | 0.586 | 0.583 | 0.669 | 0.785 | 0.686 | 0.736 | 0.797 |
| −1.48[b] | | 0.235 . . . . . . . | 0.550 | 0.579 | 0.671 | 0.707 | 0.855 | 0.807 | 0.908 | 0.849 | 0.897 | 0.954 |
| −0.47[c] | | 0.247 . . . . . . . | 0.758 | 0.880 | 1.08 | 1.12 | 1.14 | 1.29 | 1.27 | 1.36 | 1.47 | 1.59 |
| 0.00 | 0.017 | 0.270 . . . . . . . | 0.905 | 1.11 | 1.24 | 1.45 | 1.48 | 1.57 | 1.64 | 1.77 | 1.73 | 1.80 |
| 0.38 | 0.04 | 0.270 . . . . . . . | 1.22 | 1.61 | 1.76 | 1.86 | 1.99 | 2.19 | 2.19 | 2.33 | 2.48 | 2.45 |
| 0.58 | 0.06 | 0.270 . . . . . . . | 1.43 | 1.86 | 1.98 | 2.10 | 2.30 | 2.38 | 2.56 | 2.56 | 2.69 | 2.82 |
| 0.43 | 0.04 | 0.339 . . . . . . . | 1.18 | 1.54 | 1.66 | 1.80 | 2.02 | 2.00 | 2.15 | 2.31 | 2.26 | 2.37 |
| 0.71 | 0.06 | 0.450 . . . . . . . | 1.35 | 1.68 | 1.79 | 1.88 | 2.07 | 2.22 | 2.19 | 2.31 | 2.45 | 2.60 |

[a] [O/Fe] = 0.50    [b] [O/Fe] = 0.60    [c] [O/Fe] = 0.23

TABLE 4

MAIN SEQUENCE TURNOFF MASSES

| [Fe/H] | Z | Y | Age (Gyr) | | | | | | | | | |
|---|---|---|---|---|---|---|---|---|---|---|---|---|
| | | | 2 | 4 | 6 | 8 | 10 | 12 | 14 | 16 | 18 | 20 |
| −2.26[a] | | 0.235 . . . . . . . | 1.34 | 1.10 | 0.98 | 0.91 | 0.85 | 0.81 | 0.78 | 0.75 | 0.73 | 0.71 |
| −1.48[b] | | 0.235 . . . . . . . | 1.29 | 1.09 | 0.97 | 0.91 | 0.86 | 0.82 | 0.79 | 0.76 | 0.74 | 0.72 |
| −0.47[c] | | 0.247 . . . . . . . | 1.44 | 1.11 | 1.00 | 0.95 | 0.91 | 0.87 | 0.85 | 0.82 | 0.80 | 0.78 |
| 0.00 | 0.017 | 0.270 . . . . . . . | 1.50 | 1.22 | 1.05 | 1.00 | 0.96 | 0.92 | 0.89 | 0.87 | 0.85 | 0.83 |
| 0.38 | 0.04 | 0.270 . . . . . . . | 1.69 | 1.36 | 1.23 | 1.10 | 1.05 | 1.02 | 1.00 | 0.97 | 0.95 | 0.93 |
| 0.58 | 0.06 | 0.270 . . . . . . . | 1.76 | 1.41 | 1.27 | 1.18 | 1.09 | 1.06 | 1.03 | 1.01 | 1.00 | 0.98 |
| 0.43 | 0.04 | 0.339 . . . . . . . | 1.48 | 1.18 | 1.02 | 0.95 | 0.92 | 0.90 | 0.88 | 0.86 | 0.84 | 0.82 |
| 0.71 | 0.06 | 0.450 . . . . . . . | 1.20 | 0.95 | 0.86 | 0.81 | 0.78 | 0.75 | 0.72 | 0.70 | 0.68 | 0.67 |

[a] [O/Fe] = 0.50    [b] [O/Fe] = 0.60    [c] [O/Fe] = 0.23

Finally, Table 6 lists values of $\dot{n}_0(t)$ for each composition and age, expressed in units of number of stars

TABLE 5

RED GIANT TIP MASSES

| [Fe/H] | $Z$ | $Y$ | Age (Gyr) | | | | | | | | | |
|---|---|---|---|---|---|---|---|---|---|---|---|---|
| | | | 2 | 4 | 6 | 8 | 10 | 12 | 14 | 16 | 18 | 20 |
| − 2.26[a] | | 0.235 . . . . . . . | 1.42 | 1.16 | 1.03 | 0.95 | 0.89 | 0.85 | 0.81 | 0.78 | 0.76 | 0.74 |
| − 1.48[b] | | 0.235 . . . . . . . | 1.44 | 1.17 | 1.04 | 0.96 | 0.90 | 0.86 | 0.83 | 0.80 | 0.78 | 0.76 |
| − 0.47[c] | | 0.247 . . . . . . . | 1.54 | 1.27 | 1.13 | 1.05 | 0.99 | 0.94 | 0.90 | 0.87 | 0.85 | 0.82 |
| 0.00 | 0.017 | 0.270 . . . . . . . | 1.63 | 1.35 | 1.20 | 1.11 | 1.05 | 1.00 | 0.97 | 0.93 | 0.90 | 0.88 |
| 0.38 | 0.04 | 0.270 . . . . . . . | 1.81 | 1.49 | 1.34 | 1.25 | 1.18 | 1.12 | 1.08 | 1.05 | 1.01 | 0.99 |
| 0.58 | 0.04 | 0.270 . . . . . . . | 1.87 | 1.56 | 1.38 | 1.29 | 1.23 | 1.17 | 1.13 | 1.09 | 1.06 | 1.04 |
| 0.43 | 0.06 | 0.339 . . . . . . . | 1.58 | 1.30 | 1.17 | 1.09 | 1.03 | 0.98 | 0.95 | 0.91 | 0.89 | 0.87 |
| 0.71 | 0.06 | 0.450 . . . . . . . | 1.30 | 1.07 | 0.96 | 0.89 | 0.85 | 0.80 | 0.78 | 0.75 | 0.73 | 0.71 |

[a] [O/Fe] = 0.50  [b] [O/Fe] = 0.60  [c] [O/Fe] = 0.23

TABLE 6

STARS LEAVING MAIN SEQUENCE / GYR / $L_V^\odot$  ($\dot{n}_0(t)$)

| [Fe/H] | $Z$ | $Y$ | Age (Gyr) | | | | | | | | | |
|---|---|---|---|---|---|---|---|---|---|---|---|---|
| | | | 2 | 4 | 6 | 8 | 10 | 12 | 14 | 16 | 18 | 20 |
| − 2.26[a] | | 0.235 . . . . . . . | 0.032 | 0.030 | 0.031 | 0.033 | 0.036 | 0.031 | 0.029 | 0.035 | 0.032 | 0.029 |
| − 1.48[b] | | 0.235 . . . . . . . | 0.028 | 0.034 | 0.031 | 0.032 | 0.031 | 0.033 | 0.030 | 0.039 | 0.035 | 0.032 |
| − 0.47[c] | | 0.247 . . . . . . . | 0.031 | 0.039 | 0.034 | 0.035 | 0.039 | 0.034 | 0.041 | 0.039 | 0.037 | 0.040 |
| 0.00 | 0.017 | 0.270 . . . . . . . | 0.070 | 0.076 | 0.065 | 0.042 | 0.042 | 0.045 | 0.044 | 0.041 | 0.051 | 0.047 |
| 0.38 | 0.04 | 0.270 . . . . . . . | 0.072 | 0.055 | 0.065 | 0.058 | 0.045 | 0.030 | 0.050 | 0.050 | 0.052 | 0.062 |
| 0.58 | 0.06 | 0.270 . . . . . . . | 0.076 | 0.078 | 0.059 | 0.069 | 0.056 | 0.050 | 0.039 | 0.035 | 0.059 | 0.061 |
| 0.43 | 0.04 | 0.339 . . . . . . . | 0.082 | 0.063 | 0.060 | 0.051 | 0.052 | 0.072 | 0.054 | 0.057 | 0.069 | 0.070 |
| 0.71 | 0.06 | 0.450 . . . . . . . | 0.110 | 0.094 | 0.078 | 0.074 | 0.072 | 0.077 | 0.084 | 0.080 | 0.082 | 0.086 |

[a] [O/Fe] = 0.50  [b] [O/Fe] = 0.60  [c] [O/Fe] = 0.23

per Gyr per unit solar $V$-luminosity[4]. These numbers have been derived from the mass function and from numerical differentiation of the isochrones. This operation is prone to numerical noise, and the contents of Table 6 are probably reliable to no better than ±20%.

## 6.2 Advanced Evolutionary Stages

Evolutionary sequences for advanced (post-RGB) stages are discussed in detail in Paper I. Briefly, for a given composition, the models have fixed core mass and varying $M_{env}^0$, thus implicitly assuming different degrees of mass loss prior to reaching the ZAHB. The models concentrate on the EHB mass range, in which stars do not reach the P-AGB, and explore in detail the AGB-manqué-type behavior that we have here associated with Galactic field sdO stars. Since our evolutionary sequences for red HB models generally do not include the P-AGB phase, we have adopted the model calculations of Schönberner (1979, 1983) to estimate this whenever appropriate.

---

[4]    The Kurucz solar model gives $(L_V/L_{bol})_\odot = 0.1179$, i.e., $L_V^\odot = 4.511 \times 10^{32}$ erg s$^{-1}$. The solar $V$-luminosity per unit wavelength is $5.176 \times 10^{29}$ erg s$^{-1}$ Å$^{-1}$.

TABLE 7

INTEGRATED ENERGIES [$E_\lambda(M_{env}; Y, Z)$] FOR LATE EVOLUTIONARY STAGES

(log $L_V^\odot$ Gyr Å$^{-1}$)

(a) [Fe/H] = −2.26 $Y$ = 0.245 $M_c$ = 0.495

| $M_{env}$ | $E_{1500}$ | $E_{2500}$ | $E_{3300}$ | $E_V$ | $E_{1500}{}^{HB}$ | $E_{2500}{}^{HB}$ | $E_{3300}{}^{HB}$ | $E_V{}^{HB}$ |
|---|---|---|---|---|---|---|---|---|
| 0.003 | −1.71 | −2.43 | −2.85 | −3.57 | −1.90 | −2.62 | −3.02 | −3.71 |
| 0.005 | −1.66 | −2.38 | −2.78 | −3.49 | −1.89 | −2.60 | −2.99 | −3.66 |
| 0.010 | −1.57 | −2.27 | −2.67 | −3.33 | −1.90 | −2.56 | −2.94 | −3.57 |
| 0.015 | −1.56 | −2.20 | −2.56 | −3.14 | −1.90 | −2.54 | −2.90 | −3.50 |
| 0.020 | −1.57 | −2.13 | −2.42 | −2.75 | −1.91 | −2.53 | −2.88 | −3.45 |
| 0.025 | −1.65 | −2.16 | −2.38 | −2.54 | −1.91 | −2.50 | −2.85 | −3.39 |
| 0.035 | −1.72 | −2.19 | −2.35 | −2.38 | −1.92 | −2.47 | −2.80 | −3.30 |
| 0.045 | −1.79 | −2.23 | −2.42 | −2.50 | −1.92 | −2.44 | −2.75 | −3.22 |
| 0.085 | −1.91 | −2.29 | −2.38 | −2.31 | −1.95 | −2.36 | −2.61 | −2.92 |
| 0.105 | −1.93 | −2.27 | −2.36 | −2.28 | −1.97 | −2.33 | −2.54 | −2.78 |
| 0.145 | −2.04 | −2.25 | −2.29 | −2.18 | −2.10 | −2.30 | −2.44 | −2.52 |
| 0.225 | −2.78 | −2.33 | −2.18 | −2.06 | −3.24 | −2.39 | −2.30 | −2.31 |
| 0.285 | −2.93 | −2.54 | −2.23 | −2.05 | −4.07 | −2.64 | −2.35 | −2.28 |
| 0.405 | −2.96 | −2.78 | −2.32 | −2.02 | −5.30 | −2.97 | −2.48 | −2.23 |

(b) [Fe/H] = −1.48 $Y$ = 0.247 $M_c$ = 0.485

| $M_{env}$ | $E_{1500}$ | $E_{2500}$ | $E_{3300}$ | $E_V$ | $E_{1500}{}^{HB}$ | $E_{2500}{}^{HB}$ | $E_{3300}{}^{HB}$ | $E_V{}^{HB}$ |
|---|---|---|---|---|---|---|---|---|
| 0.003 | −1.71 | −2.42 | −2.83 | −3.54 | −1.91 | −2.61 | −3.01 | −3.68 |
| 0.005 | −1.67 | −2.38 | −2.78 | −3.48 | −1.91 | −2.60 | −2.99 | −3.64 |
| 0.010 | −1.59 | −2.28 | −2.67 | −3.33 | −1.91 | −2.56 | −2.93 | −3.54 |
| 0.015 | −1.55 | −2.19 | −2.56 | −3.16 | −1.91 | −2.53 | −2.88 | −3.46 |
| 0.020 | −1.54 | −2.12 | −2.45 | −2.96 | −1.91 | −2.50 | −2.85 | −3.40 |
| 0.025 | −1.57 | −2.06 | −2.33 | −2.67 | −1.91 | −2.48 | −2.81 | −3.34 |
| 0.035 | −1.72 | −2.15 | −2.31 | −2.35 | −1.94 | −2.46 | −2.77 | −3.24 |
| 0.045 | −1.84 | −2.21 | −2.34 | −2.35 | −1.93 | −2.42 | −2.71 | −3.13 |
| 0.055 | −1.89 | −2.25 | −2.32 | −2.28 | −1.93 | −2.39 | −2.66 | −3.04 |
| 0.075 | −1.90 | −2.27 | −2.36 | −2.24 | −1.94 | −2.33 | −2.55 | −2.83 |
| 0.105 | −2.02 | −2.25 | −2.28 | −2.15 | −2.07 | −2.29 | −2.43 | −2.53 |
| 0.125 | −2.29 | −2.26 | −2.24 | −2.10 | −2.40 | −2.31 | −2.36 | −2.37 |
| 0.145 | −2.71 | −2.35 | −2.21 | −2.06 | −3.07 | −2.41 | −2.32 | −2.32 |
| 0.165 | −2.89 | −2.53 | −2.28 | −2.13 | −3.70 | −2.61 | −2.37 | −2.30 |
| 0.215 | −2.96 | −2.90 | −2.39 | −2.04 | −5.60 | −3.15 | −2.59 | −2.29 |
| 0.295 | −2.96 | −3.02 | −2.49 | −2.08 | −6.26 | −3.32 | −2.66 | −2.26 |
| 0.415 | −2.96 | −3.01 | −2.45 | −2.01 | −6.43 | −3.36 | −2.67 | −2.23 |

(c) [Fe/H] = −0.47 $Y$ = 0.257 $M_c$ = 0.475

| $M_{env}$ | $E_{1500}$ | $E_{2500}$ | $E_{3300}$ | $E_V$ | $E_{1500}{}^{HB}$ | $E_{2500}{}^{HB}$ | $E_{3300}{}^{HB}$ | $E_V{}^{HB}$ |
|---|---|---|---|---|---|---|---|---|
| 0.003 | −1.71 | −2.39 | −2.79 | −3.48 | −1.92 | −2.57 | −2.96 | −3.62 |
| 0.005 | −1.66 | −2.33 | −2.72 | −3.40 | −1.92 | −2.55 | −2.92 | −3.55 |
| 0.010 | −1.61 | −2.24 | −2.61 | −3.24 | −1.92 | −2.51 | −2.86 | −3.44 |
| 0.015 | −1.57 | −2.16 | −2.51 | −3.08 | −1.93 | −2.49 | −2.81 | −3.36 |
| 0.020 | −1.55 | −2.09 | −2.40 | −2.92 | −1.94 | −2.46 | −2.77 | −3.27 |
| 0.025 | −1.56 | −2.03 | −2.29 | −2.69 | −1.95 | −2.44 | −2.73 | −3.20 |
| 0.035 | −1.76 | −2.12 | −2.25 | −2.31 | −1.97 | −2.41 | −2.67 | −3.07 |
| 0.045 | −1.92 | −2.27 | −2.30 | −2.19 | −2.00 | −2.40 | −2.62 | −2.95 |
| 0.055 | −1.95 | −2.31 | −2.35 | −2.14 | −2.02 | −2.36 | −2.55 | −2.79 |
| 0.065 | −2.01 | −2.30 | −2.34 | −2.16 | −2.06 | −2.33 | −2.48 | −2.64 |
| 0.085 | −2.41 | −2.38 | −2.29 | −2.07 | −2.56 | −2.43 | −2.41 | −2.41 |
| 0.125 | −2.96 | −3.15 | −2.52 | −2.06 | −6.17 | −3.50 | −2.74 | −2.35 |
| 0.185 | −2.96 | −3.31 | −2.63 | −2.06 | −7.61 | −4.00 | −2.95 | −2.35 |
| 0.225 | −2.96 | −3.31 | −2.63 | −2.05 | −7.60 | −3.98 | −2.94 | −2.34 |
| 0.305 | −2.96 | −3.30 | −2.62 | −2.05 | −7.58 | −3.96 | −2.92 | −2.32 |
| 0.425 | −2.96 | −3.30 | −2.61 | −2.04 | −7.54 | −3.93 | −2.88 | −2.28 |

Mass loss during the later evolution, although certainly significant on the AGB and in red HB sequences, has been neglected in these calculations. We argued in Paper I that this omission does not strongly prejudice our results for integrated fluxes. Our reasoning was as follows: for sufficiently large $M_{env}^0$, the total energy

TABLE 7 *continued*

$(\log L_V^{\odot} \ \mathrm{Gyr} \ \text{Å}^{-1})$

(d) [Fe/H] = 0.00 $Y = 0.29$ $M_c = 0.469$

| $M_{\mathrm{env}}$ | $E_{1500}$ | $E_{2500}$ | $E_{3300}$ | $E_V$ | $E_{1500}{}^{HB}$ | $E_{2500}{}^{HB}$ | $E_{3300}{}^{HB}$ | $E_V{}^{HB}$ |
|---|---|---|---|---|---|---|---|---|
| 0.002 | −1.73 | −2.38 | −2.77 | −3.47 | −1.96 | −2.57 | −2.95 | −3.60 |
| 0.004 | −1.68 | −2.31 | −2.70 | −3.37 | −1.95 | −2.54 | −2.90 | −3.52 |
| 0.006 | −1.65 | −2.27 | −2.65 | −3.29 | −1.95 | −2.53 | −2.87 | −3.47 |
| 0.011 | −1.61 | −2.19 | −2.53 | −3.12 | −1.96 | −2.49 | −2.81 | −3.35 |
| 0.016 | −1.58 | −2.11 | −2.42 | −2.96 | −1.96 | −2.46 | −2.74 | −3.24 |
| 0.021 | −1.57 | −2.05 | −2.31 | −2.77 | −1.98 | −2.44 | −2.70 | −3.16 |
| 0.026 | −1.64 | −2.03 | −2.22 | −2.51 | −2.01 | −2.44 | −2.67 | −3.08 |
| 0.031 | −1.81 | −2.13 | −2.23 | −2.32 | −2.03 | −2.43 | −2.64 | −3.00 |
| 0.041 | −2.00 | −2.32 | −2.31 | −2.21 | −2.07 | −2.41 | −2.58 | −2.84 |
| 0.051 | −2.05 | −2.34 | −2.34 | −2.14 | −2.13 | −2.39 | −2.52 | −2.67 |
| 0.061 | −2.28 | −2.40 | −2.31 | −2.06 | −2.40 | −2.44 | −2.48 | −2.52 |
| 0.071 | −2.66 | −2.52 | −2.31 | −2.06 | −2.58 | −2.46 | −2.46 | −2.42 |
| 0.081 | −2.91 | −2.76 | −2.40 | −2.14 | −3.86 | −2.85 | −2.51 | −2.39 |
| 0.091 | −2.96 | −3.09 | −2.46 | −2.06 | −5.27 | −3.34 | −2.67 | −2.40 |
| 0.111 | −2.97 | −3.41 | −2.70 | −2.15 | −7.91 | −4.28 | −2.99 | −2.44 |
| 0.151 | −2.97 | −3.43 | −2.72 | −2.15 | −8.24 | −4.43 | −3.01 | −2.41 |
| 0.191 | −2.97 | −3.43 | −2.70 | −2.14 | −8.22 | −4.41 | −3.00 | −2.40 |
| 0.231 | −2.97 | −3.43 | −2.70 | −2.13 | −8.21 | −4.39 | −2.98 | −2.38 |
| 0.431 | −2.97 | −3.43 | −2.69 | −2.12 | −8.17 | −4.35 | −2.94 | −2.34 |

(e) $Z = 0.04$ $Y = 0.29$ $M_c = 0.464$

| $M_{\mathrm{env}}$ | $E_{1500}$ | $E_{2500}$ | $E_{3300}$ | $E_V$ | $E_{1500}{}^{HB}$ | $E_{2500}{}^{HB}$ | $E_{3300}{}^{HB}$ | $E_V{}^{HB}$ |
|---|---|---|---|---|---|---|---|---|
| 0.003 | −1.74 | −2.35 | −2.73 | −3.41 | −1.99 | −2.56 | −2.92 | −3.55 |
| 0.004 | −1.71 | −2.32 | −2.69 | −3.36 | −1.99 | −2.55 | −2.89 | −3.50 |
| 0.006 | −1.68 | −2.27 | −2.63 | −3.27 | −2.00 | −2.53 | −2.85 | −3.44 |
| 0.011 | −1.65 | −2.20 | −2.53 | −3.11 | −2.00 | −2.49 | −2.77 | −3.29 |
| 0.016 | −1.62 | −2.14 | −2.44 | −2.97 | −2.02 | −2.47 | −2.72 | −3.18 |
| 0.021 | −1.61 | −2.09 | −2.35 | −2.80 | −2.06 | −2.46 | −2.68 | −3.09 |
| 0.026 | −1.63 | −2.06 | −2.28 | −2.64 | −2.08 | −2.45 | −2.64 | −2.99 |
| 0.036 | −1.77 | −2.06 | −2.18 | −2.33 | −2.16 | −2.44 | −2.58 | −2.80 |
| 0.046 | −2.16 | −2.35 | −2.30 | −2.15 | −2.34 | −2.46 | −2.53 | −2.61 |
| 0.056 | −2.59 | −2.53 | −2.37 | −2.09 | −2.91 | −2.60 | −2.50 | −2.47 |
| 0.066 | −2.92 | −2.82 | −2.42 | −2.04 | −4.00 | −2.94 | −2.57 | −2.42 |
| 0.076 | −2.97 | −3.23 | −2.61 | −2.09 | −5.96 | −3.60 | −2.81 | −2.44 |
| 0.096 | −2.97 | −3.46 | −2.86 | −2.13 | −8.74 | −4.74 | −3.26 | −2.48 |
| 0.116 | −2.97 | −3.46 | −2.86 | −2.14 | −8.73 | −4.72 | −3.24 | −2.46 |
| 0.136 | −2.97 | −3.46 | −2.85 | −2.12 | −8.71 | −4.71 | −3.23 | −2.45 |
| 0.236 | −2.97 | −3.46 | −2.82 | −2.08 | −8.67 | −4.67 | −3.19 | −2.40 |
| 0.336 | −2.97 | −3.45 | −2.80 | −2.07 | −8.65 | −4.64 | −3.16 | −2.38 |

(f) $Z = 0.06$ $Y = 0.29$ $M_c = 0.458$

| $M_{\mathrm{env}}$ | $E_{1500}$ | $E_{2500}$ | $E_{3300}$ | $E_V$ | $E_{1500}{}^{HB}$ | $E_{2500}{}^{HB}$ | $E_{3300}{}^{HB}$ | $E_V{}^{HB}$ |
|---|---|---|---|---|---|---|---|---|
| 0.003 | −1.72 | −2.30 | −2.67 | −3.34 | −2.02 | −2.55 | −2.88 | −3.49 |
| 0.005 | −1.69 | −2.25 | −2.60 | −3.24 | −2.02 | −2.53 | −2.84 | −3.41 |
| 0.007 | −1.67 | −2.21 | −2.54 | −3.14 | −2.02 | −2.51 | −2.80 | −3.34 |
| 0.012 | −1.65 | −2.13 | −2.42 | −2.94 | −2.05 | −2.48 | −2.73 | −3.20 |
| 0.017 | −1.64 | −2.08 | −2.31 | −2.75 | −2.09 | −2.48 | −2.68 | −3.08 |
| 0.022 | −1.72 | −2.06 | −2.23 | −2.52 | −2.15 | −2.49 | −2.66 | −2.98 |
| 0.032 | −2.18 | −2.36 | −2.31 | −2.22 | −2.32 | −2.53 | −2.62 | −2.78 |
| 0.042 | −2.42 | −2.51 | −2.44 | −2.16 | −2.75 | −2.60 | −2.57 | −2.59 |
| 0.052 | −2.90 | −2.80 | −2.46 | −2.09 | −3.86 | −2.92 | −2.59 | −2.49 |
| 0.062 | −2.98 | −3.23 | −2.66 | −2.17 | −5.80 | −3.58 | −2.82 | −2.50 |
| 0.082 | −2.98 | −3.48 | −2.99 | −2.21 | −9.16 | −5.01 | −3.41 | −2.56 |
| 0.102 | −2.98 | −3.48 | −2.98 | −2.21 | −9.14 | −5.00 | −3.40 | −2.55 |
| 0.142 | −2.98 | −3.47 | −2.90 | −2.11 | −9.11 | −4.96 | −3.37 | −2.52 |
| 0.242 | −2.98 | −3.47 | −2.90 | −2.11 | −9.07 | −4.93 | −3.33 | −2.48 |
| 0.342 | −2.98 | −3.47 | −2.89 | −2.11 | −9.05 | −4.90 | −3.30 | −2.45 |
| 0.442 | −2.98 | −3.47 | −2.89 | −2.10 | −9.03 | −4.88 | −3.28 | −2.43 |

TABLE 7 (continued)

$(\log L_V^{\odot} \, \mathrm{Gyr} \, \text{Å}^{-1})$

(g) $Z = 0.04 \; Y = 0.36 \; M_c = 0.454$

| $M_{env}$ | $E_{1500}$ | $E_{2500}$ | $E_{3300}$ | $E_V$ | $E_{1500}^{HB}$ | $E_{2500}^{HB}$ | $E_{3300}^{HB}$ | $E_V^{HB}$ |
|---|---|---|---|---|---|---|---|---|
| 0.003 . . . . . . . | −1.75 | −2.36 | −2.73 | −3.41 | −2.00 | −2.56 | −2.91 | −3.54 |
| 0.005 . . . . . . . | −1.72 | −2.31 | −2.68 | −3.33 | −1.99 | −2.53 | −2.87 | −3.47 |
| 0.007 . . . . . . . | −1.70 | −2.29 | −2.65 | −3.28 | −2.00 | −2.53 | −2.85 | −3.43 |
| 0.011 . . . . . . . | −1.68 | −2.25 | −2.59 | −3.20 | −2.01 | −2.51 | −2.81 | −3.35 |
| 0.016 . . . . . . . | −1.65 | −2.20 | −2.52 | −3.08 | −2.02 | −2.49 | −2.76 | −3.26 |
| 0.021 . . . . . . . | −1.63 | −2.15 | −2.44 | −2.93 | −2.02 | −2.47 | −2.71 | −3.18 |
| 0.026 . . . . . . . | −1.62 | −2.12 | −2.39 | −2.82 | −2.01 | −2.43 | −2.66 | −3.09 |
| 0.036 . . . . . . . | −1.59 | −2.06 | −2.30 | −2.67 | −1.98 | −2.36 | −2.56 | −2.93 |
| 0.046 . . . . . . . | −1.58 | −2.01 | −2.22 | −2.51 | −2.01 | −2.33 | −2.48 | −2.74 |
| 0.056 . . . . . . . | −1.62 | −1.99 | −2.15 | −2.34 | −2.11 | −2.36 | −2.44 | −2.58 |
| 0.066 . . . . . . . | −1.75 | −2.02 | −2.10 | −2.18 | −2.25 | −2.43 | −2.46 | −2.48 |
| 0.076 . . . . . . . | −2.22 | −2.27 | −2.16 | −2.03 | −2.46 | −2.51 | −2.47 | −2.39 |
| 0.086 . . . . . . . | −2.63 | −2.58 | −2.35 | −2.00 | −2.95 | −2.66 | −2.50 | −2.33 |
| 0.096 . . . . . . . | −2.92 | −2.81 | −2.43 | −2.01 | −3.83 | −2.92 | −2.56 | −2.30 |
| 0.121 . . . . . . . | −2.97 | −3.43 | −2.74 | −2.03 | −8.12 | −4.41 | −3.05 | −2.32 |
| 0.246 . . . . . . . | −2.97 | −3.45 | −2.76 | −2.02 | −8.54 | −4.53 | −3.05 | −2.27 |
| 0.346 . . . . . . . | −2.97 | −3.45 | −2.75 | −2.01 | −8.51 | −4.51 | −3.03 | −2.24 |
| 0.446 . . . . . . . | −2.97 | −3.44 | −2.74 | −2.00 | −8.49 | −4.49 | −3.01 | −2.23 |

(h) $Z = 0.06 \; Y = 0.46 \; M_c = 0.434$

| $M_{env}$ | $E_{1500}$ | $E_{2500}$ | $E_{3300}$ | $E_V$ | $E_{1500}^{HB}$ | $E_{2500}^{HB}$ | $E_{3300}^{HB}$ | $E_V^{HB}$ |
|---|---|---|---|---|---|---|---|---|
| 0.006 . . . . . . . | −1.78 | −2.36 | −2.73 | −3.39 | −2.04 | −2.57 | −2.90 | −3.51 |
| 0.011 . . . . . . . | −1.74 | −2.31 | −2.66 | −3.29 | −2.02 | −2.53 | −2.84 | −3.41 |
| 0.016 . . . . . . . | −1.71 | −2.27 | −2.61 | −3.22 | −2.02 | −2.51 | −2.81 | −3.36 |
| 0.021 . . . . . . . | −1.67 | −2.22 | −2.55 | −3.14 | −1.97 | −2.45 | −2.74 | −3.28 |
| 0.026 . . . . . . . | −1.65 | −2.19 | −2.51 | −3.10 | −1.93 | −2.41 | −2.69 | −3.22 |
| 0.036 . . . . . . . | −1.60 | −2.13 | −2.43 | −2.97 | −1.87 | −2.33 | −2.59 | −3.07 |
| 0.046 . . . . . . . | −1.59 | −2.10 | −2.37 | −2.80 | −1.87 | −2.30 | −2.52 | −2.88 |
| 0.056 . . . . . . . | −1.58 | −2.08 | −2.34 | −2.68 | −1.88 | −2.30 | −2.49 | −2.75 |
| 0.066 . . . . . . . | −1.56 | −2.07 | −2.32 | −2.59 | −1.88 | −2.29 | −2.48 | −2.65 |
| 0.076 . . . . . . . | −1.55 | −2.05 | −2.29 | −2.52 | −1.89 | −2.29 | −2.46 | −2.58 |
| 0.086 . . . . . . . | −1.54 | −2.03 | −2.27 | −2.45 | −1.90 | −2.30 | −2.45 | −2.51 |
| 0.106 . . . . . . . | −1.51 | −1.99 | −2.21 | −2.34 | −1.91 | −2.29 | −2.42 | −2.40 |
| 0.116 . . . . . . . | −1.49 | −1.95 | −2.17 | −2.28 | −1.90 | −2.28 | −2.40 | −2.35 |
| 0.141 . . . . . . . | −1.62 | −2.00 | −2.13 | −2.09 | −1.96 | −2.29 | −2.37 | −2.25 |
| 0.166 . . . . . . . | −1.73 | −2.06 | −2.14 | −2.03 | −2.07 | −2.35 | −2.36 | −2.16 |
| 0.216 . . . . . . . | −2.69 | −2.48 | −2.23 | −1.87 | −3.21 | −2.77 | −2.45 | −2.03 |
| 0.266 . . . . . . . | −2.98 | −3.45 | −2.71 | −1.90 | −8.63 | −4.48 | −2.88 | −2.03 |
| 0.366 . . . . . . . | −2.98 | −3.45 | −2.69 | −1.88 | −8.60 | −4.46 | −2.86 | −2.01 |
| 0.366 . . . . . . . | −2.98 | −3.45 | −2.69 | −1.88 | −8.60 | −4.46 | −2.86 | −2.01 |

radiated in any passband will change little with mass since the evolutionary tracks are almost coincident. If the ZAHB envelope mass is small enough that the outer envelope changes from convective to radiative structure as a result of mass loss (thus evolving rapidly blueward on the HR diagram), then the lifetime UV energy emitted by the model will be enhanced over its value for the non-mass losing case. However, since mass loss rates are greatest at higher luminosity, and mass loss is unobserved in stars with luminosities typical for the HB phase (Dupree 1986), this is likely to occur only in the post-HB stages. Thus the resultant UV energy is limited to that produced by the post-EAGB models in their transit to the white dwarf cooling track. Finally, hot radiative stars may also lose mass, by different processes than those important in cool stars. In this case, the total UV radiation will be within the range emitted by constant-mass EHB models that evolve into AGB-manqué objects. In all possible cases, therefore, we have firm limits to the UV radiation. For stars reaching the P-AGB, the energy computed from Schönberner (1983) 0.546 $M_{\odot}$ sequences ($M_{bol} \sim -3.1$) has been added to that calculated from the HB/AGB tracks. By definition, this track is actually a bright P-EAGB sequence as it does not reach the thermal pulsing regime of the AGB (see Castellani & Tornambé 1991 and Paper I). Since its luminosity is only slightly less than that of the thermally-pulsing regime ($M_{\mathrm{bol}}^{\mathrm{TP}} \sim -3.3$), we can take the energy from this sequence to be an upper bound for P-AGB stars.

The energy radiated in the required bandpasses has been computed with a procedure slightly simplified over that used for the isochrones. As much of the UV radiation arises from EHB stars with high surface gravity, and also because the Kurucz model grid does not cover the entire gravity range needed for the models, we have fixed the value of $\log g$ used for the flux calculations to be $\log g = 5$. Calculations of $\alpha_{\lambda}(T_{\mathrm{eff}}, g; Z)$ for different surface gravities indicated that it varies by only $10 - 15\,\%$ even for large changes in the surface gravity. In addition, for $T_{\mathrm{eff}} > 50000\,\mathrm{K}$ we selected stellar atmospheres from the models given by Clegg & Middlemass (1987), with gravities lying along the evolutionary path of the tracks in the $(\log T_{\mathrm{eff}}, \log g)$ diagram ranging from $\log g = 5$ to 6.5. The location of the continuum for models from this set matches well that produced by those from Kurucz database of similar $T_{\mathrm{eff}}$ and $\log g$. However, even at wavelengths as short as 1500Å the flux is not very sensitive to stars at these high temperatures, partly because the evolution is rapid and partly because $\alpha_{1500}$ is decreasing in this temperature range.

The results for the integrated energies of individual tracks are given in Table 7 (a) – (h), for each of the composition choices in the model grid. The first column gives $M_{\mathrm{env}}^0$, and the next four columns give respectively the lifetime integrated energy per unit wavelength from the evolutionary sequences in the passbands $1500 \pm 300$, $2500 \pm 300$, $3300 \pm 300$ Å and in $V$. The units of these energies are $L_V^{\odot}$ Gyr Å$^{-1}$ (see footnote 4 above for the definition of $L_V^{\odot}$). Columns 5 to 8 give the same quantities for the energy radiated in the HB phase alone. Data selected from these tables are plotted in Figure 6, which highlights the variations in far-UV and $V$-band energy with composition. In Figs. 6 (a) and (b) the energies are shown for [Fe/H] $= -2.26$ and 0 respectively. Figs. 6 (c) and 6 (d) illustrate the same for [Fe/H] $= 0.58, 0.71$, which have similar envelope abundance ($Z = 0.06$), but different $Y$. The first has $Y = 0.29$ (assuming $\Delta Y/\Delta Z = 0$ for $Z > Z_{\odot}$) and the second has 0.46 ($\Delta Y/\Delta Z = 4$). Thick lines show the total energy, while the thinner lines show the energy from the HB phase alone. Table 8 gives the integrated energies calculated from the Schönberner (1979, 1983) P-AGB tracks. Note that the values presented in Table 7 include the flux from the $0.546\,M_{\odot}$ sequence.

The key points that can be seen in these graphs are:

(1) The HB and post-HB contributions of EHB stars in the UV are similar in magnitude, and the peak flux at 1500 Å is reached at a value of $M_{\mathrm{env}}^0 \sim 0.02 M_{\odot}$. Therefore, our tracks, despite the fact that they do not quite reach $M_{\mathrm{env}}^0 = 0$, can produce reliable estimates of far-UV contribution from advanced stages of evolution[5].

(2) The far-UV energy produced by a population of EHB stars is approximately 20–30 times greater than can be produced by a similar number of P-AGB stars, a difference of 3–4 magnitudes in $15 - V$.

(3) The 1500 Å energy from the models with the smallest envelope masses varies relatively little with metallicity: $E_{1500}(M_{\mathrm{env}}^0)$ is, however, somewhat greater for the metal-rich, high-helium models (Fig. 6d).

(4) For the lowest metallicity models, there is a significant range of $M_{\mathrm{env}}^0$ in which the far-UV energy is dominated by the HB phase. These are the blue HB stars that evolve to the AGB after core helium exhaustion. For metal-rich compositions, only EHB stars (with large post-HB UV flux) contribute to the UV.

(5) For the solar and super-solar abundance tracks, the far-UV energy output changes dramatically (by factors $\lesssim 5$) for very small changes in $M_{\mathrm{env}}^0$ amounting to only $\sim 0.05 M_{\odot}$. This means that UV fluxes can be strongly sensitive to the precise nature of the mass-loss process.

---

[5] This is not necessarily true at shorter wavelengths: for example, $E_{1000}$ is still rising for our least massive model.

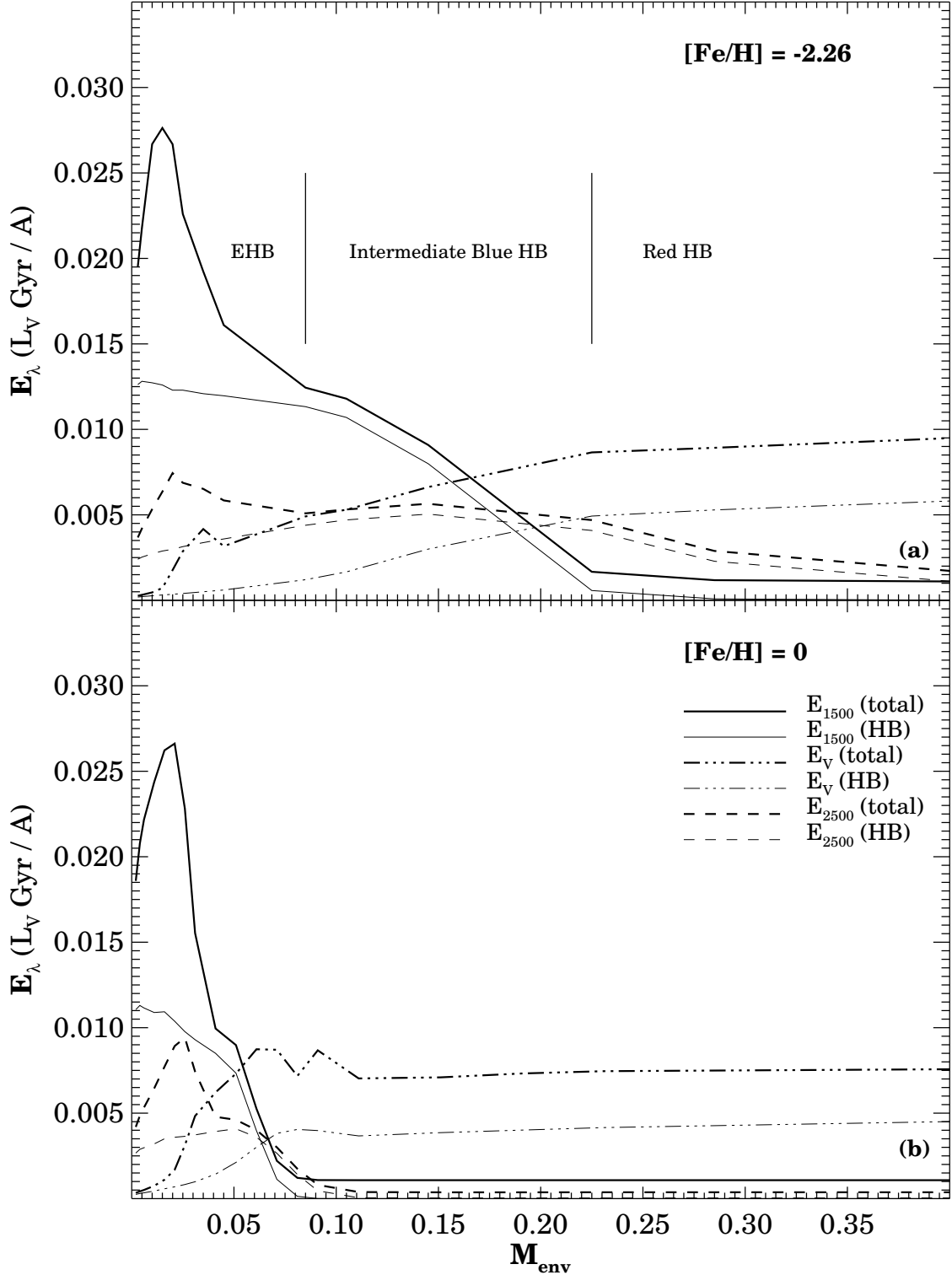

**Figure 6:** *Integrated (lifetime) energy output $E_\lambda(M_{env}^0)$ from HB/post-HB evolutionary tracks as a function of envelope mass, in units $L_V^\odot\,Gyr\,\text{Å}^{-1}$. Energies are plotted for the passband $1500\pm300$ Å (solid line), $2500\pm300$ Å (dashed line), and the $V$-filter (dashed-dotted line). The curves plotted show the total energy emitted during the HB and post-HB by thicker lines of each type, and from the HB phase alone by thinner lines. The compositions plotted in the different panels are (a) [Fe/H] = $-2.26$; (b) [Fe/H] = $-0.47$.*

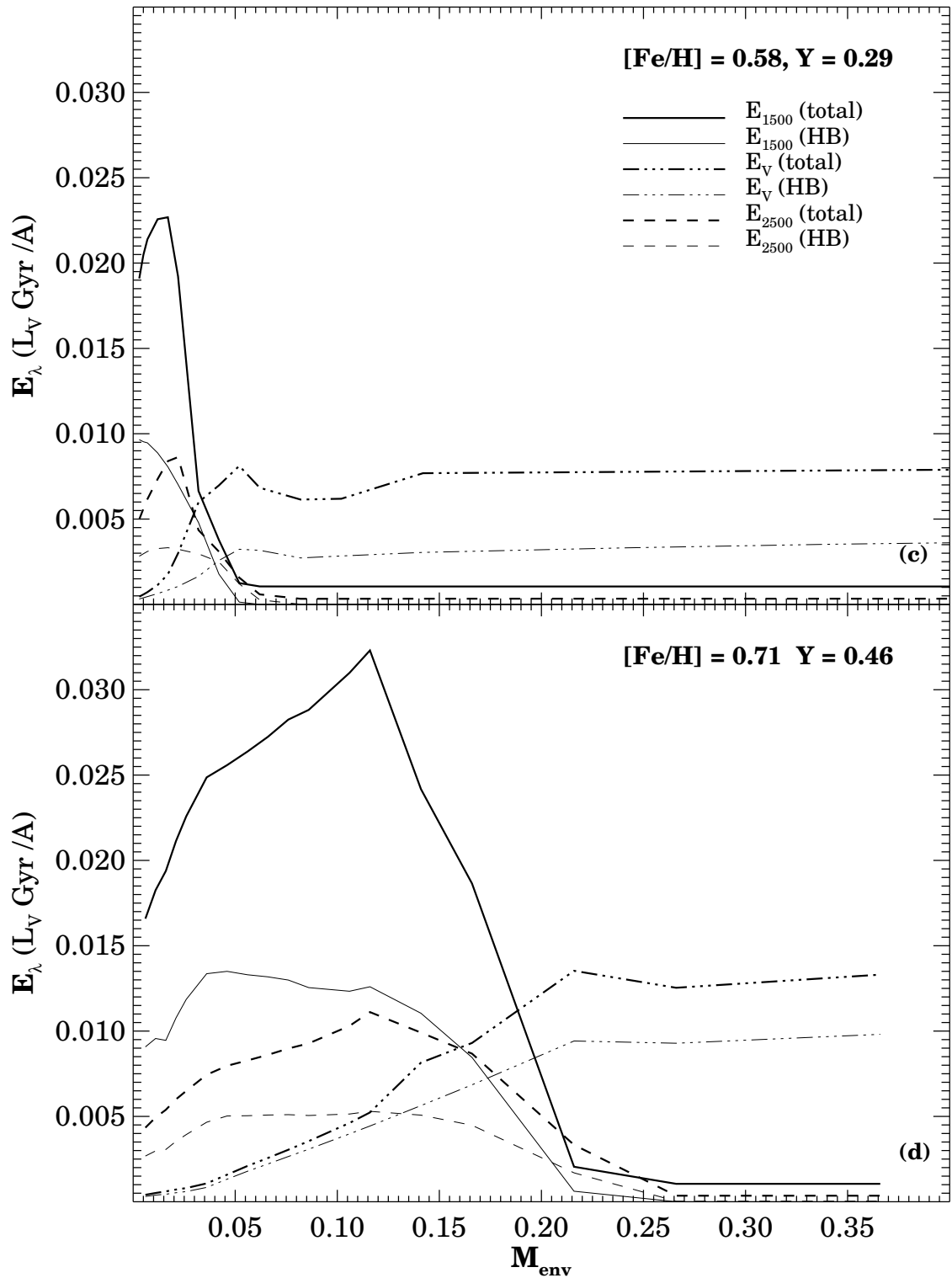

**Figure 6** (continued:) *(c)* [Fe/H] = +0.58, *Y* = 0.29; *(d)* [Fe/H] = 0.71, *Y* = 0.46.

TABLE 8

INTEGRATED ENERGIES [$E_\lambda(M_c)$] FOR P-AGB TRACKS

($\log L_V^\odot$ Gyr Å$^{-1}$)

| $M_c$ | $E_{1500}$ | $E_{2500}$ | $E_{3300}$ | $E_V$ |
|---|---|---|---|---|
| 0.546...... | −2.97 | −3.52 | −3.77 | −4.07 |
| 0.565...... | −3.80 | −4.40 | −4.65 | −4.84 |
| 0.598...... | −4.04 | −4.61 | −4.79 | −4.86 |
| 0.644...... | −4.46 | −5.02 | −5.01 | −4.86 |

(6) In each diagram, the $V$-band energy is virtually unchanged for a significant range in $M_{env}^0$ : this corresponds to the stars at the red end of the HB, and comprises nearly the entire range of $M_{env}^0$ in Figs. 6 (b) and (c).

(7) For the range of $M_{env}^0$ in which the $V$-band output is constant, the ratio of the integrated lifetime output in the $V$-band emitted during HB evolution only to the total for post-RGB stages is $\sim 0.5 \pm 0.1$.

$E_{1500}$ is extremely sensitive to $M_{env}^0$, despite the fact that the fuel consumed during the HB phase varies relatively little with envelope mass. There are two reasons for this. First, the core helium burning phase produces strong UV flux only if the stars are hot; this means the envelope must contain very little mass. Second, the post-HB phases are strong in the UV only if the stars do not evolve back to the AGB: by definition, this happens only in EHB stars which have strong UV flux in the HB phase. In contrast, the flux from P-AGB stars is much easier to estimate from fuel consumption considerations.

The most important difference between the metal-poor and metal-rich compositions is that for low abundance the HB is populated at intermediate temperatures (item 4 above). In contrast, for metal-rich compositions most stars will be either very hot (EHB) or very cool (red HB); only for a very small range of masses do metal-rich stars spend significant fractions of their lifetimes between these extremes as IBHB stars. As is readily apparent from the globular cluster record, IBHB stars tend to dominate the HB population at low metallicity. Fig. 6a indicates the ranges of masses occupied by EHB, IBHB, and red HB stars. The UV output from the HB is clearly seen to be a strong function of HB morphology.

It is clear from item (2) that EHB stars are excellent candidates for the source of the observed UV upturns. In addition, item (3) indicates that it is not necessary to invoke large degrees of helium enhancement at high metallicity to obtain large UV fluxes, provided that mass loss rates are sufficiently large at high $Z$. We will discuss the issues related to the value of $Y$ in §8. For the next sections, we adopt the lower $Y$ set of models to represent high metallicity populations.

## 6.3    Envelope Mass Distribution and Model Construction

We now have all the necessary ingredients to produce theoretical colors for old stellar populations with arbitrary ZAHB mass distributions. The expression for the integrated luminosity of a population at any wavelength is obtained by combining equations (5.1), (5.8), and (5.10):

$$\mathcal{L}_\lambda = \mathcal{L}_\lambda^E(t, Y, Z) + \dot{n}_0(t)\mathcal{L}_V^E \int_0^{M_{RGB}-M_c} P(m)E_\lambda(m; Y, Z)\,dm. \qquad (6.1)$$

The remaining problem is how to characterize the $P(m)$ function that describes the distribution over envelope mass, $M_{env}^0$. At present, there is no theoretical basis for predicting a realistic $P(m)$, which is determined by mass loss on the RGB. Rood (1973), Lee, Demarque, & Zinn (1990, 1994), and Catelan (1993) have all used modifications of a Gaussian mass distribution to model cluster HBs. However, as noted by Rood (1991), the assumed mass distribution *was completely arbitrary* and was motivated by a vague allegiance to the central limit theorem and by the fact that it worked for the clusters in question. In clusters with significant populations of very hot HB stars there has been to date only limited information. For M79, which has a relatively rich HB observed out to the hot end of the ZAHB, Dixon *et al.* (1994) find that the HUT far-UV spectrum is inconsistent with a symmetrical mass distribution, but consistent with a distribution that is distinctly skewed towards the cool end ($T_{\rm eff} \approx 8500\,{\rm K}$). In general, the mass distribution may be quite complex. Crocker, Rood & O'Connell (1988), and Rood & Crocker (1989) note that many clusters with extended blue HB's have bimodal distributions which could arise from a bimodal mass distribution. Perhaps the most extreme example of a bimodal HB is that of NGC 2808. Rood *et al.* (1993) note that the NGC 2808 HB can be explained if the two groups of stars have a mass loss efficiency differing by $\sim 40\%$.

Given this absence of hard information on the mass-distribution function, we have chosen two simple, but arbitrary, ways of characterizing $P(m)$ in our models. For both, we simplify the treatment by noting that the $E_\lambda$ functions for a given composition in Fig. 6 become roughly constant for envelope masses larger than some critical value. For [Fe/H]= 0 and +0.58, this critical envelope mass $M_{\rm e,crit}^0$ is $\sim 0.1$. It is $\sim 0.2$ for the [Fe/H] = $-2.26$ and 0.71 compositions. We can therefore approximate the contribution of all large envelope stars for a given composition by a single $E_\lambda$ function. We denote this as $\mathcal{E}_\lambda^{C}$, where the $C$ stands for "cool" ZAHB stars. The fraction of the ZAHB population which is so represented is

$$f_C = \int_{\rm M_{e,crit}}^{M_{\rm RGB} - M_c} P(m)dm.$$

Note that although stars in the $C$ component are relatively cool on the ZAHB, their $\mathcal{E}_\lambda^{C}$ s do include contributions from the subsequent hot PAGB phase, as described in §6.2.

The remainder of the population is represented in two different ways:

    (1) By a single evolutionary track ("$\delta-$function" models).

    (2) By a uniform distribution of EHB stars ("mean EHB" models).

In the first type of model, a fixed fraction $f_H$ of the stars passing through the helium flash is given a Dirac delta function mass distribution at a particular value of $M_{\rm env}^0$, which is allowed to vary. These $\delta-$function models are obviously unphysical, but they allow us to explore the range of color effects possible as a function of $M_{\rm env}^0$. They provide a bound to the location in the UV color-color plane of more realistic stellar populations, which are linear combinations of $\delta-$function models. In the "mean EHB" models, we assume that $P(m)$ is a constant for $M_{\rm env}^0 \sim 0$ to $M_{e,crit}$. The integrated energy from $f_H$ in either type of model is denoted $\mathcal{E}_\lambda^{H}$.

A model of the first type has two parameters, $f_H$ and $M_{\rm env}^0$, while the second type only has $f_H$. In general, of course, both the number fraction of UV-bright stars and their envelope mass distribution will both vary among systems. These simple idealizations are designed to separate the effects of the size of the hot population from its characteristic temperature.

With this characterization, the luminosity at any wavelength becomes

$$\mathcal{L}_\lambda = \mathcal{L}_\lambda^{\rm E}(t, Y, Z) + \dot{n}_0(t)\mathcal{L}_{\rm V}^{\rm E}\,[f_H\mathcal{E}_\lambda^{H} + (1 - f_H)\mathcal{E}_\lambda^{C}], \tag{6.2}$$

and the corresponding ratio to the V-band luminosity is

$$\mathcal{L}_\lambda/\mathcal{L}_V = \frac{(\mathcal{L}_\lambda/\mathcal{L}_V)^E + \dot{n}_0(t)[f_H\mathcal{E}_\lambda^{H} + (1 - f_H)\mathcal{E}_\lambda^{C}]}{1 + \dot{n}_0(t)[f_H\mathcal{E}_V^{H} + (1 - f_H)\mathcal{E}_V^{C}]}. \tag{6.3}$$

The predicted color is then

$$m_\lambda - V = -2.5 \log_{10}[\mathcal{L}_\lambda / \mathcal{L}_V]. \tag{6.4}$$

For use in these expressions, the $\dot{n}_0(t)$ values in Table 6 must be multiplied by a factor of 872 (the width of the V-filter in Å) when used with the results given in Table 7. Although we will use these formulae to compute synthetic colors, it is useful to note that because the $C$-component will contribute very little to the 1500 Å energy, and the $H$-component will contribute little to $V$, we can approximate the far-UV color as

$$15 - V \sim -2.5 \log_{10}\left[\frac{\dot{n}_0(t) f_H \mathcal{E}_\lambda^H}{1 + \dot{n}_0(t)(1 - f_H) \mathcal{E}_V^C}\right] \tag{6.5}$$

For any given composition and age we can find $(\mathcal{L}_\lambda / \mathcal{L}_V)^E$ and $\dot{n}_0(t)$ from the isochrones and the assumed initial mass function. $\mathcal{E}_\lambda^C$ is taken from the large envelope mass limit for each composition, and $\mathcal{E}_\lambda^H$ can be derived from the HB and post-HB models after assuming a mass distribution.

## 6.4     Comparison Between Component Colors and Observations

Figures 7a and b compare the locations in the $15 - V$ vs. $25 - V$ plane of individual components of an evolved population with those of clusters and the galaxies. In the flux-ratio plane corresponding to this magnitude plane, any composite model is a weighted vector-sum of its individual components. In this plane, one can crudely visually estimate where composite models will lie, although the logarithmic distortion of the combination trajectories is considerable because of the large color range.

The shaded areas enclose all of the galaxy data and nearly all of the cluster data plotted in Figs 1–3. Hotter (bluer) objects are higher and to the right in these plots. The dotted lines joining the symbols show the colors produced by the pre-HB isochrones for ages 4–16 Gyr, taken from Table 3. The younger populations lie to the blue for each composition. These "bare" isochrones are considerably redder in $15 - V$ than the clusters and galaxies, except for the two most metal-poor sequences. Isochrones with $Z \sim 2 Z_\odot$, thought to be representative of giant elliptical galaxies, fall over 7.5 magnitudes (a factor $> 1000$ in flux ratio) redder in $15 - V$ than the corresponding observations, for all ages considered. There are two globular clusters lying outside the darker shaded region in the Figure, *viz.* 47 Tuc (from OAO-2) and NGC 6388 (from ANS). 47 Tuc falls leftward of the shaded region marking the galaxies: it has been excluded from our analysis because its flux is affected by a P-AGB star. NGC 6388 falls within the 'galaxy' region.

In the upper right of the diagrams are the integrated outputs of the HB and later stages, for each composition and $M_{env}^0$, where the colors are given by

$$m_\lambda - V = -2.5 \log_{10}[E_\lambda(M_{env}^0) / E_V(M_{env}^0)]. \tag{6.6}$$

For each composition, the curves trace the locus of the integrated output of post-RGB phases as the envelope mass is varied from nearly zero (EHB stars, extreme upper right) to the largest possible values (red HB, lying near the blue edge of the cluster and galaxy shaded areas). All of the EHB curves are coincident at the upper right of the plots ( $15 - V < -1$). The curves separate, and the $25 - V$ color changes rapidly, at the transition between envelope masses that produce P-EAGB and AGB-manqué stars, and those that produce only P-AGB objects. It is the P-AGB component that produces the red-side asymptotic behavior of $15 - V$ with $M_{env}^0$, which, for the Schönberner (1983) 0.546 $M_\odot$ model, tends to $\sim 2$. Adopting a higher mass P-AGB track would shift this asymptotic limit redward, as will be illustrated in the next section. We have plotted the colors of the Schönberner (1979, 1983) P-AGB tracks by themselves in Fig. 7b (open triangles), with the mass decreasing from left to right. Their lifetimes also increase dramatically from left to right.

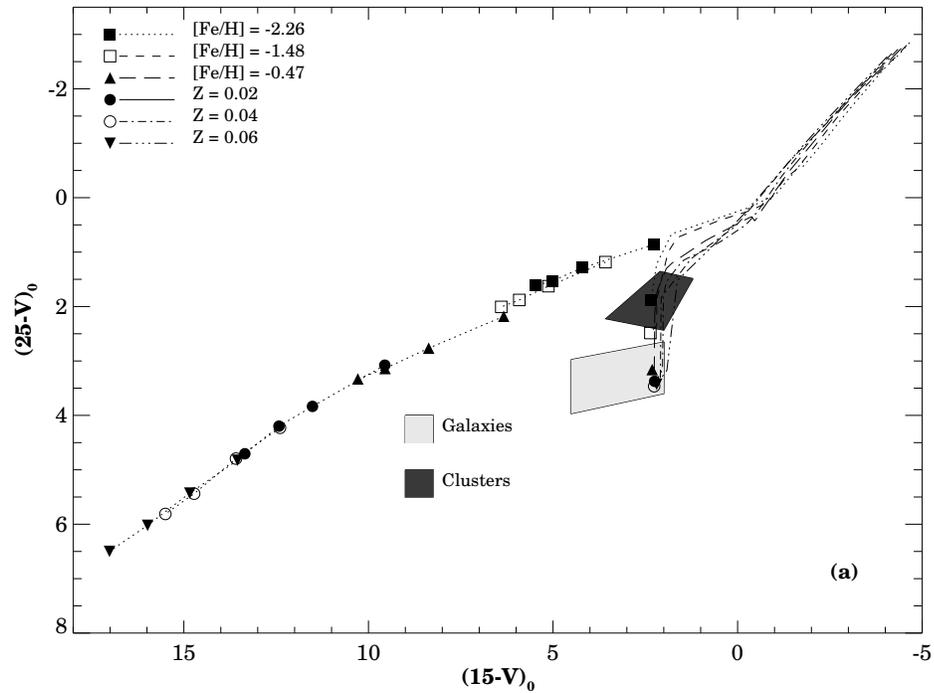

**Figure 7** *(a): Ultraviolet two-color* $(25 - V$ *vs.* $15 - V)$ *diagram showing location of theoretical components of old stellar populations. Isochrones for different metallicities are shown by symbols:* $[Fe/H] = -2.26$, *filled squares;* $[Fe/H] = -1.48$, *open squares;* $[Fe/H] = -0.47$, *filled triangles;* $[Fe/H] = 0$, *filled circles;* $[Fe/H] = 0.39$, *open circles;* $[Fe/H] = 0.58$, *inverted filled triangles, and joined by thick dotted lines. The ages plotted for each composition are 4, 8, 12, and 16 Gyr, with age decreasing left to right. Lines representing the HB/post-HB theoretical components appear in the upper right-hand corner and are denoted by the linestyles:* $[Fe/H] = -2.26$, *dotted;* $[Fe/H] = -1.48$, *short dashes;* $[Fe/H] = -0.47$, *long dashes;* $[Fe/H] = 0$, *solid;* $[Fe/H] = 0.39$, *dash-dot;* $[Fe/H] = 0.58$, *dash- three dots. The location of the data plotted in Fig. 3b is shown by the shaded regions.*

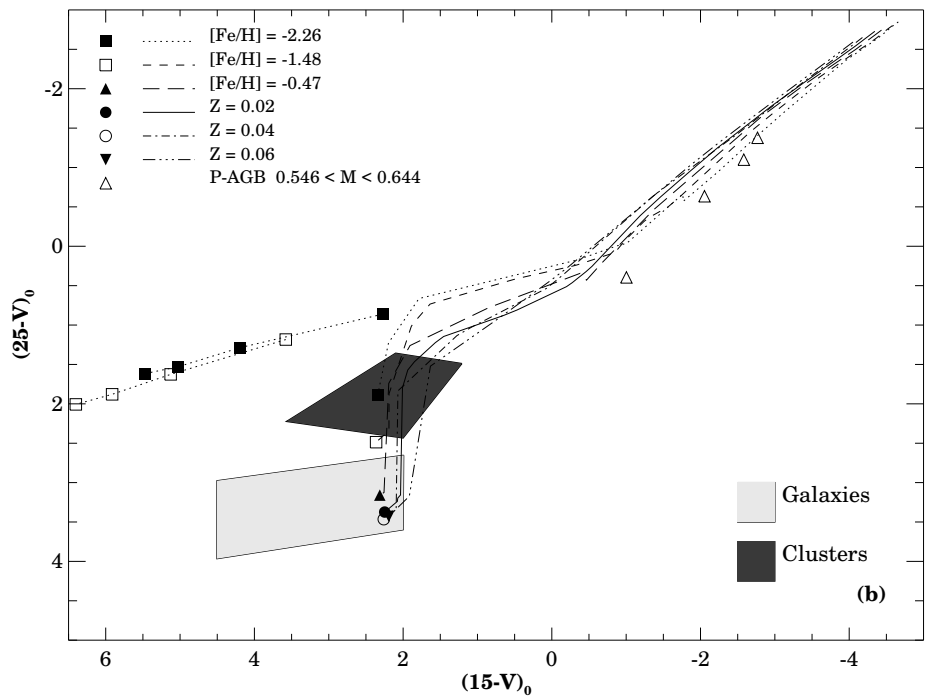

**Figure 7** *(b): as (a), except the scale focuses on the area of the diagram enclosing the data and hot components. The location of the Schönberner (1979, 1983) model colors is indicated by open triangles, with the mass decreasing from right to left.*

To repeat a conclusion from Paper I, it is clear from Fig. 7 that the UV output of post-RGB phases is not a strong function of composition, especially for small envelope masses. All compositions have the same potential for producing the UV upturn in galaxies: the determining factor is the distribution of envelope mass on the ZAHB.

To estimate the size of the hot component implied by the bluest galaxy observations, we can adopt from Table 6 the rate at which stars leave the main sequence per unit $V$ luminosity for [Fe/H] = 0 and age 10 Gyr, $\dot{n}(t) \sim 0.04$ stars/ Gyr/ $L_V^\odot$. Then from Table 7, $\log E_{1500} (L_V^\odot$ Gyr Å$^{-1}$) is typically $-1.6$ at the peak of HB integrated UV energy curves. The predicted color from equation (6.5) if $f_H = 1.0$ is thus $(15 - V)_{\min} = 0.14$, which is about 2 mag (a factor $\sim 6$ in UV output) bluer than the strongest UV upturns. This comparison implies that *the colors of the bluest galaxies can be explained if $\sim 15\%$ of the stars passing through the helium flash lose sufficient envelope mass to become strong UV-emitters.* A very similar estimate (20%) may be derived from equation (15) of GR, assuming (in their notation) $L^{\mathrm{UV}}/L_{\mathrm{T}} = 0.02$.

Other useful results can be inferred from Figure 7a. For [Fe/H] $\geq 0$, the mid-UV color of the older pre-HB isochrones is very red, with $25 - V > 4$. Since the bluest galaxies have $25 - V \lesssim 3$, this implies that the mid-UV flux from these systems is dominated by the hot component (see also B3FL). For [Fe/H] $< -1$, the pre-HB isochrones have $25 - V$ colors bluer than the galaxies. We infer that the galaxies cannot have large metal-poor populations (cf. §7.2.3). Also, isochrones with $t \geq 6$ Gyr have $15 - V$ colors redder than either the galaxies or the clusters. Therefore, all of the clusters plotted must have far-UV contributions from post-RGB stars. This is consistent with the conclusions of ABD and WC80, based on a different kind of analysis. Another inference is that *the separation between the clusters and the galaxies in $25 - V$ in Figs 4 and 7 is primarily a consequence of the composition dependence of the pre-HB light and not an effect of the hot post-RGB phases.* This suggests that the $25 - V$ color is a useful metallicity indicator, provided the hot stellar component can be reliably modeled.

# 7    Synthesis Models for Clusters and Galaxies

## 7.1    Models for Globular Clusters

We are now in a position to consider the ranges of parameters for hot stellar components that are consistent with the observations. We first compare metal-poor models to the globular cluster UV data. As noted in §6, there is an important qualitative difference between the globular clusters and the galaxies. Because of their smaller mass, the UV output of clusters is strongly affected by the stochastic effects of small numbers of UV-bright objects, e.g. P-AGB or even AGB-manqué stars, which cannot be adequately modeled by our technique. As a consequence, our models are not intended to reproduce the UV colors of specific clusters. Because of its stochastic nature, we have removed the P-AGB component from our models for globular clusters. The UV contribution from cool HB/AGB stars is thus neglected, although the $V$ contribution is included.

We use the $\delta-$function models described earlier to compare with the observations. Figures 8a and 8b compare different representations of the models with the data. The range of colors occupied by the cluster data is indicated by the shaded polygon. In Fig. 8a the lines trace out the colors produced by populations of age 14 Gyr and $f_H = 1$ as $M_{\mathrm{env}}^0$ is varied from the red HB (at the left) to the EHB (at the right). The three curves represent sets of models with [Fe/H] = $-2.26, -1.48,$ and $-0.47$. The colors of the 14 Gyr isochrones without the post-RGB phases are represented by the large symbols. Variations in age of up to 3 Gyr among clusters are consistent with current observational evidence (VandenBerg *et al.* 1990); these would cause small shifts in the $25 - V$ colors in the sense that bluer mid-UV colors correspond to younger systems (cf. Fig. 7a).

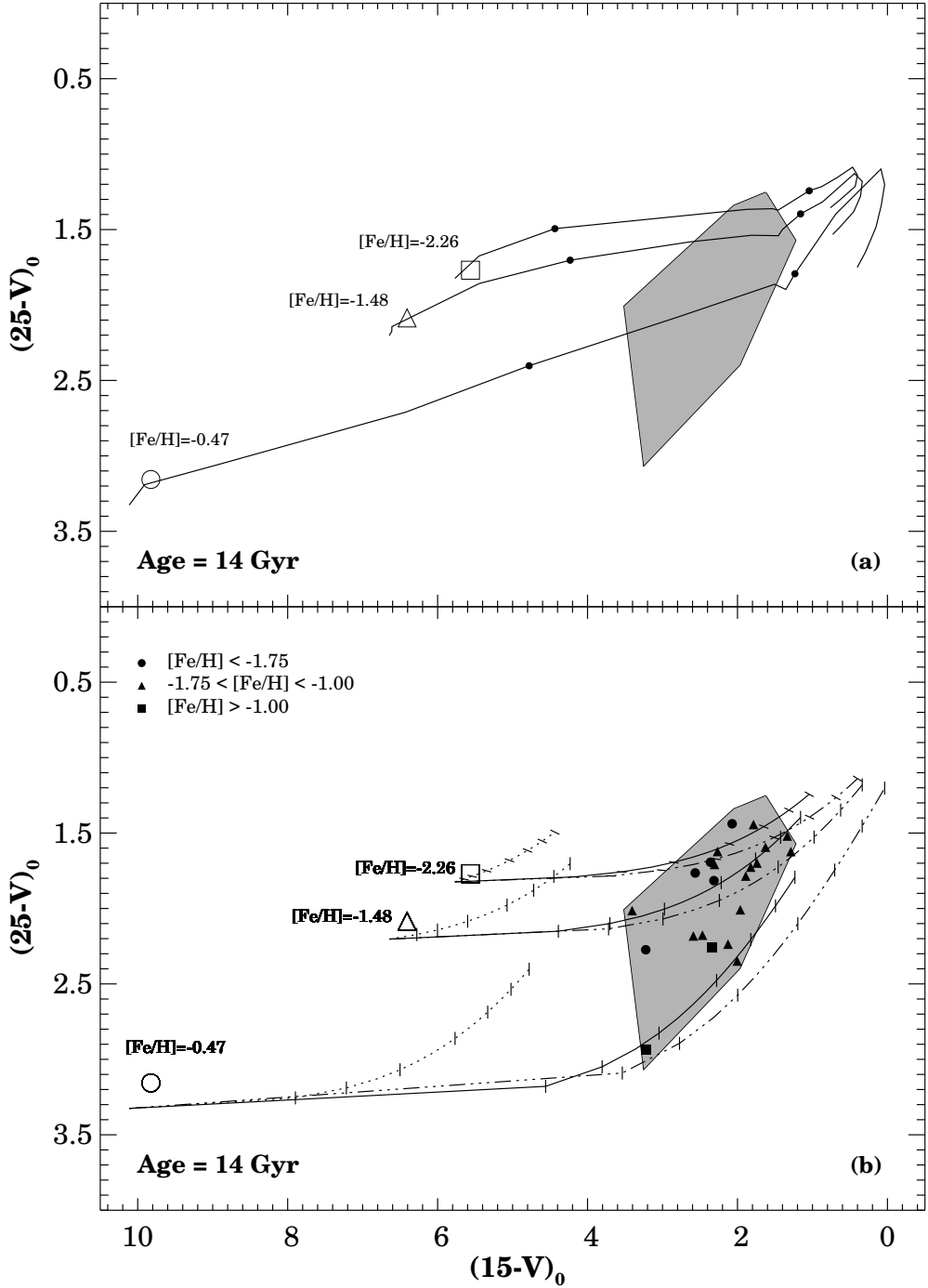

**Figure 8:** *Synthetic models for 14 Gyr populations in the $15-V$ vs. $25-V$ two-color diagram superposed on the cluster data. In both panels, the area enclosed by the cluster data is represented by the shaded polygon. (a): Models with $f_H = 1$, in which the HB component is represented by the "$\delta$-function" models, for metallicities [Fe/H] = $-2.26$, $-1.48$, and $-0.47$. $M_{env}^0$ in these models is allowed to vary over the full range of the HB, from red (lower left) to blue. The location of a pure "isochrone" component is denoted by the large symbols. The dots represent the color location of models with selected $M_{env}^0$ for each metallicity, as follows: for [Fe/H] = $-2.26$, $M_{env}^0 = 0.045$, $0.225$ $M_\odot$ (left to right); for [Fe/H] = $-1.48$, $M_{env}^0 = 0.045$, $0.145$ $M_\odot$; for [Fe/H] = $-1.48$, $M_{env}^0 = 0.045$, $0.195$ $M_\odot$. (b): The variation in the color for $0 < f_H < 1$. The solid and dotted curves represent models constructed with the same $M_{env}^0$ as in Panel (a). The dashed-dotted curve represents models with $M_{env}^0$ chosen to give the maximum possible UV flux, and so represents the limiting $15-V$ for choices of $f_H$ at 14 Gyr and for each metallicity. Vertical tickmarks show the location of models with $f_H = 0.05$, 0.10, 0.20, 0.40, 0.60, 0.80, and 1.00. The globular cluster observations are divided into metallicity classes and are represented by filled symbols: [Fe/H] < $-1.75$, circles; $-1.75 <$ [Fe/H] < $-1.00$, triangles; [Fe/H] > $-1.00$, squares.*

The shape of the solid curves can best be understood as a mapping of the energy curves of Figure 6 into this color-color plane. The red extreme of the model curves lies below and to the left of the isochrone location. The population represented by this point has an entirely red HB, and so contributes only to $V$. As the envelope mass of the model track is decreased, the models become bluer until a peak corresponding to the maximum possible integrated output is reached. Beyond this there is a shift to slightly redder colors as the stars become still hotter but less luminous (see Fig. 6 and Paper I). Dots indicate the location on each sequence for two envelope masses selected to roughly enclose the color range of the IBHB. The least massive of these is one of the cooler EHB models, while the most massive is near the red edge of or inside the RR Lyrae instability strip. For [Fe/H] $= -2.26$ the dots represent the models (right to left) with $M_{\mathrm{env}}^0 = 0.045$ and $0.225\ M_\odot$; for [Fe/H] $= -1.48$ the dots are at $M_{\mathrm{env}}^0 = 0.045$ and $0.145\ M_\odot$; while for [Fe/H] $= -0.47$ they are $M_{\mathrm{env}}^0 = 0.045$ and $0.095\ M_\odot$. As the metallicity is increased the mass range producing IBHB models shrinks appreciably from 0.18 to 0.10 to 0.05 $M_\odot$. Clearly, for [Fe/H] $= -0.47$, only a mass distribution with very small dispersion will produce an integrated UV color similar to what is commonly seen in more metal-poor clusters, requiring some degree of "tuning" of the mass loss rate.

Despite the artificiality of these "$\delta$-function" models, the curves are important because the colors of populations consisting of realistic mass distributions with fraction $f_H < 1.0$ will lie below the curves plotted in Fig. 8a. Simply, they will always have a color redder than the *bluest* stars in the distribution in each passband.

In the lower panel we show the effect of varying the size of the hot stellar component for selected $M_{\mathrm{env}}^0$s. Curves are plotted for the metallicities [Fe/H] $= -1.48$, close to the abundance of the bulk of the globular clusters with blue HB morphology, and for [Fe/H] $= -0.47$, illustrating the behavior of metal-rich systems. The dataset has been divided into 3 metallicity groups: [Fe/H] $< -1.75$, $-1.75 \leq$ [Fe/H] $< -1$, and [Fe/H] $> -1$, which are indicated by different symbols. The curves have been constructed by taking $\delta$−function models with the masses used in the upper panel, and varying $f_H$ from 0.05 to 1.0. We have added a third curve that joins models with $M_{\mathrm{env}}^0$ chosen to represent EHB stars with maximal $E_{1500\mathrm{s}}$. Vertical tickmarks from right to left indicate $f_H = 1.00$, 0.80, 0.60, 0.40, 0.20, 0.10, and 0.05.

This diagram needs to be interpreted with care since it is evident that a given color point is consistent with a variety of combinations of metallicities and hot star populations. Constraints on any particular hot component, such as EHB stars, can be improved only by considering the available information on abundance and HB structure for individual clusters. However, without doing this, we can comment on the overall consistency of the models and the data. Clusters with red HB morphology and no UV-bright sources would lie redward of the curves traced out by the most massive model plotted; no such objects are above the ANS/OAO detection limits. Clusters with "blue HB" morphology such as M13 will lie between the two lefthand curves, which approximately span the IBHB but do not include EHB populations. It is clear that the models are generally in good agreement with the data, both in terms of the $25 - V$ color, which is most sensitive to age and abundance, and $15 - V$, which measures the hot star population. Most of the clusters known to have rich IBHB populations fall in the appropriate location. In a few cases (e.g., M53 and M80) both colors appear to be too red for the HB morphology or metallicity, possibly indicating observational error. The metal-rich clusters with predominantly red HB's have UV colors consistent with the effects of small numbers of UV-bright stars. None of the intermediate metallicity or metal-poor clusters *require* the presence of EHB stars to explain their UV properties, although the models cannot rule out a significant EHB contribution, depending on the temperature distribution of the IBHB stars.

With the exception of $\omega$ Cen, clusters that have been adequately searched for EHB sources (see Whitney *et al.* 1994 and §2) are generally found to have sparse EHB populations. Fig. 8 suggests that other clusters with rich IBHB populations which have yet to be searched will likewise have only a small population of EHB stars. In the metal-rich clusters, detected far-UV flux will often indicate the presence of EHB-type objects.

The most metal rich cluster for which the data are reliable, NGC 6388, falls near the dotted EHB limiting curve at $f_H = 0.10$. For an HB population size of $\sim 100 - 200$, this represents a hot stellar component of order ten EHB stars.

For $\omega$ Cen itself the far-UV photometry from UIT (Whitney *et al.* 1994) shows that the HB contains $\sim 20 - 25\%$ EHB stars, with most of the remainder being in the IBHB category. $\omega$ Cen is the bluest object in the cluster sample in both $15 - V$ and $15 - 25$ (M62 has a similar $15 - 25$ color but has a large and uncertain reddening correction). It is redder in $25 - V$ than several other clusters, because of its higher metallicity. Based on its location in Fig. 8b, as well as its optical CM diagram, NGC 6752 is likely to be the only other cluster with an EHB population $\gtrsim 20\%$.

## 7.2    Models for Elliptical Galaxies

### 7.2.1 Large Envelope (P-AGB Only) Models

We begin our consideration of models relevant to the galaxies by illustrating metal-rich models that contain only main sequence, RGB, cool HB (i.e., large envelope), and P-AGB stars. No hot HB stars or their progeny are included (i.e., $f_C = 1$). The flux from the assumed P-AGB component, taken from the tracks of Schönberner (1979, 1983), is varied. Figure 9 compares the galaxy data to these models with [Fe/H] $= -0.47$, 0, and 0.38 and ages in the range from 4 to 16 Gyr. The galaxy data, excluding M32, is represented by a shaded region. The vertical lines in the figure represent the predicted colors for systems with various assumed P-AGB masses. Smaller ages are at the top of each line. The jaggedness in the lines reflects noise in the numerical derivatives of the isochrones used to calculate the specific evolutionary fluxes.

As expected, the most important effect of age on the models is to change the $25 - V$ contribution of the isochrone component. Age will also affect the masses of P-AGB stars formed, since we expect that younger populations will produce more luminous AGB and thus P-AGB stars. This general trend is at present impossible to quantify, but it is appropriate to assume slightly more massive P-AGB stars for intermediate-age populations. In most cases here, we adopt $M_{\text{P-AGB}} = 0.565 M_\odot$ for ages of 6 Gyr or below and $M_{\text{P-AGB}} = 0.546 M_\odot$ for older models.

The bluest set of lines is for the least massive P-AGB sequences of Schönberner (1983), $M_{\text{P-AGB}} = 0.546 M_\odot$. Recall (§6.2) that these objects are really better regarded as P-EAGB stars rather than true P-AGB stars. At the left of the diagram are models with the 0.565 $M_\odot$ P-AGB track. At fixed age, models of higher metallicity are redder in $25 - V$, although their $15 - V$ colors may be slightly bluer because the predicted evolutionary flux $\dot{n}(t)$ can be larger.

Although some of the weaker UV upturn galaxies are consistent with this kind of model, the figure demonstrates that *most of the stronger UV upturns cannot be produced by models whose only hot components are P-AGB stars.* This confirms the conclusions of Brocato *et al.* (1990), GR, Castellani & Tornambè (1991), and Magris & Bruzual (1993) that, despite their high luminosity, the available P-AGB models evolve too rapidly to produce the stronger UV upturns. The implication is that contributions from other hot phases such as the EHB and their progeny are required to match the UV colors of galaxies.

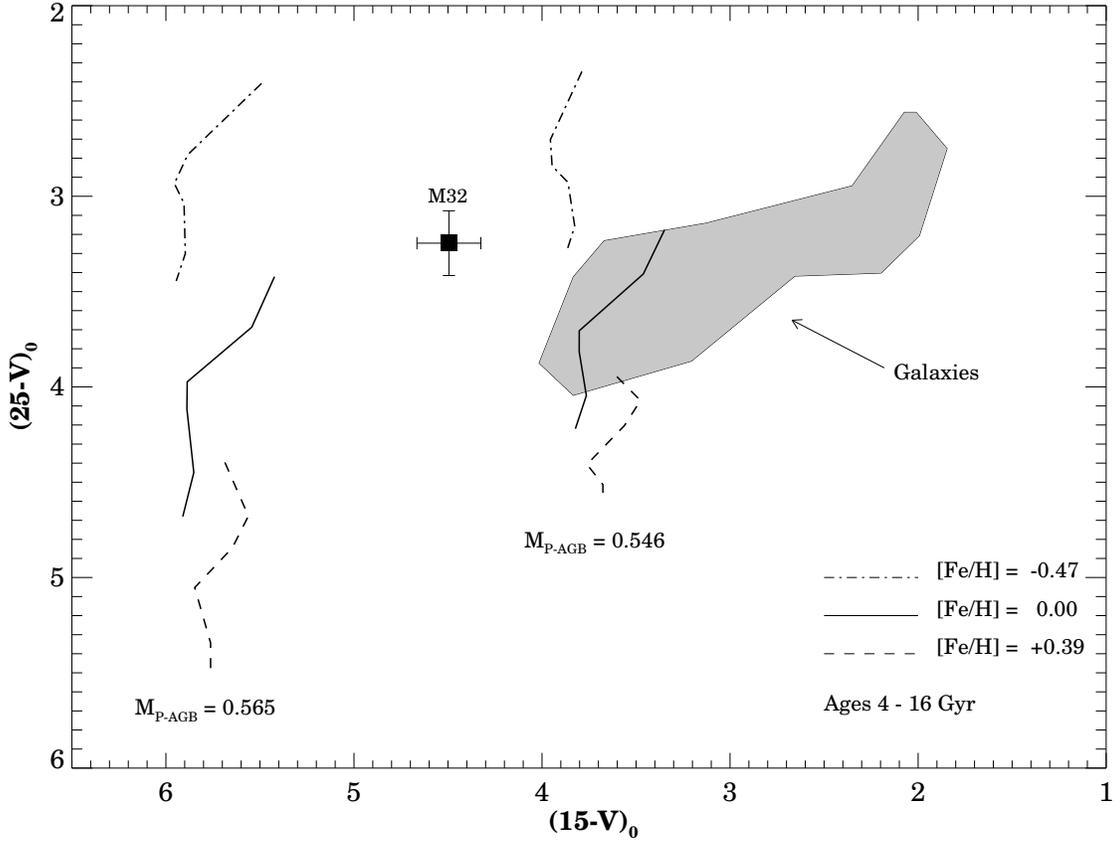

**Figure 9:** *Limits to the $15 - V$ color in hot components consisting purely of P-AGB stars. The location of the galaxy data is indicated by the shaded area, with M32 shown separately. The UV flux from the P-AGB is represented by the Schönberner (1983) sequences as indicated in the frame. The jagged lines show the limit to the UV colors at different metallicities, for ages (top to bottom) 4 to 16 Gyr. The "jaggedness" reflects variations in our computed specific evolutionary fluxes (Table 6).*

### 7.2.2 MODELS WITH EHB COMPONENTS

Figure 10 illustrates complete models including all hot components. The area enclosing the galaxy data is again shaded. The datapoints themselves are marked. The selected objects discussed in more detail below are indicated by larger filled circles. The solid lines show the locus of the $\delta$−function models for number fractions $f_H = 0.10$, $0.20$ and $0.30$ and the full range of $M_{\mathrm{env}}^0$ s. The dashed line is the locus of models with the 'mean EHB' component described earlier for $0 \leq f_H \leq 0.30$, with tickmarks placed at $f_H = 0.05$, $0.10$, $0.20$, and $0.30$. Panel (a) shows models for [Fe/H] = 0, age 10 Gyr, with P-AGB mass $M_{\mathrm{P-AGB}} = 0.546 M_\odot$. Panels (b) and (c) show models with the same parameters but metallicities [Fe/H] = +0.39, +0.58 respectively. Panel (d) shows younger models with [Fe/H] = 0, ages 4 and 6 Gyr, and $M_{\mathrm{P-AGB}} = 0.565 M_\odot$.

The shape of the solid curves again represents a mapping of the UV energy curves, this time from Figure 6b and 6c, into the color-color plane. In contrast to the metal-poor cases considered in §6.5, however, significant 1500 Å output from the metal rich models of Fig. 6b and 6c occurs *only* for very small envelope masses, $\lesssim 0.06 M_\odot$.

The near-vertical part of the curves at the left end corresponds to situations where all HB stars have large envelope masses (i.e., are red HB stars). Their far-UV color reflects the P-AGB evolutionary sequence

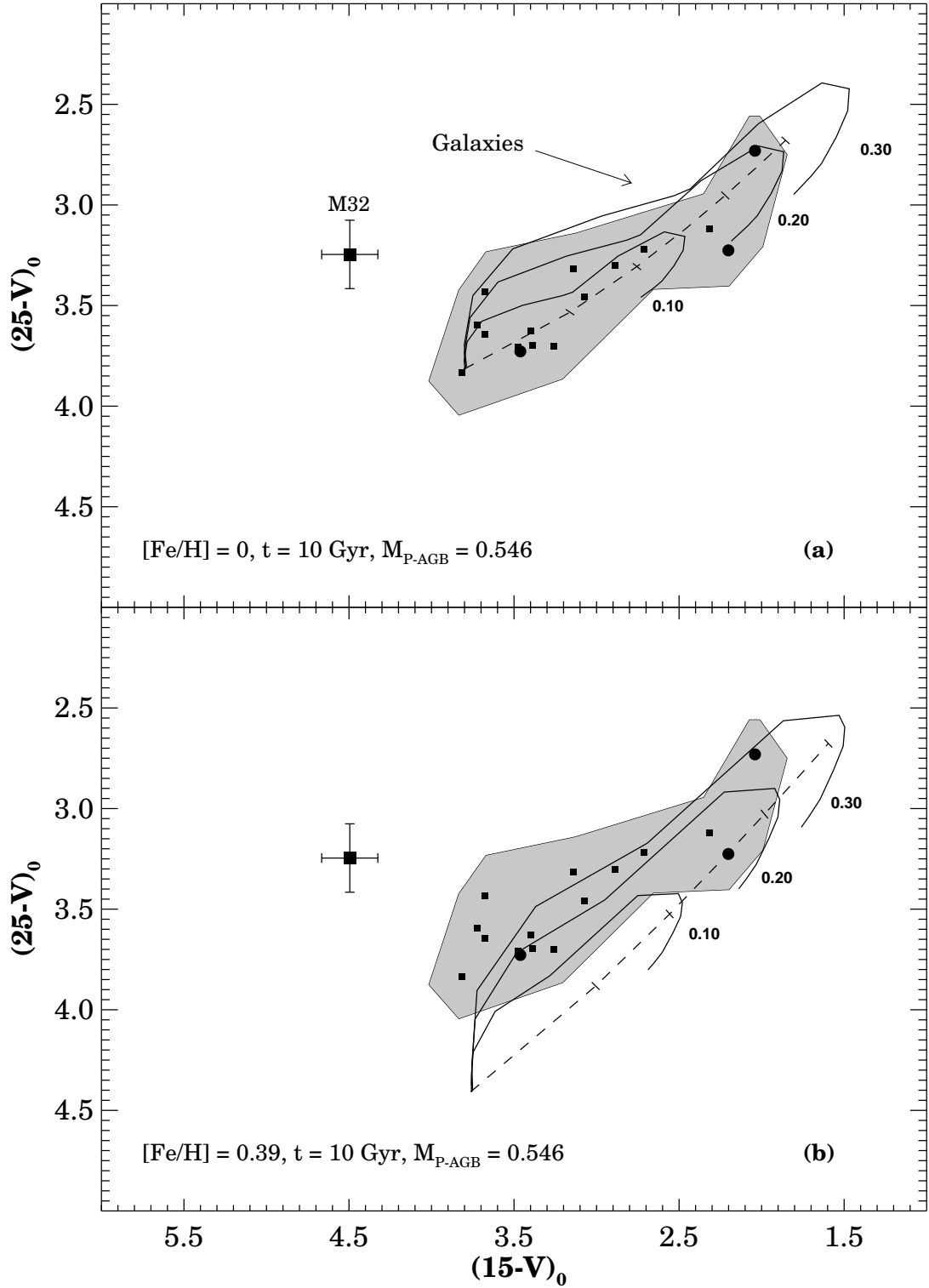

**Figure 10:** *Models for evolved stellar populations superposed on the area occupied by the galaxy data. The solid curves represent δ−function models in which the envelope mass $M_{env}^0$ varies from the red (left) to the extreme blue end of the HB. Values of $f_H$ are marked by each curve; models for more realistic envelope mass distributions lie under the curve. The dashed curve represents the mean-EHB models with values of $f_H$ less than or equal to the outermost δ−function model curve plotted. In both cases, the UV flux from the cool component (fraction $1 - f_H$) is taken from the Schönberner (1983) 0.546 $M_\odot$ model sequence. Large filled circles in panels (a), (b), and (c) show the location of galaxies for which models are listed in Table 8. These are (left to right) M31, NGC 4649, and NGC 1399. Panels (a), (b), and (c) show models with $f_H \leq 0.3$. Panel (a): Models with [Fe/H] = 0, t = 10 Gyr $M_{P-AGB} = 0.546$; (b) [Fe/H] = +0.39, t = 10 Gyr, $M_{P-AGB} = 0.546$.*

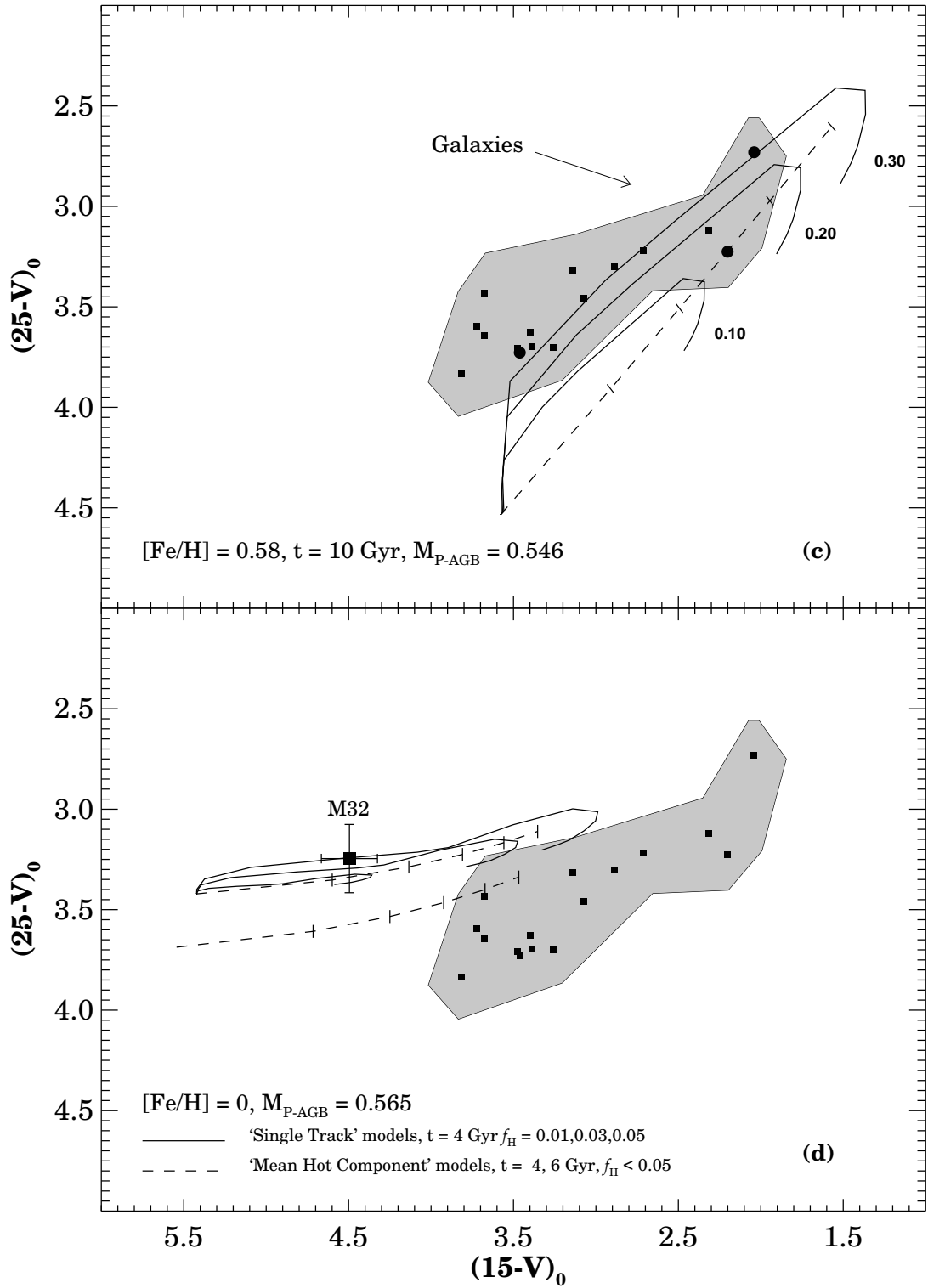

**Figure 10** continued: *Panel (c):* [Fe/H] = +0.58, *t = 10 Gyr*, $M_{P-AGB}$ = 0.546. *Panel (d): model fits to M32 with* [Fe/H] = 0, $M_{P-AGB}$ = 0.565. *The solid curves shows the $\delta$-function models for (left to right) $f_H$ = 0.01, 0.03, and 0.05 for t = 4 Gyr. The dashed curves illustrate the mean-EHB models for t = 4 and 6 Gyr, and for $f_H \leq 0.05$.*

adopted (*cf.* Fig. 9). At the blue end of these curves are systems whose far-UV flux is dominated by EHB stars and their AGB-manqué progeny, even though only 20–30% of the total HB population is in such objects. The colors of models containing EHB contributions $\gtrsim 5\%$ are nearly independent of the choice of P-AGB track. That is, the shaded region is also consistent with larger mean P-AGB masses (*cf.* Fig. 11). The observed red boundary for the bulk of the galaxy sample is, however, consistent with a mean P-AGB flux close to that predicted by the Schönberner $0.546\ M_\odot$ sequence. If the boundary is not a selection effect, it suggests that the upper bound to real P-AGB fluxes is well represented by current models.

All of the data points, with the notable exception of M32, are consistent with the solar abundance models in panel (a) for some $f_H \lesssim 0.2$ and some distribution of EHB stars. Most of the data points lie within the curve described by the models with $f_H = 0.10$. Only galaxies with the strongest UV upturns require $f_H > 0.1$. Therefore, *only a modest fraction of the stars evolving through the helium flash need become EHB stars to explain all of the observed UV upturns.* As panels (b) and (c) show, this result is *independent of the metallicity assumed* for the galaxies, at least insofar as the $15 - V$ color is concerned.

### 7.2.3 AGE AND ABUNDANCE EFFECTS

What limits do the UV observations place on galaxy ages and metal abundance? The effect on these parameters on model colors is further explored in Figure 11. To avoid confusion in the diagrams, we use the models with the 'mean EHB' component, again with $f_H \leq 0.30$. The upper two panels contain models with ages 4 and 6 Gyr, with $M_{\rm P-AGB} = 0.565 M_\odot$, and models with ages 6, 10, and 14 Gyr, $M_{\rm P-AGB} = 0.546 M_\odot$. In the lower panels, the models are seen to be almost insensitive to age, because the turnoff is too red to contribute much to the mid-UV at these higher metallicities.

As is evident from Figs. 10 and 11, only the $25 - V$ color is directly sensitive to age and metallicity. The hot populations which control $15 - V$ depend on these parameters only through the processes (mainly mass-loss) that determine $M_{\rm env}^0$. Without an accurate physical prescription for connecting mass-loss to age and abundance, the $15 - V$ colors therefore cannot be used to infer these. The $25 - V$ color, on the other hand, is very sensitive to abundance. It exhibits a much larger *range* relative to achievable photometric precision than most optical-IR indices (e.g. Worthey 1994). In our 10 Gyr models near solar abundance, $\partial(25-V)/\partial \log Z \sim 2.7$. However, $25 - V$ also suffers to some extent from the well-known age-abundance ambiguity effect (e.g., O'Connell 1986; Silva 1991; Worthey 1994). This is visible as overlaps in $25 - V$ between different model sequences in Figs. 9, 10, and 11. The sense is the same as for the optical region: models with higher abundances yield colors similar to models with lower abundances but larger ages. The shaded region of the galaxy data is consistent with solar abundance models with $t \sim 6 - 14$ Gyr but also models with [Fe/H] $= \pm 0.2$ at ages $10 \mp 5$ Gyr.

The sensitivity of $25 - V$ is, nevertheless, sufficient to place some limits on galaxy abundances. Mean metallicities of [Fe/H] $\leq -0.5$ are excluded for the galaxies for any age less than a Hubble time. This is not a new result, since many optical-IR studies have reached the same conclusion, but it is a useful consistency check on our models. The mid-UV color can also be used to place a strong upper bound on the fraction of light contributed by a minority metal poor population. For instance, Figs. 7 and 10 imply that metal poor globular cluster-type populations (age 16 Gyr, [Fe/H] $= -2.26$) can contribute no more than 15% of the V light in typical elliptical galaxies.

More interesting is the result from Figs 11c and 11d that *metal-rich models, with* [Fe/H] $\sim 0.4 - 0.6$, *cannot reproduce the colors of most of the galaxies.* They are consistent only with the galaxies with the strongest far-UV upturns. Metal rich models with "delta function" hot components (as in Fig 10b and c) can fit some of the intermediate UVX galaxies, but these models are less plausible since they would require a small mass dispersion among the EHB stars present and thus a very finely tuned mass loss mechanism. To state this another way, the mid-UV colors of objects like M31 or NGC 4472 ($25 - V \sim 3.7$), which constitute

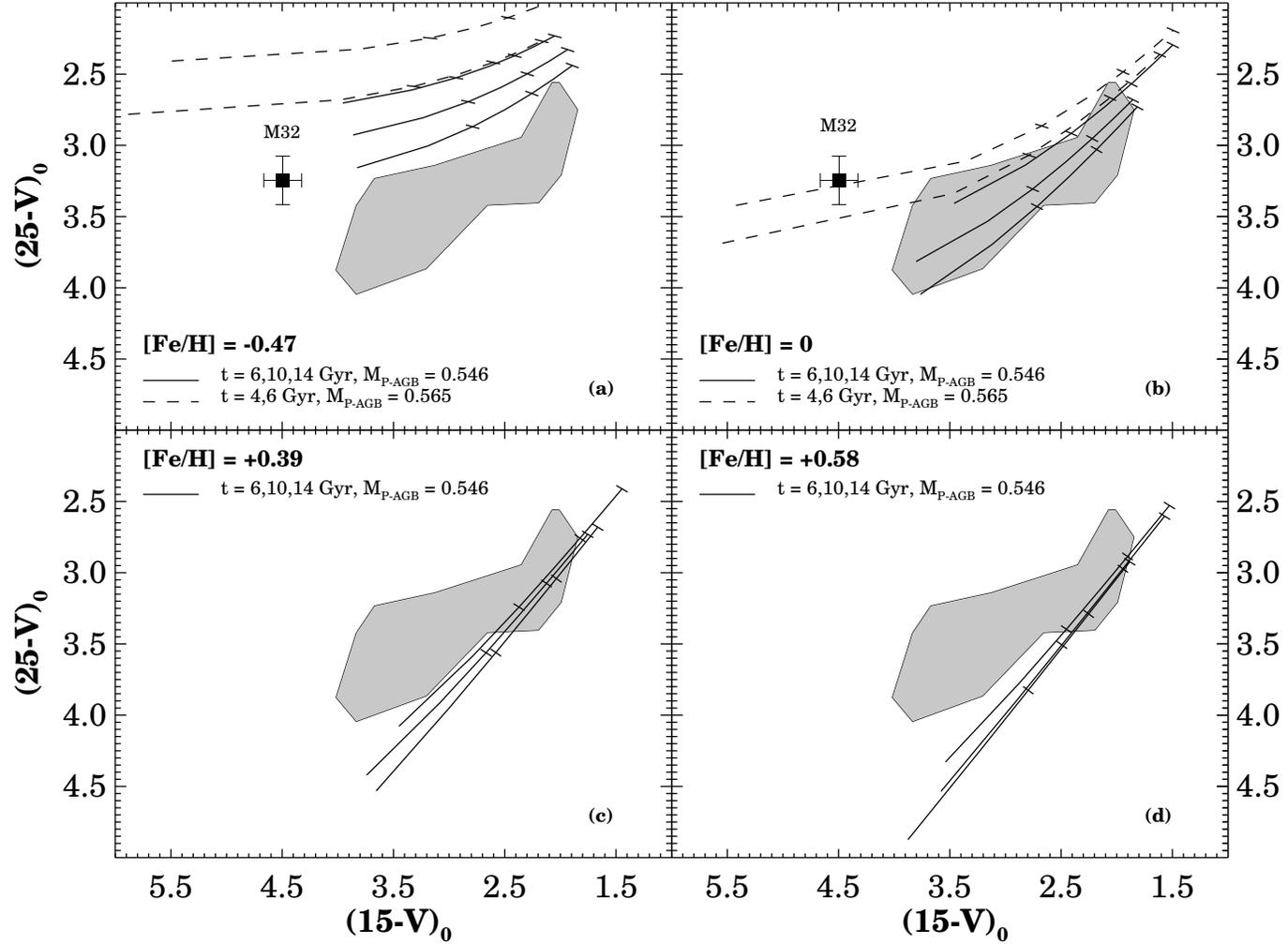

**Figure 11:** *Models illustrating the effect of varying age at four different metallicities. All the models in these panels are of the 'mean-EHB' type. In each panel, tickmarks represent the location of models with (left to right) $f_H = 0.1$, $0.2$, and $0.3$. Solid lines represent models computed with the flux from the Schönberner track with $M_{P-AGB} = 0.546\,M_\odot$ while dashed lines give models with flux assumed equal to his $M_{P-AGB} = 0.565\,M_\odot$ sequence. The data are indicated as in Figs. 9 and 10. Panel (a) solid line – [Fe/H] = −0.47, t = 6, 10, and 14 Gyr; dashed line — t = 4, and 6 Gyr. Model fits to M32 are only possible with the larger-mass P-AGB flux. Panel (b) [Fe/H] = 0; The solid curves show mean-EHB models for ages 6, 10, and 14 Gyr. The dashed curves correspond to the mean-EHB models illustrated in Fig 10d, except that they are extended to larger $f_H$. Panels (c) and (d): mean-EHB models for [Fe/H] = +0.39, +0.58 respectively, t = 6, 10, and 14 Gyr*

the bulk of the leftward half of the shaded region, are too blue to be consistent with metallicities higher than [Fe/H] $\sim 0.2$ for any age. Since the $Mg_2$ strengths of these objects (Fig. 1) indicate Mg abundances considerably higher than solar, and the mid-UV atmospheric opacity is contributed mostly by iron-peak elements, this suggests that [Mg/Fe]$> 0$ in elliptical galaxies and spiral bulges. There is increasingly good evidence that this is true from optical line strengths, and we return to this issue in §8.

Regardless of the details of this particular set of models, however, it is clear that the mid-UV spectral region has considerable potential as an age-abundance disciminant, particularly when narrow-band, high precision UV data are combined with optical and infrared observations.

### 7.2.4 M32

M32, the lowest luminosity galaxy in the sample, occupies a unique position in our color-color diagrams[6]. M32's stellar population had been the subject of some controversy (O'Connell 1986, Renzini 1986, Freedman 1992) because early integrated light studies (O'Connell 1980, Rose 1985) suggested a mean age of only $\sim 5$ Gyr, much smaller than expected for an elliptical galaxy. However, a considerable number of integrated light and color-magnitude diagram studies now support the intermediate-age interpretation of M32 (Burstein *et al.* 1984, Boulade, Rose, & Vigroux 1988, Bica, Alloin, & Schmidt 1990, Davidge & Nieto 1992, Faber, Worthey, & Gonzalez 1992, Elston & Silva 1992, Freedman 1992, Hardy *et al.* 1994, Rose 1994). Optical line strength modeling by Faber *et al.* (1992), Gonzales (1993), and Worthey (1994) further indicates that a significant fraction of elliptical galaxies may have populations with mean ages of 3-8 Gyr, as does a large body of other evidence (O'Connell 1994).

In Fig. 10d we compare the UV colors of M32 to intermediate age models. The $15 - V$ colors are only consistent with models in which the mean P-AGB luminosity is higher than in the other galaxies. We have used the flux from the Schönberner $M_{\rm P-AGB} = 0.565 M_\odot$ track for these models. $M_{\rm bol}$ for such tracks is $\sim -4.3$, which is in accord with stars near the tip of M32's AGB as found in the near-infrared color-magnitude diagram of Freedman (1992). The models are for [Fe/H] = 0, which is the best estimate of M32's abundance based on optical line strengths (O'Connell 1980, Boulade et al. 1988, Faber *et al.* 1992, Rose 1994). The solid curves are based on isochrones for 4 Gyr and $f_H \le 0.05$, and represent the outer envelope of the possible colors of models with $f_H = 0.01$, 0.03 and 0.05. The dashed lines showing the 'mean' hot component have ages 4 and 6 Gyr. The 4 Gyr models provide good fits to M32's colors for some $f_H < 0.05$. The 'mean' model for 4 Gyr implies $0.01 < f_H < 0.02$. Even with this small fraction of RGB tip stars reaching the EHB, the EHB component will dominate the far-UV light. The 6 Gyr models, however, appear to be somewhat too red in $25 - V$, whatever mass distribution is assumed.

High quality IUE spectra are available for the 2500-3200 Å region of M32. These are also well fit by intermediate age, solar abundance models (Kjaergaard 1987, Rocca-Volmerange & Guiderdoni 1987, O'Connell 1990, Magris & Bruzual 1993). Thus, the UV properties of M32 are consistent with an intermediate age population in agreement with the optical-IR studies. However, the low-resolution UV colors alone cannot be used to exclude alternative interpretations. As shown in Fig. 11, M32's UV colors would also be consistent with old [Fe/H] = $-0.47$ models if $M_{\rm P-AGB} = 0.565 M_\odot$. Although such core masses are not thought to be consistent with large ages, this emphasizes the fact that UV colors by themselves are not sufficient to separate age and metallicity unambiguously near solar abundance (as discussed in the last section). Higher resolution observations in the mid-UV and optical regions are necessary to place good limits on the allowable metallicity range.

---

[6] The original B3FL survey contained two other objects with similar colors: NGC 4111 and 4382. However, NGC 4111 does not have an available mid-UV flux, and 4382 shows evidence of recent star formation and internal extinction (§4.2). The photometry for M32 is good, despite its luminosity and red far-UV color, because it is nearby. Most other objects of comparable luminosity are too faint for IUE observations.

### 7.3    Models for Selected Galaxies

In Table 9 we list models with the "mean EHB" component chosen to match approximately the observed colors for the galaxies NGC 1399, NGC 4649, M31, and M32 (highlighted in Figure 10 with larger symbols). These are not intended to produce best estimates of parameters but rather to illustrate the range of models lying near selected objects. We have not interpolated in the grid to optimize the fits. For each choice of age and abundance, we give $f_H$, the predicted colors, and the fraction of the light at three wavelengths arising from the EHB/post-EHB and P-AGB components.

The UV light fractions from these two components are useful in predicting the extent to which the UV images of such galaxies would be resolved by HST. P-AGB stars, with absolute magnitudes at 1500 Å of $M_{15} \sim -5$ are sufficiently bright to be detected by HST in nearby galaxies with practicable integration times, (e.g. King et al. 1993), but EHB stars, with $M_{15} \sim 0$, are too faint.

As described in the last section, the colors of M32 are consistent with intermediate-age, solar abundance models or old, metal-poor models (if a larger P-AGB mass is adopted). In either case the UV output is dominated by EHB/post-EHB stars. This is partly an artifact of the limited P-AGB model grid, since M32 is too red to be consistent with the 0.546 $M_\odot$ track but considerably bluer than the 0.565 $M_\odot$ track (cf. Fig. 9). Note also that the mid-UV flux is dominated by the turnoff population because the hot component is very small.

For objects like M31, which is representative of the galaxies with weaker UVX, over 60% of the 1500 Å light is produced by P-AGB objects; at 2500 Å this is reduced to below 25%. In a strong UVX source like NGC 4649, the models predict that 70–80% of the 1500 and 2500 Å light should originate in EHB stars. Note the wide range of ages and abundances which are capable of yielding the observed NGC 4649 colors.

The unusually blue $25 - V$ color of NGC 1399 cannot be fit well with super metal-rich, old models. The closest "mean EHB" match has [Fe/H] = 0, $t = 8$ Gyr, and 85% of the far-UV light emanating from EHB stars and their progeny. However, according to Fig. 10b, these colors can also be reproduced by a population with [Fe/H] = 0.58, $t \sim 10$ Gyr, $f_H \sim 0.3$, and a narrower EHB mass distribution in which $M_{\rm env}^0 \lesssim 0.03 M_\odot$. In either case, the fraction of P-AGB light is expected to be small.

These estimates of the P-AGB vs. EHB/post-EHB fraction as a function of $15 - V$ color are similar to those made by Ferguson & Davidsen (1993) based on their model fits to HUT spectroscopy (cf. §2.1).

### 7.4    Galaxies versus Clusters

Our final diagram, Fig. 12, revisits the $15 - V$ vs. $15 - 25$ two-color diagram shown in Fig. 3a. Recall that the galaxies and clusters form two parallel sequences. We can now interpret this as being caused primarily by the difference in their metallicity ranges. The galaxies are bluer in $15 - 25$ at a given $15 - V$ mainly because the mid-UV energy distribution of their turnoff and RGB components is depressed by greater line blanketing and they contain fewer stars on the IBHB, which are bright in the mid-UV.

The boxes in the figure enclose the area occupied by models of different metallicity ranges for two selected $f_H$'s. The upper box contains the $\delta$−function models with $0 \geq$ [Fe/H] $\geq 0.58$ for 10 Gyr, with $f_H = 0.2$. The lower box contains models with [Fe/H] $\leq -1.48$ for 14 Gyr with $f_H = 0.75$. These $f_H$'s were chosen so that the bluest models coincide approximately with the bluest objects; larger (smaller) choices for $f_H$ would shift the boxes rightward (leftward) in this figure. The shaded areas in the boxes represent the color range of models containing EHB stars.

The lower box shows that the slope of the cluster data is reproduced as the temperature of the ZAHB stars shifts from higher values (EHB, upper right) to lower ones (red HB). The diagram illustrates the point made earlier that few clusters appear to be dominanted by EHB components, whereas most galaxies are.

TABLE 9

MODELS FOR SELECTED GALAXIES

'MEAN EHB' MODELS

| Object | $(15-V)_0$ | $(25-V)_0$ | [Fe/H] | $t$ | $f_H$ | $15-V$ | $25-V$ | $EHB_{15}$ % | $PAGB_{15}$ % | $EHB_{25}$ % | $PAGB_{25}$ % | $EHB_V$ % | $PAGB_V$ % |
|---|---|---|---|---|---|---|---|---|---|---|---|---|---|
| M 32 | 4.50 | 3.25 | $0.00^a$ | 4 | 0.012 | 4.49 | 3.34 | 58.1 | 41.3 | 7.2 | 3.7 | 0.2 | 0.1 |
| | | | $-0.47^a$ | 14 | 0.025 | 4.45 | 3.21 | 74.2 | 25.3 | 8.2 | 2.0 | 0.3 | 0.0 |
| M 31 | 3.46 | 3.73 | 0.00 | 10 | 0.020 | 3.44 | 3.71 | 29.4 | 70.6 | 12.7 | 25.1 | 0.1 | 0.2 |
| NGC 4649 | 2.20 | 3.23 | 0.00 | 16 | 0.170 | 2.20 | 3.16 | 80.7 | 19.3 | 66.0 | 13.0 | 0.9 | 0.2 |
| | | | 0.39 | 10 | 0.160 | 2.18 | 3.20 | 79.8 | 20.2 | 69.2 | 14.9 | 0.9 | 0.2 |
| | | | 0.58 | 6 | 0.140 | 2.20 | 3.17 | 74.2 | 25.8 | 68.0 | 18.5 | 1.3 | 0.3 |
| | | | 0.58 | 16 | 0.240 | 2.19 | 3.23 | 84.8 | 15.2 | 83.2 | 11.6 | 1.5 | 0.2 |
| | | | 0.58 | 16 | 0.240 | 2.19 | 3.23 | 84.8 | 15.2 | 83.2 | 11.6 | 1.5 | 0.2 |
| NGC 1399 | 2.04 | 2.73 | 0.00 | 8 | 0.210 | 1.99 | 2.79 | 84.4 | 15.6 | 59.4 | 9.0 | 1.1 | 0.2 |
| | | | 0.39 | 6 | 0.160 | 1.88 | 2.89 | 79.8 | 20.2 | 68.1 | 14.6 | 1.2 | 0.3 |
| | | | 0.58 | 6 | 0.200 | 1.90 | 2.89 | 81.5 | 18.5 | 76.1 | 13.4 | 1.9 | 0.3 |

[a] Models used $M_{P-AGB} = 0.565 M_\odot$; all others, $M_{P-AGB} = 0.546 M_\odot$.

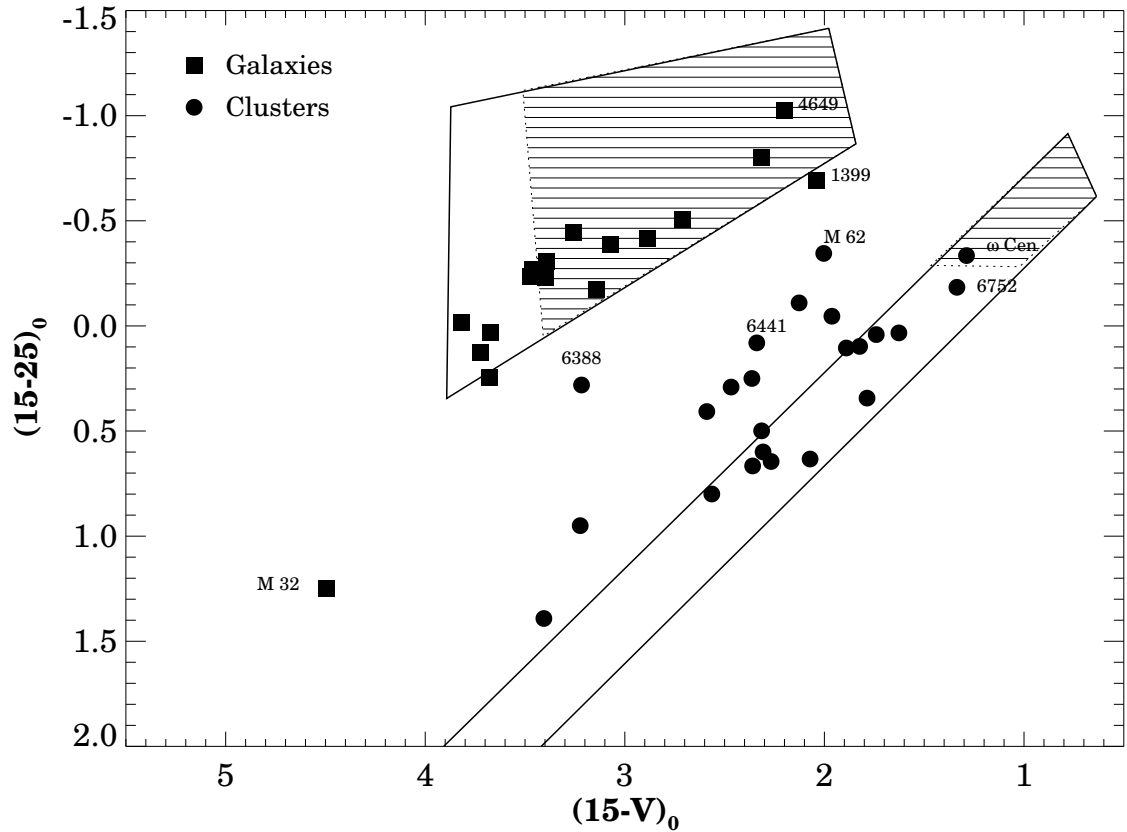

**Figure 12:** *Models of metal-rich and metal-poor populations superposed on the galaxy and Galactic globular cluster data. The parameter $f_H$ has been chosen to fit the data for canonical choices for the age and abundance of galaxies and clusters. Both boxes enclose "$\delta-function$" models as described in §6. The upper box encloses models with $f_H = 0.2$, with abundances [Fe/H] $\geq 0$ at $t = 10$ Gyr. The lower box encloses models with [Fe/H] $= -2.26$ and $-1.48$, with $f_H = 0.75$ at $t = 14$ Gyr. The shaded regions represent models in which the hot stellar component consists of EHB stars.*

The galaxies lie at the lower edge of the box (near solar metallicity models); only the point bluest in $15 - 25$, NGC 4649, is fit well by models with very high metallicity (*cf.* Table 8).

As was clear from §7.1 and 7.2, adjustments in models parameters can explain the various objects lying away from the main distributions. Some of the clusters with relatively blue $15 - 25$ color falling outside the lower box have higher metallicity (e.g., NGC 6388); others may suffer from observational uncertainty (e.g., M62).

Figure 12 summarizes the conclusions of this section because it shows that the models can reproduce both the cluster and galaxy data with a small number of input parameters which take plausible values. The agreement between theory and observation is very good, considering the simplicity of the scheme.

# 8    Discussion and Summary

Our models have provided quantitative estimates of the size and type of hot stellar component that is able to reproduce the far-UV observations of globular clusters and elliptical galaxy nuclei. We have found that EHB stars and their progeny are the likely origin of the UV upturn in most galaxies and that only a small fraction of all red giants need become such objects to explain the data. Our conclusions are, of course, qualified by various uncertainties in the modeling, which require a more careful discussion. In this section, we consider the effects of model-dependence and complications such as non-solar abundance ratios on our results. We also emphasize the central role which stellar mass loss plays in interpreting the UV spectra of galaxies.

## 8.1    The Color-Line Strength Correlation Revisited

One of the most striking features of the UVX phenomenon in the galaxy sample is the correlation between central UV color and the $Mg_2$ index, which measures the strength of the Mg I "triplet" feature at 5175 Å. The available information on UV color gradients *within* galaxies (§2.1) likewise suggests a strong correlation with abundance, which declines outwards in galaxies. Before we address possible origins for the correlation there are several complications to consider.

First, magnesium is a so-called $\alpha$-element, with atomic weight intermediate between the light metals (CNO) and the iron peak. $\alpha$-elements do not themselves have an important influence on stellar interiors, although they do affect the colors of stellar evolutionary sequences somewhat (Weiss, Peletier, & Matteucci 1994). Rather, it is the effects of the iron-peak elements on opacity and the effects of the CNO group and helium on energy generation that control the evolution of low mass stars. Those $\alpha$-elements which have low ionization potentials affect the line structure of the emergent stellar spectra, but as a whole they have small effects on atmospheric blanketing.

Furthermore, the $Mg_2$ index may not accurately reflect the mean abundances of the metals in galaxies. Early evidence for variations in the ratio of light to heavy elements among galaxies was reported by O'Connell (1976), Peterson (1976), Pritchet & Campbell (1979), and O'Connell (1980). This case has been considerably strengthened by recent measures of Fe and Mg features in galaxy nuclei and exteriors (Peletier 1989; Gorgas, Efstathiou, & Aragon Salamanca 1990; Worthey, Faber & Gonzalez 1992; Davies, Sadler & Peletier 1993; Gonzales 1993; Worthey 1994). The data suggest that [Fe/H] may be $\sim 0$ in most ellipticals but that at least Mg, and perhaps all light elements, are enhanced relative to Fe in the more luminous ellipticals. As pointed out in §7, the UV colors of our elliptical sample are all consistent with [Fe/H] $\sim 0$ models but most of the objects are not consistent with [Fe/H] $\gtrsim 0.2$.

The conventional assumption that elliptical galaxies have iron peak element abundances considerably higher than solar may therefore be incorrect. This could seriously alter the various astrophysical mechanisms proposed for the UV color-$Mg_2$ correlation. An enhancement relative to the iron peak of $\alpha$-elements alone should have little effect on the UVX or galaxy spectra, apart from individual spectral features. However, an overall enhancement of most of the light elements, including CNO, could significantly affect the observations. Worthey *et al.* (1992) have outlined plausible scenarios whereby IMF differences could influence the mixture of metals in galaxies.

A second factor is that the apparent correlation between UV colors and $Mg_2$ may not reflect smoothly varying properties but instead might arise from several discrete groups of galaxies. There is a hint of this in the UV color data (Figs. 1, 2, and 4). The bulk of the galaxy $15 - V$ colors are in the range $3.5 \pm 0.5$. Excluding objects with internal extinction or star formation, the low luminosity elliptical M32 is the only system with $15 - V > 4$. As noted in §2, it also shows a radial UV color gradient which is distinct from the more luminous systems. At the other extreme, there are 5 objects with $15 - V < 3$. In Figs. 1 and 4, these galaxies stand out as a distinct group. Thus, instead of a continuum with $Mg_2$ strength, Figs. 1, 2, and 4 may be consistent with three distinct classes of galaxies: (1) galaxies that are faint in the UV, like M32; (2) galaxies like M31 and NGC 4472 with a "normal" UVX component and only a modest dependence of UV colors on $Mg_2$; and (3) strong-lined systems with an enhanced UVX component which may vary independently of $Mg_2$. It is unclear as yet whether the internal UV color gradients can be similarly interpreted.

Thirdly, it is possible that mean metal abundance is not the sole controlling factor in determining UV properties. M32 is the only example of the first group close enough to be observed by IUE, and it appears (§7.2.4) to have a significant intermediate age population. Current calibrations of optical line strengths (Worthey 1994, Gonzalez 1993) suggest that there is a positive correlation between luminosity and mean age among ellipticals. Age could therefore be an important factor in UVX properties. As discussed in §7, the low resolution UV colors we have modeled do not strongly constrain the ages of most of the galaxies.

Other parameters, such as total mass or the dynamical environment, could also affect the UVX phenomenon. One hint of this is the preferential appearance of stronger upturns in objects with "boxy" isophotes (Longo *et al.* 1989 and §2), which are possibly merger products (Bender 1988). As noted above, these objects tend to form a distinct group in the color-$Mg_2$ diagrams. The unusual UV properties of M32 might also be attributed to a special dynamical environment, as it is the only dwarf elliptical in the observed sample. The possibility that early mergers could influence the IMF (Worthey *et al.* 1992), hence the ratio of light to iron-peak elements, and hence the UVX ought to be investigated further.

As well as being a direct consequence of single-star evolution, UV-bright objects can be produced in binaries either by mergers ("two-core" origin) or by stripping in close binary systems ("single-core" origin). The latter possibility (Mengel *et al.* 1976) seems to be much more common, based on the small dispersion of surface gravities for the field sdB stars and the lack of faint pre-merger products observed (Heber 1992). However, there is unlikely to be a strong *direct* effect of galaxy abundance on the binary parameter range giving rise to EHB stars.

On the other hand, the dynamical environment could affect the production of hot subdwarfs via stellar encounters or by changing the frequency of binary systems. There is some evidence for dynamical effects in clusters (see §3.1), although galaxies have much lower mean densities and much longer crossing times than clusters. Alternatively, the dynamics may change the initial conditions (e.g., through the Initial Mass Function) or the chemical history of the galaxy such that the hot population is affected.

These latter possibilities certainly deserve further study. For the present, however, we will explore the more straightforward effects of the basic mean properties of the stellar population—that is, age and abundance—on the UVX-$Mg_2$ correlation. We have shown in Paper I that the qualitative behavior of metal-rich HB's is very sensitive to $M_{env}^0$ but insensitive to metallicity (compare Figs. 6b, 6c). Thus, UV output is not directly related to metallicity itself. Nor does it depend strongly on the core mass (and therefore the age) of the ZAHB stars. Instead, the dominant factor is the distribution of envelope masses on the ZAHB, which in turn (at least in single-star evolution) is determined by mass loss on the giant branch. The mass loss processes are likely to be affected directly by metal abundance; age enters indirectly through the decrease in the RGB tip mass with time. In addition, $Y$ affects both the RGB tip mass and the UV output during the advanced stages of evolution. Therefore, in the next sections, we turn our attention to effects of mass loss and helium abundance.

## 8.2 Giant Branch Mass Loss

In single-star evolution, mass loss on the RGB is the dominant physical process governing the distribution of $M_{env}^0$. All the modeling of post-RGB evolution of low mass stars, including Schönberner's (1983) and ours (see Fig. 4b and c), indicates that changes of only a few $0.01\,M_\odot$ in RGB mass loss can dramatically effect the consequent UV output, so that a thorough understanding of mass loss is critical. Unfortunately, it is not at hand. Here we discuss the viability of our modeling in the context of what is known about mass loss.

In metal-poor systems, stars with significant UV flux are produced with modest ($\lesssim 0.2 M_\odot$) amounts of mass loss. These objects are the ordinary IBHB stars discussed above, which star for star contribute up to half the output of the EHB stars (*cf.* Table 7). The HBs, and thus UV flux, of many metal-poor clusters are dominated by such stars. Even from our simplistic models, it is clear that a large population of EHB stars is not necessary to explain the UV colors in most clusters. To detect these stars in integrated light requires more spectral coverage than available in our colors (e.g., Ferguson & Davidsen 1993).

In contrast, for galaxy populations we find that practically all the UV-radiating stars must be EHB stars and their descendants. The presence of EHB stars requires that an efficient mass loss mechanism operates at the bright end of the RGB. At a constant age of 14 Gyr and with slowly varying $Y$, the mass loss required to reach the EHB is 0.34 $M_\odot$ at [Fe/H] $= -2.26$, rising to 0.52 $M_\odot$ at [Fe/H] $= 0.0$ and 0.69 $M_\odot$ at [Fe/H] $= 0.58$[7]. This large a mass loss need apply to only a fraction of the population. As discussed in §6 and 7, the UV colors of the bluest galaxies require that $\sim 15 - 20\%$ of stars lose sufficient mass on the RGB to become EHB objects (comparable to the fraction of EHB stars in the bluest globular clusters). On the other hand, colors of the bulk of the galaxy sample (objects like M31) are consistent with this much mass loss in only $\lesssim 5\%$ of the HB population. Hot subdwarfs in the Galactic field, presumably near solar abundance, are about 2% of the total HB population (see §3.3). This value is in reasonable agreement with the EHB fraction necessary to explain most of the galaxies in Figs. 11 and 12. Other evidence on the AGB (integrated light studies, surface brightness fluctuations, and CMD's) and post-AGB populations (planetary nebulae) in elliptical galaxies seems consistent with an upper bound of $\sim 20\%$ on the fraction of the RGB with large mass loss, although a critical review of such evidence would be desirable.

As discussed in the introduction and §6.3, there is little empirical or theoretical information on the RGB mass loss process. Observational evidence indicates only that mass loss increases with luminosity and with decreasing temperature (see Dupree 1986 for a review), so that its rate is greatest close to the helium flash that terminates the RGB evolution. The widely-used Reimers (1975) formula is scaled by a free parameter, $\eta_R$, which for typical red giants in globular clusters is estimated to be $\eta_R \sim 0.25 - 0.5$ (Renzini 1981 and Paper I). The formula includes composition and age effects only through their effect on stellar temperature and gravity. If $\eta_R$ is a constant the total mass lost at constant age varies by only 10–20% over the range of compositions we consider. Thus in the context of the Reimers formula, high $Z$ EHB stars can be produced only if $\eta_R$ is larger by a factor of 2–3 over the value required to produce EHB stars in globular clusters (see also Jorgensen & Thejll 1993).

To see whether such a variation is reasonable or not consider the dispersion of mass required to produce observed cluster HBs. For a cluster with [Fe/H] $= -1.48$ (similar to M79) stars in the RR Lyrae strip have $\eta_R = 0.49$, those at $B - V = 0.0$ have $\eta_R = 0.55$, and EHB stars $\eta_R \sim 0.75$[8]. Thus within a single cluster like M79 $\eta_R$ varies by 50% if one insists on describing mass loss in terms of the Reimers law. Considering that we do not even know what physical parameter leads to this "observed" variation within a cluster, a factor of 2 change in $\eta_R$ between the metallicities of the clusters and the galaxies seems entirely plausible.

Mass-loss mechanisms in which the radiation field couples to molecules or grains[9] in the atmospheres of cool stars will operate below some threshold temperature. If processes of this kind are important, then the dependence of mass loss on $Z$ will be non-linear. The observational data on H$\alpha$ emission and other indicators of mass loss from cool stars show an increase with decreasing temperature, supporting an expectation of increased mass loss with metallicity.

---

[7]  The total mass loss required to reach the EHB can be approximated by subtracting $\sim 0.5\ M_\odot$ from the RGB masses given in Table 5.

[8]  The values of $\eta_R$ required for globular clusters are strongly dependent on the assumed [O/Fe] (e.g., Rood & Crocker 1985).

[9]  The material coupling to the radiation field must be fairly close to the surface in order to produce large mass loss rates, which favors molecules rather than grains as the most important mechanism; see Holzer & MacGregor (1985).

In the case where the galaxies are enhanced in the light elements relative to the iron peak, it is also plausible that RGB mass loss would be increased. As long as CNO is involved, the formation in RGB atmospheres of molecules and grains including graphite and silicates would be facilitated.

It is also possible, however, to increase the production of EHB stars without increasing RGB mass loss. One way is to adopt a large value for the enhancement $\Delta Y/\Delta Z$ of the helium abundance with metallicity. (see Fig. 1 of GR and §8.3); a second is to assume larger ages (e.g., Lee 1993). In these cases, less mass loss is required because the stars leaving the main sequence are less massive. For metal-rich populations, age variations alone are not viable unless $\Delta Y/\Delta Z$ is also large. For example, at constant $Y$ the amount of mass loss required to reach the EHB increases by 0.12 $M_\odot$ when going from $Z = 0.017$ to 0.04. The compensating age increase required would be $\sim 8$ Gyr, so that without a large $\Delta Y/\Delta Z$ a very large range of galaxy ages is implied. In any event, the presence of EHB stars in the Galactic disk in amounts capable of explaining the bulk of the E-galaxy UV colors suggests that large ages are not necessary.

Other descriptions of mass loss from cool stars than the Reimers formulation may provide a better description of the physical process (see Trimble & Leonard 1994 for a review; Han, Podsiadlowski, & Eggleton 1994). Given the fact that both a He-flash and a small residual envelope are needed to produce EHB stars, it may seem that the mass loss process has to be implausibly finely tuned. However, the existence of EHB stars in globular clusters and the Galactic disk gives concrete evidence that such a combination occurs in nature. If reproducing this observation requires a fine tuning of the Reimers parameters, then it is reasonable to suppose that the formulation does not properly represent the process.

To summarize: what little is known about RGB mass loss, based largely on observed globular cluster HB's, does not appear inconsistent with the moderately increased rates required to produce appropriate numbers of EHB stars in elliptical galaxies. The existence of an EHB population in the Galactic disk sufficient to explain the UV outputs of most galaxies encourages this view. The alternative, which is to require that mass loss rates are never higher than exhibited by globular clusters and that therefore the $Y$'s or ages of galaxies must be large, is itself an assumption, no better justified by the evidence and perhaps less plausible in its implications (see next section).

## 8.3    Effects of Helium Abundance

An alternative to increased mass loss with metallicity is a large increase in the helium abundance in high metallicity stars, as discussed by GR. Horch *et al.* (1992) suggested that only high metallicity populations can produce EHB stars. However, we have shown (Paper I) that this is not the case. Their result depended on assuming that $Y$ increases strongly with $Z$. They adopted $\Delta Y/\Delta Z = 3$ or 4. At the high helium abundances implied by this relation, our models likewise produce large UV outputs (but so do models of all abundances with sufficiently small $M_{\mathrm{env}}^0$).

The higher helium abundances can lead to more UV radiation for three reasons. First, there is a greater range of masses producing UV radiation because the HB evolution produces wider and longer blueward loops with higher surface temperatures (see also Sweigart & Gross 1976). Second, for a given $M_{\mathrm{env}}^0$ the higher $Y$ stars burn more hydrogen during the core helium burning phase, and produce AGB-manqué type behavior for a larger range of masses. Thirdly, the mass loss rates can be smaller because the RGB masses with high $Y$ at fixed age are smaller by $\sim 0.1 - 0.2 M_\odot$.

These "$Y$-effects" each require a very large fractional He abundance: $Y \gtrsim 0.40$. If $Y$ is strictly correlated with $Z$, then even with a large $\Delta Y/\Delta Z$ the mean metallicity must be considerably greater than solar. For example, to reach $Y \geq 0.40$ with $\Delta Y/\Delta Z = 4$, we need $Z \sim 0.05$. The assumption that RGB mass loss does not increase with metal abundance and that increased hot star production arises only from enhanced $Y$ would therefore predict few EHB stars at solar metallicity. It *requires* that the UVX phenomenon should be confined to truly metal rich objects. But this is apparently inconsistent with the population of EHB stars in the Galactic disk (§4).

The abundance of helium for metallicities above solar has not been studied in as much detail as has the extremely metal-poor regime. Values of $\Delta Y/\Delta Z \sim 3 - 4$ have been obtained from observations designed to determine the primordial value of $Y$ (Pagel *et al.* 1992, Wilson & Rood 1994). Skillman *et al.* (1993) and Campbell (1992) have both suggested systematic effects that might lower the slope further even at low $Z$. The "metal abundance" in these determinations refers almost exclusively to the light elements and not the iron peak. High values of $\Delta Y/\Delta Z$ at $Z \sim Z_\odot$ are not easy to explain on the basis of the usually accepted stellar nucleosynthesis ideas (Wilson & Matteucci 1992): for $Z > Z_\odot$, "standard" models of chemical evolution give $\Delta Y/\Delta Z \sim 1$. Based on accepted solar abundances, the average of $\Delta Y/\Delta Z$ from $Z = 0$ to $Z_\odot$ is approximately 2. This suggests that the slope at high metal abundances is lower than at $Z \sim 0$. Only a few chemical evolution models have been developed which produce higher $\Delta Y/\Delta Z$. Maeder (1992) has recently obtained $\Delta Y/\Delta Z \sim 3 - 4$ by varying the lower mass boundary where stars collapse into black holes. This requires stars with $M > 20–25\,M_\odot$ to collapse rather than explode as supernovae, thus burying the heavy elements they have produced. Giovagnoli & Tosi (1994) have used the Maeder (1992) yields in chemical evolution models and, while they confirm the steeper slopes, they find that they cannot satisfy some of the usual constraints applied to such models.

It should also be noted that high $Y$ will dramatically affect many aspects of populations besides EHB production. The self-consistent picture of stellar evolution has been developed using models with "normal" $Y$: it is not at all clear that the same can be achieved with high $Y$. For example, the AGB luminosity-core mass relation will certainly be affected (Paper I). This in turn will affect lifetimes of P-AGB stars and luminosities of planetary nebulae. Because they bring their own difficulties, we are not inclined to adopt the high helium abundance alternative simply in order to keep the required RGB mass loss constant.

## 8.4    Uncertainties in the Models

Relatively few sets of models exist for the EHB and subsequent phases of evolution, and it is inevitable that deficiencies will be revealed in them. While the gross parameters that drive the luminosities and gravities of our models, the core mass and the stellar structure of stars with helium burning cores, seem secure, there are several ways in which the models disagree quantitatively with existing observations. First, UIT observations of the globular clusters M79 (Hill *et al.* 1992) and $\omega$ Cen (Whitney *et al.* 1994) show that for $\log T_{\rm eff} > 4.2$ the observed stars lie $\sim 0.4$ mags below the theoretical ZAHB, and a few stars are hotter than any of the models, even for relatively metal-poor compositions. It is unclear as yet whether the modeling or the photometry is at fault (or whether a distinct new population has been detected here). Second, the temperature range of the field sdOB population extends to $T_{\rm eff} \geq 30000\,{\rm K}$, which is too hot for our models that are close to or above solar metallicity.

These problems could be related to known deficiencies in the models, which include the boundary conditions and the temperature-UV color transformations for hot subluminous stars. The solution may lie in the use of accurate stellar atmosphere calculations, where available. The surface pressures of the models of Paper I – indeed, all models to date of EHB type stars – have been computed with a very rudimentary methods, by integrating a T-$\tau$ relationship from the top of the atmosphere until the photosphere is reached. Comparison with the Kurucz (1991) model atmospheres of similar $T_{\rm eff}$ and $\log g$ shows that our derived surface pressures are too small by about 0.2 dex. The obvious inference is that the models are too cool, possibly by a significant factor. A new set of models is necessary to quantify this effect, but this problem should not seriously change the conclusions of this paper. Its effect would be to stretch ZAHB sequences to higher temperatures, so that EHB models at a given temperature would have somewhat larger $M_{\rm env}^0$s than predicted here.

Similarly, although the UV opacities used in the Kurucz atmospheres are the best available at present, they are subject to significant refinement, which will alter the predicted colors for a given effective temperature, gravity, and abundance. In addition, since hot subdwarfs often show abundance anomalies, their integrated UV output may be quantitatively different from model predictions based on standard mixtures.

Finally, an improved grid of P-AGB models is highly desirable. For objects such as M32, P-AGB models of the appropriate mean core mass can obviously account for the UV flux without any EHB stars, but the model grid for low mass P-AGB stars needs to be calculated at intervals in mass $\lesssim 0.005 M_\odot$.

## 8.5 Summary

We have presented an analysis of the far-ultraviolet upturn phenomenon (UVX) observed in elliptical galaxies, based on the grid of evolutionary models for HB and post-HB stars in Paper I. The grid covers an age and [Fe/H] range of 2 to 16 Gyr and $-2.26$ to $+0.58$ with and without helium enhancements. We use a new set of main sequence and giant branch isochrones to determine the rate, $\dot{n}(t)$ at which RGB stars pass through the He flash. We have supplemented our models with the set of P-AGB models from Schönberner (1979, 1983). The ultraviolet output of old populations is governed primarily by the distribution of envelope mass ($M_{env}^0$), $P(M_{env}^0)$, on the Zero Age Horizontal Branch. We investigate the range of $P$, using a simple description, which is consistent with observed broad-band UV colors. Because it is not well understood physically, we choose to leave mass loss on the giant branch as an implicit free parameter in our models.

We first review what is known about the UVX phenomenon in elliptical galaxies. We then re-derive the broad-band UV colors $15 - V$ and $25 - V$ for globular clusters and galaxies from the available satellite data. We use newer values for the integrated $V$ band magnitudes of the clusters, which results in some changes with respect to the compilation of de Boer (1985). We investigate the color-color and color-line strength correlations of both types of objects. There are several important distinctions between clusters and galaxies. They do not occupy a single $Mg_2$ color sequence. Clusters can be bluer than any galaxy in $15 - V$ and $25 - V$, implying larger proportional hot star populations, but galaxies are significantly bluer than clusters in $15 - 25$ at a given $15 - V$. We attribute this primarily to the effect of metal abundance on the mid-UV (2500 Å) light, because the energy distributions of the turnoff and RGB components in the galaxies are depressed by line blanketing and they contain fewer stars on the intermediate temperature HB, which is bright in the mid-UV. The difference also implies that the UVX in galaxies is not produced by metal poor subpopulations similar to the clusters.

We define "extreme HB" (EHB) stars to be objects with such small $M_{env}^0$s that they do not reach the thermally pulsing stage of the AGB. Although the boundary depends on abundance, objects with $M_{env}^0 \lesssim 0.05 M_\odot$ are usually classified EHB; these have ZAHB temperatures $T_{eff} \gtrsim 16,000$ K. Such objects exist in many of the globular clusters which have been searched carefully (usually in the UV). In the Galactic disk, most of the hot subdwarf sdB stars appear to be EHB stars, and many sdO objects are probably the AGB-manqué progeny of EHB stars; EHB objects constitute about 2% of all field HB stars. EHB stars therefore appear to be normal constituents of both old, metal-poor and intermediate-age, near-solar abundance populations.

We have developed a simple spectral synthesis formulation that requires only one or two parameters for each choice of age and abundance. UV properties of the model grid are predicted using the Kurucz (1991) atmospheres. Maximum lifetime UV output is produced by EHB stars with $M_{env}^0 \sim 0.02 M_\odot$ and can be up to 30 times higher than for post-asymptotic-giant-branch (P-AGB) stars. UV output can change very rapidly with $M_{env}^0$, meaning that it is extremely sensitive to the precise nature of giant branch mass loss. We find that the UV output of post-RGB phases is not a strong function of age or $Z$. Rather, the UV properties of a stellar population are determined primarily by $P(M_{env}^0)$.

Our models for $t = 14$ Gyr accurately predict the range of UV colors observed for the globular clusters, given known constraints on their abundances and HB morphologies. The clusters brightest in the UV are consistent with models with strong IBHB components, which is in accord with their optical color magnitude diagrams. Clusters with such "blue HB" morphologies do not require the hotter EHB stars to explain their UV colors, although a small EHB population is consistent with our models. The largest known population of these stars in a cluster, as a fraction of the total HB, is $\sim 20\%$ in $\omega$ Cen. One cluster with higher metallicity, NGC 6388 has UV colors that are consistent with a $\sim 5 - 10\%$ EHB population. Good measures of UV output in most such metal rich clusters are not available.

By contrast with the clusters, we find that models which do not contain EHB stars cannot reproduce the colors of most of the galaxies, which is largely a consequence of the fact that IBHB stars will be rare at galaxy metallicities. However, the potential far-UV output of an old population is far greater than what is observed, implying that only a small fraction of EHB stars is required, independent of the assumed [Fe/H]. The "moderate UVX" group, with a $15 - V$ color of $3.5 \pm 0.5$, contains most of the B3FL galaxy sample and is consistent with an EHB fraction $\lesssim 5\%$. This agrees roughly with the estimated size of the subdwarf population in the Galactic disk, which considerably strengthens the case for EHB populations as the source of elliptical galaxy UV light. Some systems at the lower end of this range can be modeled with "pure" P-AGB populations. The 4–5 galaxies with the strongest UV upturns require an EHB fraction of $\lesssim 20\%$. This may be a distinct group of systems which do not participate in the generally mild UV color-line strength correlation of the remaining objects.

We find that $25 - V$, but not $15 - V$, is strongly sensitive directly to the age and abundance, though these cannot always be unambiguously distinguished. For [Fe/H] = +0.58, the colors become insensitive to age since the turnoff and giant branch are too cool to contribute much mid-UV flux. The observed galaxy UV colors place strong limits of $\langle$[Fe/H]$\rangle > -0.5$, for any age less than a Hubble time, and $< 15\%$ on the contribution of globular cluster-type populations to the $V$ light. The galaxy colors are all consistent with solar-abundance models with ages in the range 4 – 14 Gyr. However, the $25 - V$ colors of the galaxies other than the strong UVX systems are too blue to be consistent with [Fe/H] $> 0.2$ for any age. Because of its bright mid-UV, the bluest galaxy in the sample, NGC 1399, is best fit at solar metallicity and $t = 6$ Gyr. However, the strong UVX systems have strong Mg$_2$ indices, and this may be additional evidence that [Mg/Fe]$> 0$ in elliptical galaxies. The dwarf elliptical M32 is faintest in the UV and may again be representative of a class of galaxies distinct from most E's. Its $15 - V$ requires models in which the mean P-AGB mass is higher than in the other galaxies, suggestive of a brighter AGB. Its colors are consistent with the solar abundance, intermediate age (4– 6 Gyr) population inferred from optical/IR observations.

The models predict that the fraction of the far-UV light from P-AGB stars, which are spatially resolvable in nearby galaxies, is $\sim 70\%$ for moderate UVX and $\sim 20\%$ for strong UVX systems.

Finally, we discuss several implications of the observations and the models, including the question of light metal vs. iron peak enhancements in galaxies, whether the UV color-Mg$_2$ correlation is continuous or discrete, and the effects of the dynamical environment. We focus on the central obstacle to any attempt to explain the UVX with post-RGB stars, which is the absence of a good physical understanding of giant branch mass loss. We argue that observations of the HB in globular clusters and the Galactic field provide support for the hypothesis that RGB mass loss in metal rich systems is sufficient to produce the EHB populations we need to explain the UVX in galaxies. Alternatives, such as enhanced helium abundance or larger ages, seem less plausible.

We are grateful to Dave Burstein and his collaborators for supplying the LWP spectrum of NGC 1399 in advance of publication. Many thanks also go to Detlef Schönberner for providing his calculations in machine readable form. Thanks also to Jon Whitney and Bob Hill for computing integrated cluster fluxes from UIT images. We would also like to thank Guy Worthey, Rex Saffer, Wayne Landsman, and Ulrich Heber for useful discussions. We made use of the Astrophysics Data System (ADS), the Simbad database operated at CDS, Strasbourg, France, and the NASA Extragalactic Database (NED). This research was supported by NASA Long Term Space Astrophysics Research Program grant NAGW-2596.